\documentclass[aps,prb,amsmath,amssymb,footinbib,showpacs,twocolumn,superscriptaddress,longbibliography]{revtex4-2}
\usepackage{amsmath}
\usepackage{amssymb}
\usepackage{amsthm}
\usepackage{balance}
\usepackage{setspace}
\usepackage{graphicx}
\allowdisplaybreaks[4]
\usepackage{braket}
\usepackage{mathrsfs}
\usepackage{float}
\usepackage[colorlinks = true,linkcolor = red,urlcolor  = blue,citecolor = blue,anchorcolor = blue]{hyperref}
\usepackage[utf8]{inputenc}
\usepackage[english]{babel}
\usepackage{commath}
\usepackage{bm}
\usepackage[caption=false]{subfig}

\begin{document}

\title{Chiral Anomaly Induced Non-linear Transverse Planar Transport Phenomena in Three Dimensional Spin-Orbit Coupled Metals}

\author{Rishi G. Gopalakrishnan}
\affiliation{Department of Physics and Astronomy, Clemson University, Clemson, SC 29634, USA}

\author{Binayyak B. Roy}
\affiliation{Department of Physics and Astronomy, Clemson University, Clemson, SC 29634, USA}

\author{Gargee Sharma}
\affiliation{Department of Physics, Indian Institute of Technology Delhi, Hauz Khas, New Delhi-110016, India}

\author{Sumanta Tewari}
\affiliation{Department of Physics and Astronomy, Clemson University, Clemson, SC 29634, USA}

\begin{abstract}
We investigate the nonlinear transverse planar transport phenomena (viz., nonlinear Hall, thermal Hall, and Nernst coefficients) induced by chiral anomaly in three-dimensional spin-orbit coupled metallic systems. Unlike Weyl semimetals, these systems do not possess multiple Weyl nodes located at isolated points in the momentum space but instead host a pair of Fermi surfaces characterized by opposite Berry curvature fluxes enclosing the same band-degeneracy point. Using semiclassical Boltzmann transport formalism within the relaxation time approximation, we derive the second-order transverse planar transport coefficients induced by electrical and thermal gradients in the presence of an in-plane magnetic field. Our analysis reveals distinctive angular dependencies of the non-linear transport coefficients, along with characteristic scaling behavior with the magnetic field strength. Furthermore, we demonstrate that the anomaly-induced transport coefficients exhibit an exponential temperature dependence. This unconventional behavior leads to the violations of the Wiedemann-Franz law and Mott relation, highlighting unique thermoelectric signatures that can be probed experimentally in 3D spin-orbit coupled metallic systems.
 \end{abstract}

\maketitle

\section{Introduction}
Recent advancements in condensed matter physics have demonstrated that certain classes of materials, known as topological materials, can host exotic quasiparticles whose properties resemble those predicted in high-energy physics~\cite{Hasan2010, Qi2011, Armitage2018}. These systems are characterized by topologically protected band degeneracies and nontrivial electronic structures, giving rise to phenomena analogous to relativistic effects in particle physics. Examples include topological insulators, which possess gapless edge or surface states protected by time-reversal symmetry~\cite{Hasan2010, Qi2011}; Weyl semimetals, featuring pairs of topologically stable Weyl nodes that act as sources or sinks of Berry curvature in momentum space~\cite{ Armitage2018, Jia:2016wal, PhysRevB.83.205101, Burkov2014,Bauer_2024, Peskin:1995ev, PhysRevLett.107.186806, PhysRevB.84.235126, PhysRevLett.107.127205, PhysRevLett.108.266802, PhysRevLett.108.046602, Saha2018}; Dirac semimetals, characterized by fourfold degenerate nodes stabilized by symmetries like rotation or inversion~\cite{Young2012, Wang2012, Yangnature, PhysRevB.107.L241101, Wieder2020, RevModPhys.81.109, Sato2011, Liu_2014, Jeon_2014, PhysRevB.89.235127, PhysRevB.96.195119}; and topological superconductors, promising candidates for realizing Majorana fermions with potential applications in fault-tolerant quantum computing~\cite{Fu2008, Sato_2017, PhysRevB.110.115436, Kitaev2003, Nayak2008, sau2010generic, sau2021topological}.

The intriguing correspondence between high-energy and condensed matter physics has not only deepened our understanding of fundamental physics but has also opened pathways to realizing phenomena once thought exclusive to particle physics experiments \cite{doi:10.1080/01422419808240874, Schober_2024, PhysRevD.22.3080}. For instance, phenomena such as the chiral anomaly, originally discussed in the context of quantum field theory \cite{jackiw2008axial, claude1980quantum}, have now been realized experimentally in Weyl and Dirac semimetals \cite{huang2015observation, wang2016gate, dos2016search, arnold2016negative, wu2018probing, deng2019quantum}; such realizations allow condensed matter systems to serve as experimental platforms to probe quantum anomalies and their associated transport signatures \cite{nielsen1983adler, aji2012adler, son2013chiral, parameswaran2014probing, burkov2015negative, ong2021experimental}.
Chiral anomaly refers to the non-conservation of chiral charge in the presence of parallel electric and magnetic fields (\(\mathbf{E} \cdot \mathbf{B} \neq 0\)). In Weyl semimetals, this leads to distinct experimental signatures such as negative longitudinal magnetoresistance and the planar Hall and Nernst effects~\cite{volovik2003universe,xu2011chern,zyuzin2012weyl,son2013chiral,goswami2013axionic,goswami2015axial,zhong2015optical,kim2014boltzmann,lundgren2014thermoelectric,cortijo2016linear,zyuzin2017magnetotransport,kundu2020magnetotransport,knoll2020negative,bednik2020magnetotransport,he2014quantum,liang2015ultrahigh,zhang2016signatures,li2016chiral,xiong2015evidence,hirschberger2016chiral, Sharma2019, Wang2022, Arouca_2022, PhysRevB.108.L161106, Ong_2021, LI2016107,nandy2017chiral,sharma2016nernst,sharma2017nernst,sharma2023decoupling,ahmad2021longitudinal,ahmad2023longitudinal,PhysRevResearch.2.033511,sharma2017chiral,sharma2020sign,ahmad2021longitudinal,Sharma2019, ahmad2021longitudinal,ahmad2023longitudinal,sharma2016nernst,sharma2017chiral,sharma2017nernst,sharma2020sign,sharma2023decoupling,nandy2017chiral,ahmad2023longitudinal,ahmad2021longitudinal,goswami2015optical,PhysRevResearch.2.033511,PhysRevResearch.2.013088,ahmad2024geometry,ahmad2025chiral}. The presence of chiral anomaly in these materials can be intuitively understood in terms of the Berry curvature monopoles associated with Weyl nodes, acting effectively as sources and sinks of Berry flux in momentum space~\cite{Armitage2018, Jia:2016wal}. While initially explored theoretically, these Chiral anomaly induced transport effects have now been robustly confirmed experimentally \cite{PhysRevX.5.031023, zhang2016signatures}, highlighting their practical significance as probes of the fundamental topological character in semimetallic systems.

In this paper, we extend the analysis beyond Weyl semimetals to investigate anomaly-induced transport in three-dimensional metallic systems characterized by strong spin-orbit coupling (SOC) \cite{Gao_2022, PhysRevB.105.L180303, PhysRevB.108.045405, varma2024magnetotransport, ahmad2024nonlinearanomaloushalleffect}. Unlike Weyl semimetals, these systems do not host pairs of separated Weyl nodes (see Fig.~\ref{Fermi_surface}); instead, the system studied here features a single degeneracy point accompanied by distinct Fermi surfaces, each characterized by nonzero Berry curvature flux, while the net flux integrated over all Fermi surfaces vanishes, as expected from the Nielsen-Ninomiya theorem~\cite{NIELSEN1981219, NIELSEN1983389}. Recently, it was demonstrated that such systems can support both the chiral anomaly and its thermal analog, the gravitational chiral anomaly, under parallel electric (or thermal gradient) and magnetic fields~\cite{Gao_2022, PhysRevB.105.L180303, PhysRevB.108.045405, varma2024magnetotransport}.

While the linear responses were studied earlier in Ref.~\cite{PhysRevB.108.045405}, here we extend the framework to derive the corresponding nonlinear response in the same physical setting. In particular, we demonstrate the existence of a mixed thermoelectric response that arises when electric and thermal driving fields are applied simultaneously (see Eqs.(\ref{quad_response}) and (\ref{quadQ_response})).Our principal contribution lies in the nonlinear transport regime, where we identify experimentally testable predictions (Eqs. (\ref{2.2.6}), (\ref{2.2.7}) and (\ref{2.2.8})). These include the angular dependence with respect to the applied magnetic field relative to the driving electric field and thermal gradient (Fig.~\ref{sig_quad}, top), and the scaling of the nonlinear transport coefficients with the (in-plane) magnetic field strength (Fig.~\ref{sig_quad}, bottom). Heatmaps of the chiral-anomaly-induced nonlinear electric and thermoelectric transverse transport coefficients are shown in Figs.~(5,6). 

\begin{figure}[t]
\centering
\subfloat{%
    \includegraphics[scale=0.10]{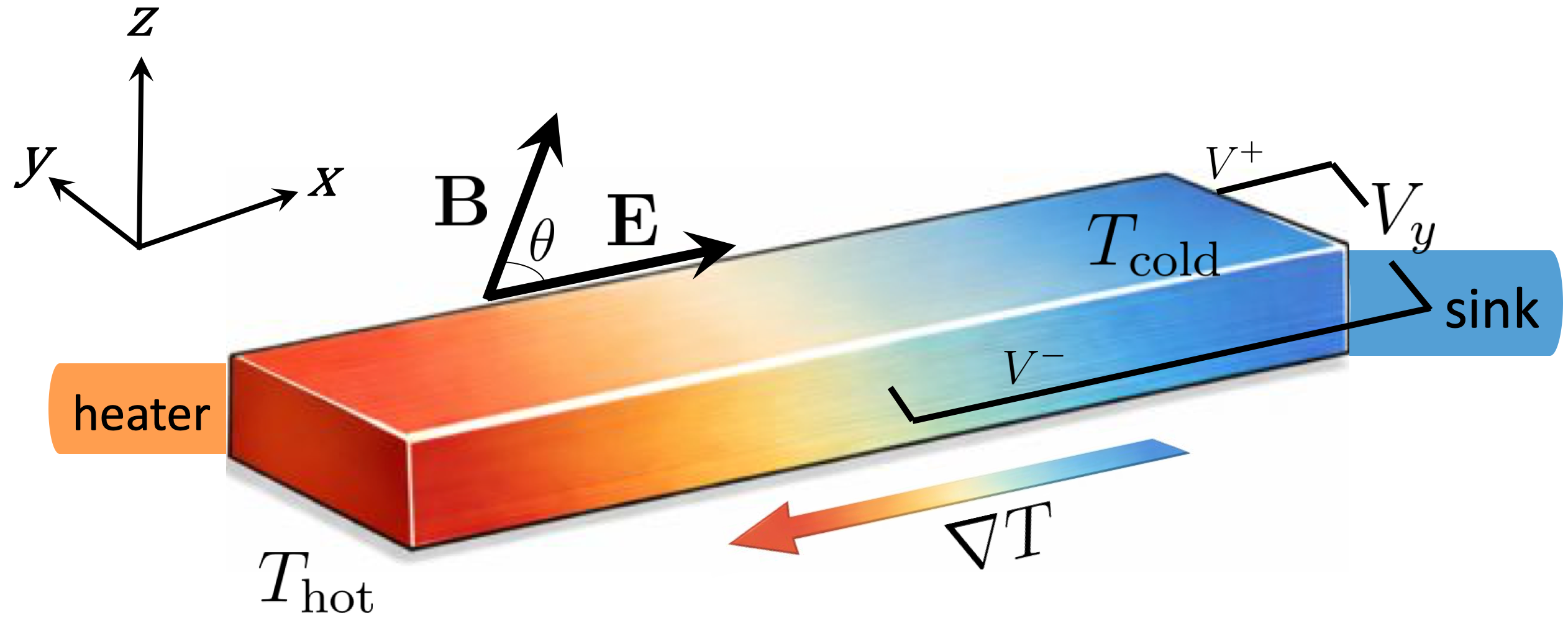}%
}

\caption{Schematic of experimental geometry to measure planar transverse responses generated by an in-plane magnetic field ($\bm{B}$) in the presence of a thermal gradient ($\bm{\nabla} T$) and/or an electric field ($\bm{E}$).}
\label{Schematic}
\end{figure}

To derive the relevant transport properties, we employ the semiclassical Boltzmann transport formalism \cite{ashcroft1976solid, de1984non, 492de033-74af-3791-807c-34957550404d, Girvin_Yang_2019} and examine the behavior in the low temperature limit at both linear and second order in electric field/ thermal gradient. We further investigate distinctive characteristics of these transport coefficients by examining their dependence on $\theta$ (the angle between the applied magnetic field and the electric field/ thermal gradient). Specifically, we show that for angles below $\pi/2$, the charge and thermal transport coefficients attain an extremum value at $\theta = \arctan(1/\sqrt{2})$ (refer to Fig.\ref{sig_quad} top and the surrounding discussion in section IV B, page 8). This angle is smaller than the corresponding extremum angle for the linear counterpart that occurs at $\theta = \pi/4$ (refer to Fig.\ref{sig_linear} top and the surrounding discussion in section IV B, page 6). The dependence on magnetic field strength also changes, scaling as $B^{3}$ at quadratic order as opposed to $B^{2}$ in the linear regime. Recent studies 
showed that chiral anomaly induced second order transport coefficients, under the Sommerfeld expansion, obey the nonlinear analogs of the Wiedemann-Franz Law and Mott relations \cite{PhysRevResearch.2.032066, PhysRevB.105.125131}.  We will show in this paper, that such relations hold in our system at very low temperatures, but quickly break down as temperature is increased as polynomial approximations to the underlying energy integrals are rendered invalid.
Moreover, we show that in the quadratic regime the mixed thermoelectric response coefficients cannot be related to the other nonlinear transport coefficients by a simple Wiedemann-Franz or Mott type relation.

\begin{figure}[t]
\centering
\includegraphics[scale=0.30]{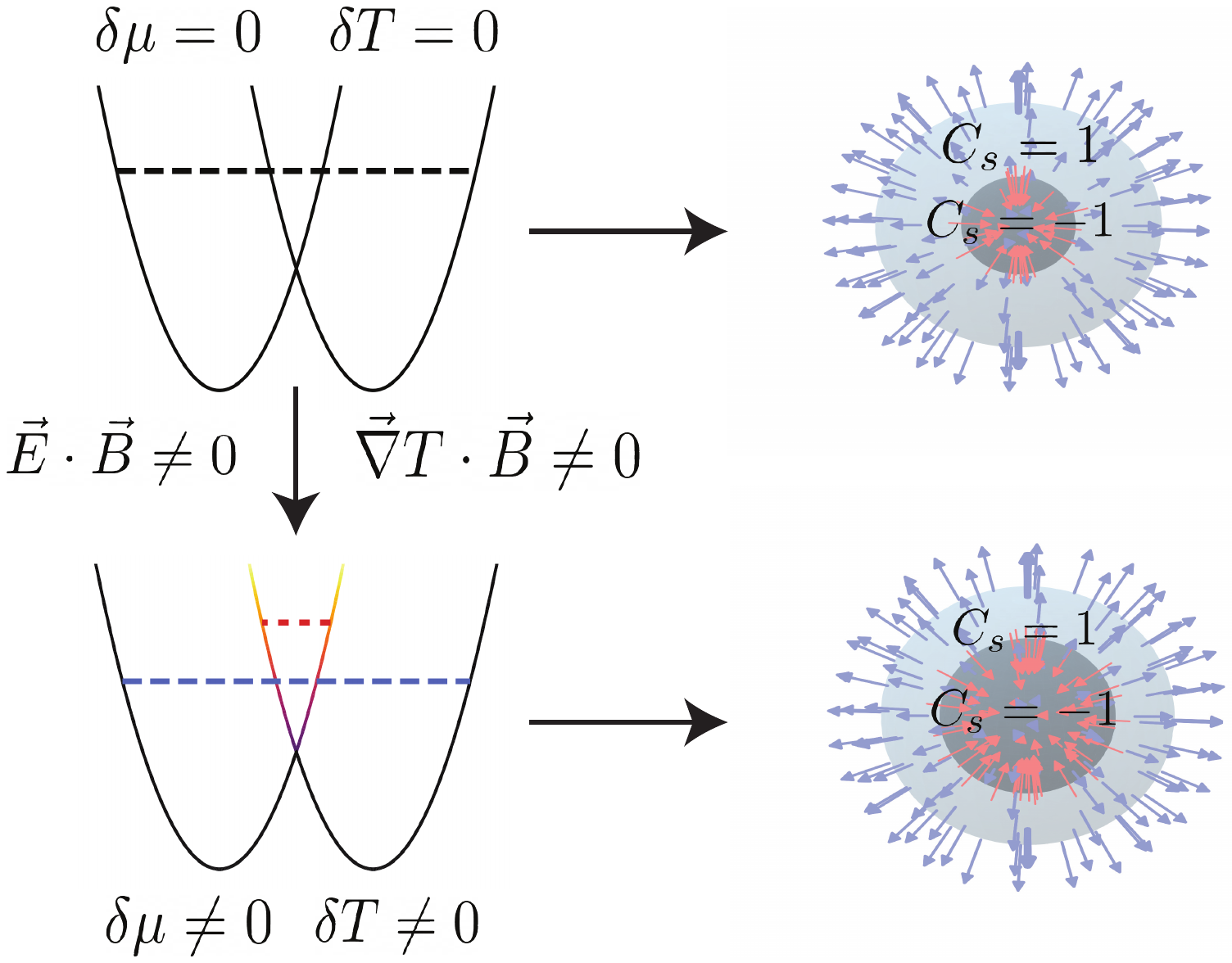}\
\captionsetup{font={stretch=1.2}}
\caption{Band structure (left) and corresponding Fermi surfaces (right) for each chirality $C_s$. The arrows indicate the direction of the Berry curvature at any given point on the respective Fermi surfaces. The red (s = 1) band corresponds to the inner Fermi surface, while the blue (s = -1) band corresponds to the outer Fermi surface. With the application of a nonzero $\bm{E}\cdot\bm{B}$ or $\bm{\nabla}T\cdot\bm{B}$ term, a chemical potential ($\delta\mu$) and temperature ($\delta T$) imbalance are created, resulting in a nonzero transverse voltage $V_H$ (see Fig.\ref{Schematic}).}
\label{Fermi_surface}
\end{figure}

The paper is organized as follows: In Section II, we introduce and derive the semiclassical Boltzmann transport equations used in our analysis. In Section III, we employ these equations to demonstrate the emergence of the chiral anomaly and to derive both linear and nonlinear anomaly induced planar transverse transport coefficients. In Section IV we apply our theoretical framework to 3D spin-orbit coupled metallic systems, illustrating the presence of distinctive transport signatures of the chiral anomaly in such systems. Finally, in Section V, we summarize our key findings and provide concluding remarks. Additional details and calculations supporting our results are provided in the Appendix.

\section{Semiclassical Charge Transport in Boltzmann Transport Equations}

In the presence of a nonzero Berry curvature as well as electric and magnetic fields, the semiclassical motion of Bloch electrons acquires anomalous terms.  Their dynamics are described by the equations of motion\,\cite{RevModPhys.82.1959}
\begin{eqnarray}
\dot{\bm r}^{s} &=& D\Bigl[\bm v^{s}
  + \frac{e}{\hbar}\,\bm E \times \bm\Omega^{s}
  + \frac{e}{\hbar}(\bm v^{s}\!\cdot\!\bm\Omega^{s})\,\bm B\Bigr],\label{1.2}\\
\dot{\bm k}^{s} &=& D\Bigl[-\frac{e}{\hbar}\,\bm E
  - \frac{e}{\hbar}\,\bm v^{s}\times\bm B
  + \frac{e^{2}}{\hbar^{2}}(\bm E\!\cdot\!\bm B)\,\Omega^{s}\Bigr],\label{1.3}
\end{eqnarray}
where $\bm\Omega^{s}$ is the Berry curvature (with band/chirality index $s=\pm1$) and  
$(D^{s})^{-1}=1+\frac{e}{\hbar}\,(\bm B\!\cdot\!\bm\Omega^{s})$
is the modified phase–space Jacobian, which reduces to unity when $\bm\Omega^{s}=0$.  
The band velocity is $\bm v^{s}=(1/\hbar)\nabla_{k}\epsilon_{k}$.  
The second term in Eq.\,\eqref{1.2} is the anomalous Hall velocity, and the last term, proportional to $(\mathbf{v}^s\cdot\mathbf{\boldsymbol{\Omega}}^s)\mathbf{B }$, is known as the \emph{chiral magnetic velocity}\,.  Although it can generate charge currents without external driving, such contributions cancel between opposite chiralities when no electric field or thermal gradient is applied, leaving the net equilibrium current zero. 

When an external electric field or a temperature gradient is applied, the non-equilibrium distribution $f^{s}_{k}(\bm r,\bm k,t)$ evolves according to\,\cite{PhysRevResearch.2.013088}
\begin{eqnarray}
\Bigl(\partial_{t} + \dot{\bm r}\!\cdot\!\nabla_{\!r}
      + \dot{\bm k}\!\cdot\!\nabla_{\!k}\Bigr)f^{s}_{k}
  &=& -\frac{f^{s}_{k}-f_{eq}(\epsilon^{s},\mu^{s},T^{s})}{\tau_{0}}\nonumber\\
  &&{}-\frac{f^{s}_{k}-f_{eq}(\epsilon^{s},\mu^{\bar s},T^{\bar s})}{\tau_{\nu}},\label{1.1}
\end{eqnarray}
where $\tau_{0}$ and $\tau_{\nu}$ are the intra- and inter-chiral center relaxation times.
Including both scattering channels is essential for reaching a steady state when charge is pumped between the two chiral centers.  
In the relaxation-time picture, the first term drives the chiral carriers toward the local equilibrium of the \emph{same} chiral center, while the second relaxes them toward the equilibrium distribution of the \emph{opposite} center. 

Substituting Eqs.\,\eqref{1.2} and \eqref{1.3} into the kinetic equation \eqref{1.1} and retaining only the term proportional to $(\bm v^{s}\!\cdot\!\bm\Omega^{s})\bm B$, the part responsible for the axial (chiral) anomaly, we obtain
\begin{eqnarray}
&&D^s\Bigl(\bm v^{s}
   +\frac{e}{\hbar}(\bm v^{s}\!\cdot\!\bm\Omega^{s})\bm B\Bigr)
   \!\cdot\!\Bigl[e\bm E+\Bigl(\frac{\epsilon^{s}-\mu}{T}\Bigr)\nabla T\Bigr]
   \Bigl(-\frac{\partial f^{s}_{k}}{\partial\epsilon^{s}}\Bigr)\nonumber\\
&&\qquad
  =-\frac{\delta f^{s}_{k}}{\tau^{*}}
   -\frac{f_{eq}(\epsilon^{s},\mu^{s},T^{s})
          -f_{eq}(\epsilon^{s},\mu^{\bar s},T^{\bar s})}{\tau_{\nu}},\label{1.4}
\end{eqnarray}
where $\tau^{*-1}= \tau_{0}^{-1}+\tau_{\nu}^{-1}$. In deriving Eq.\,\eqref{1.4} we have omitted cross-product terms such as $\bm E\times\bm\Omega^{s}$ and $\bm v^{s}\times\bm B$ in Eqs.\,\eqref{1.2} and \eqref{1.3} respectively, because they do not contribute to chiral anomaly. As a consistency check, setting $\bm E=\nabla T=0$ forces $\delta f^{s}_{k}\to0$ since the chemical potential and temperature difference between the chiral centers vanishes in the absence of the chiral anomaly, thereby recovering the Fermi–Dirac equilibrium distribution $f_{eq}=1/\bigl(e^{\beta(\epsilon_{k}-\mu)}+1\bigr)$. Even in the absence of $\bm E$ or $\nabla T$, a uniform magnetic field drives \emph{equilibrium} charge and heat currents at each chiral center through the chiral magnetic velocity in Eq.\,\eqref{1.2}:
\begin{eqnarray}
\bm j^{s}_{e,eq}=-e\bigl(\mu\,C^{0}_{s}+\,C^{1}_{s}\bigr)\bm B,\label{3.5}
\end{eqnarray}
\begin{eqnarray}
\bm j^{s}_{Q,eq}=\bigl(\frac{\mu^2}{2}\,C^{0}_{s}+\,\mu C^{1}_{s} + C^2_s\bigr)\bm B,\label{3.5}
\end{eqnarray}

\noindent where the Berry flux moments $C^{\nu}_{s}$ are defined in Eq.\,\eqref{non2.1.9}.  Because the two chiralities contribute equal and opposite values, the net equilibrium current cancels. 




When there is a nonzero $\bm{E}$ or (and) $\nabla T$ satisfying $\bm E\parallel\bm B$ or (and) $\nabla T\parallel\bm B$, the anomaly pumps charge and energy between the two chiral centers without changing the total carrier number.  This is known as chiral anomaly \cite{ Schober_2024, PhysRevD.22.3080, jackiw2008axial}. In condensed matter systems, it leads to imbalances in the chemical potential ($\mu^{s}$) and temperature ($T^{s}$) between the chiral centers. These imbalances scale linearly with $\bm{E}\cdot B$ (chiral anomaly) and $\bm{\nabla}T\cdot B$ (gravitational anomaly). We denote the deviations from equilibrium chemical potential and temperature on the chiral center $s$ by $\delta\tilde\mu^{s}$ and $\delta\tilde T^{s}$, respectively.  

\noindent We define the average deviation for each chiral center as
\[
\delta\mu^{s}=\frac{\delta\tilde\mu^{s}-\delta\tilde\mu^{\bar s}}{2},\qquad
\delta T^{s}=\frac{\delta\tilde T^{s}-\delta\tilde T^{\bar s}}{2}.
\]
This formulation is convenient, as it ensures that the deviations in chemical potential and temperature for the two chiral centers are equal in magnitude and opposite in sign. To linear order in $\bm E$ and $\nabla T$, $\delta\mu^s$ and $\delta T^s$ are determined using the transport equation (Eq. \ref{1.4})
\begin{eqnarray}
\delta\mu^{s}&=&-\frac{\tau_{\nu}}{2}\left(\frac{\mathcal{D}^{1}_{s}\mathcal{D}^{2}_{s}}
                                 {\mathcal{D}^{0}_{s}\mathcal{D}^{2}_{s}-(\mathcal{D}^{1}_{s})^{2}}\right)
\Bigl[\Bigl(\tfrac{\bm\Lambda^{0}_{s}}{\mathcal{D}^{1}_{s}}
            -\tfrac{\bm\Lambda^{1}_{s}}{\mathcal{D}^{2}_{s}}\Bigr)\!\cdot\!e\bm E \nonumber\\
&&\qquad\qquad\quad
  +\Bigl(\tfrac{\bm\Lambda^{1}_{s}}{\mathcal{D}^{1}_{s}}
          -\tfrac{\bm\Lambda^{2}_{s}}{\mathcal{D}^{2}_{s}}\Bigr)\!\cdot\!
    \tfrac{\nabla T}{T}\Bigr],\label{1.5}
\\
\frac{\delta T^{s}}{T}&=&-\frac{\tau_{\nu}}{2}\left(\frac{\mathcal{D}^{0}_{s}\mathcal{D}^{1}_{s}}
                                       {\mathcal{D}^{0}_{s}\mathcal{D}^{2}_{s}-(\mathcal{D}^{1}_{s})^{2}}\right)
\Bigl[\Bigl(\tfrac{\bm\Lambda^{1}_{s}}{\mathcal{D}^{1}_{s}}
            -\tfrac{\bm\Lambda^{0}_{s}}{\mathcal{D}^{0}_{s}}\Bigr)\!\cdot\!e\bm E \nonumber\\
&&\qquad\qquad\quad
  +\Bigl(\tfrac{\bm\Lambda^{2}_{s}}{\mathcal{D}^{1}_{s}}
          -\tfrac{\bm\Lambda^{1}_{s}}{\mathcal{D}^{0}_{s}}\Bigr)\!\cdot\!
    \tfrac{\nabla T}{T}\Bigr]\label{1.6}
\end{eqnarray}

\balance

\begin{widetext}

\noindent Using equations (\ref{1.5}) and (\ref{1.6}), we obtain the second-order corrections,

\begin{align}
f^{(2)}_k =& +(\tau^*)^2D^2\left[\bm{v}^s.\left(e\bm{E} + \left(\frac{\epsilon^s - \mu}{T}\right)\bm{\nabla} T\right)\right] \times \left(\bm{v}^s.\left(\frac{\bm{\nabla} T}{T}\right)\right)\left(\frac{\partial f_{eq}}{\partial\epsilon^s}\right) \nonumber \\
&+ (\tau^*)^2D^2\left[\bm{v}^s.\left(e\bm{E} + \left(\frac{\epsilon^s - \mu}{T}\right)\bm{\nabla} T\right)\right]^2\left(\frac{\partial^2 f_{eq}}{\partial\epsilon^{s2}}\right) \nonumber
\displaybreak[4]\\
&+ 2\left(\frac{(\tau^*)^2}{\tau_\nu}\right)D\left[\left(\frac{\delta T^s}{T}\right)\left(\frac{\partial f_{eq}}{\partial\epsilon^s}\right) \right.\left. + \left(\delta\mu^s + \left(\frac{\epsilon^s - \mu}{T}\right)\delta T^s\right)\left(\frac{\partial^2 f_{eq}}{\partial\epsilon^{s2}}\right)\right] \nonumber \\
&-2\left(\frac{\tau^*}{\tau_\nu}\right)\left[\left(\frac{\partial^2 f_{eq}}{\partial\epsilon^{s2}}\right)(\delta\mu^s)^2 + \frac{2}{T}\left(\left(\frac{\partial f_{eq}}{\partial\epsilon^{s}}\right) \right.\right. \left.\left.+ (\epsilon^s - \mu)\left(\frac{\partial^2 f_{eq}}{\partial\epsilon^{s2}}\right)\right)(\delta\mu^s)(\delta T^s)\right] \nonumber \\
&-2\left(\frac{\tau^*}{\tau_\nu}\right)\left[\frac{2}{T^2}(\epsilon^s - \mu)\left(\frac{\partial f_{eq}}{\partial\epsilon^{s}}\right) \right. \left. + \frac{1}{T^2}(\epsilon^s - \mu)^2\left(\frac{\partial^2 f_{eq}}{\partial\epsilon^{s2}}\right)\right](\delta T^s)^2
\label{A13}
\end{align}

For more details, refer Appendix \ref{appendix: Appendix A}. Here $D^{\nu}_{s}$ and $\bm\Lambda^{\nu}_{s}$ are generalized density and energy–velocity moments:
\begin{eqnarray}
\mathcal{D}^{\nu}_{s}=\int\!\frac{d^{3}k}{(2\pi)^{3}}\,
              (D^s)^{-1}\Bigl(-\frac{\partial f_{eq}}{\partial\epsilon}\Bigr)
              (\epsilon^{s}-\mu)^{\nu},\label{1.8}\\
\bm\Lambda^{\nu}_{s}=\int\!\frac{d^{3}k}{(2\pi)^{3}}
  \Bigl(-\frac{\partial f_{eq}}{\partial\epsilon}\Bigr)
  \left[\bm v^{s}+\frac{e}{\hbar}(\bm v^{s}\!\cdot\!\bm\Omega^{s})\bm B\right]
  (\epsilon^{s}-\mu)^{\nu}
=&\int\!\frac{d^{3}k}{(2\pi)^{3}}
  \Bigl(-\frac{\partial f_{eq}}{\partial\epsilon}\Bigr)
  \frac{e}{\hbar}(\bm v^{s}\!\cdot\!\bm\Omega^{s})\bm B
  (\epsilon^{s}-\mu)^{\nu},
  \label{1.7}
\end{eqnarray}

\end{widetext}

\noindent The second equality of Eq.\,\eqref{1.7} follows from the fact that the angular average of $\bm v^{s}$ vanishes.  To extract transport coefficients we expand the non-equilibrium distribution function in powers of the the electric field and thermal gradient,

 
\begin{equation}
\delta f^{s}_{k}\equiv f^{s}_{k}-f_{eq}= \sum_{n\ge1}f^{(n)}_{k},
\label{notation}
\end{equation}
where $f^{(n)}_{k}$ collects all terms of $n$th order in $\bm E$ and $\nabla T$, including every permutation.  We work in the \emph{chiral limit} \cite{PhysRevResearch.2.013088}, defined by $\tau_{\nu}\gg\tau_{0}\simeq\tau^{*}$, which ensures that the leading corrections to the transport coefficients arise solely from anomaly induced charge pumping. Expanding the inter-chiral center collision term on the right-hand side of Eq.\,\eqref{1.4} to first and second order then yields

\begin{eqnarray}
f^{(1)}_{k}&=&-2\frac{\tau^{*}}{\tau_{\nu}}
\frac{\partial f_{eq}}{\partial\epsilon^{s}}
\Bigl[\delta\mu^{s}+\frac{\epsilon^{s}-\mu}{T}\,\delta T^{s}\Bigr]
\label{1.9}
\end{eqnarray}

\noindent In the chiral limit, where $\tau_\nu \gg \tau^*$, we have,

\begin{eqnarray} 
f^{(2)}_{k}&=&-2\frac{\tau^{*}}{\tau_{\nu}}\Bigl\{
  (\delta\mu^{s})^{2}\frac{\partial^{2}f_{eq}}{\partial\epsilon^{s2}}
 +\frac{2\delta\mu^{s}\delta T^{s}}{T}
  \Bigl[\frac{\partial f_{eq}}{\partial\epsilon^{s}}\nonumber
       +\\ && (\epsilon^{s}-\mu)\frac{\partial^{2}f_{eq}}{\partial\epsilon^{s2}}\Bigr] +\frac{(\delta T^{s})^{2}}{T^{2}}
  \Bigl[2(\epsilon^{s}-\mu)\frac{\partial f_{eq}}{\partial\epsilon^{s}}\nonumber \\
       &&+(\epsilon^{s}-\mu)^{2}\frac{\partial^{2}f_{eq}}{\partial\epsilon^{s2}}\Bigr]\Bigr\}.
\label{1.10}
\end{eqnarray}

A complete derivation is provided in Appendix\,\ref{appendix: Appendix A}.  We omit orbital magnetic moment contributions, as they are negligible for Chiral anomaly driven nonlinear responses\,\cite{PhysRevB.103.045105,PhysRevB.104.205124,PhysRevB.108.045405}.  The next section discusses a model Hamiltonian and employs Eqs.\,\eqref{1.5}–\eqref{1.10} to obtain the quadratic planar response coefficients.

\section{Three-Dimensional Spin-Orbit Coupled Systems} \label{Chiral_calc}

In this section, we illustrate the general theoretical framework by applying them explicitly to a three-dimensional spin-orbit coupled metallic system. Using a model Hamiltonian, we demonstrate the emergence of chiral anomaly driven linear and nonlinear planar transport coefficients, and make several experimentally testable predictions for such systems.

\subsection{The Hamiltonian}

We model the electrons with the following effective spin–orbit coupled Hamiltonian \cite{He_2021, PhysRevB.91.134401, PhysRevB.78.144511},

\begin{eqnarray}
H = \int \frac{d^3\bm{k}}{(2\pi)^3}c_{\bm{k}}^{\dagger}\left[\frac{\hbar^2 k^2}{2m}\,\sigma_{0} \;+\;\alpha\,\bm{\sigma}\!\cdot\!\bm{k}\right]c_{\bm{k}}
\label{3.1}
\end{eqnarray}

\noindent Here, $\sigma_{0}$ is the $2\times2$ identity matrix, the $\sigma_{i}$ are the Pauli matrices, and $c_{\bm{k}}^{\dagger}=(c_{\bm{k},\uparrow}^{\dagger},c_{\bm{k},\downarrow}^{\dagger})$.  Although we have adopted a simple isotropic form for the spin–orbit interaction, the chiral‐anomaly and planar Hall/Nernst arguments presented above remain valid for more elaborate, anisotropic spin–orbit terms \cite{PhysRevB.108.045405}. The corresponding energy eigenvalues are
\begin{eqnarray}
\epsilon^{s}(k)=\frac{\hbar^{2}k^{2}}{2m}+s\,\alpha\,k,
\label{3.2}
\end{eqnarray}
where \(s=\pm1\) labels the two bands, and $k = \norm{\bm{k}}$. From Eq. \eqref{3.2}, we notice that the two bands become degenerate at \(\epsilon^{s}=0\). This is also the minimum energy for the \(s=+1\) (upper) band.  The \(s=-1\) (lower) band reaches its minimum energy at $-\epsilon_{\alpha}$, where
\(\epsilon_{\alpha}=m\alpha^{2}/2\hbar^{2}\).  We focus on the regime \(\mu\ge0\), giving each band its own Fermi surface at chemical potential \(\mu\).  The Chern number for each Fermi surface is
\begin{eqnarray}
C_{s}=\iint_{\text{FS}}dS\!\cdot\!\bm\Omega^{s}=-s,
\label{3.3}
\end{eqnarray}
\noindent The surface integral in Eq.\,\eqref{3.3} is taken over the Fermi surface of band \(s\).  For this model the Berry curvature is
\begin{eqnarray}
\bm{\Omega}_{s}(k)=-\frac{s}{2k^{3}}\,\bm{k},
\label{3.4}
\end{eqnarray}

\noindent It is instructive to contrast this system with a Weyl semimetal (WSM).  In a WSM, pairs of Weyl nodes with opposite chirality occur at distinct points in momentum space, directly linking nodal location to chirality.  By comparison, in three-dimensional spin–orbit coupled metals the two chiral sectors originate from the same momentum space node yet form separate Fermi surfaces, so that chirality coincides with the band index.  If instead \(\mu<0\), these Fermi pockets correspond only to the Fermi surface for a single band, but hosting two disjoint closed surface, one with a positive band velocity, while the other has a negative band velocity; consequently $\bm{v}^{s}\cdot\bm{\Omega}^{s}$ changes sign between the surfaces, and the net Berry flux [Eq.\,\eqref{3.3}] vanishes. Nevertheless, one can partition the band Fermi surface into two disjoint closed regions, each carrying its own nonzero Berry flux and hence a well defined chirality \cite{PhysRevB.108.045405}, and compute the flux on each region.  This procedure is identical to the separate band calculation, except that the band index no longer serves as the chirality label.  By doing so, one finds that even for \(\mu<0\) the Fermi surface contains pockets of opposite chirality, preserving the chiral anomaly.  From this viewpoint, chirality can no longer be universally associated with band index.  For the remainder of this paper we restrict to \(\mu\ge0\). However, for \(\mu<0\) by replacing band indices with pocket labels (corresponding to each disjoint surface), we find two chiral centers, hence the discussion made in the previous section applies to this case as well. Comparing the Hamiltonian for our system (Eq. (\ref{3.1})) with the Hamiltonian for the Weyl semimetal \cite{PhysRevResearch.2.013088}, we note that they differ in the $|\bf{k}|^2$ term proportional to $\sigma_0$. As a result of this, the difference in their transport properties (Eqs. (\ref{2.1.2})-(\ref{kappalin}), (\ref{2.2.6})-(\ref{lyxx})) arises from the higher momentum modes. This gives rise to differences in their scaling with chemical potential and temperature (See Appendix\,\ref{appendix: Appendix B}). 

\subsection{Transport Properties}

\begin{figure}[t]
\hspace*{-0.2 cm}
\subfloat{%
    \includegraphics[scale=0.4]{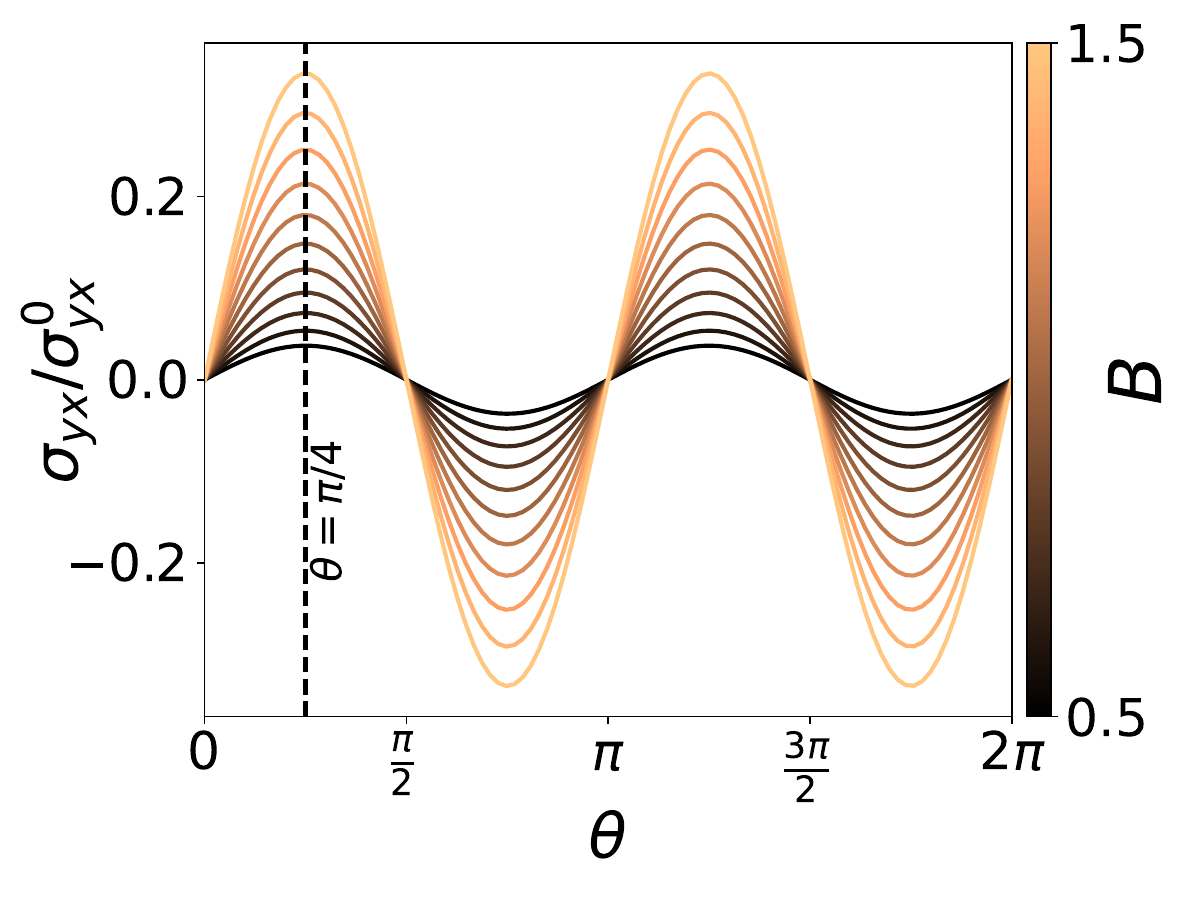}%
}
\hspace{1 cm}
\subfloat{%
    \includegraphics[scale=0.4]{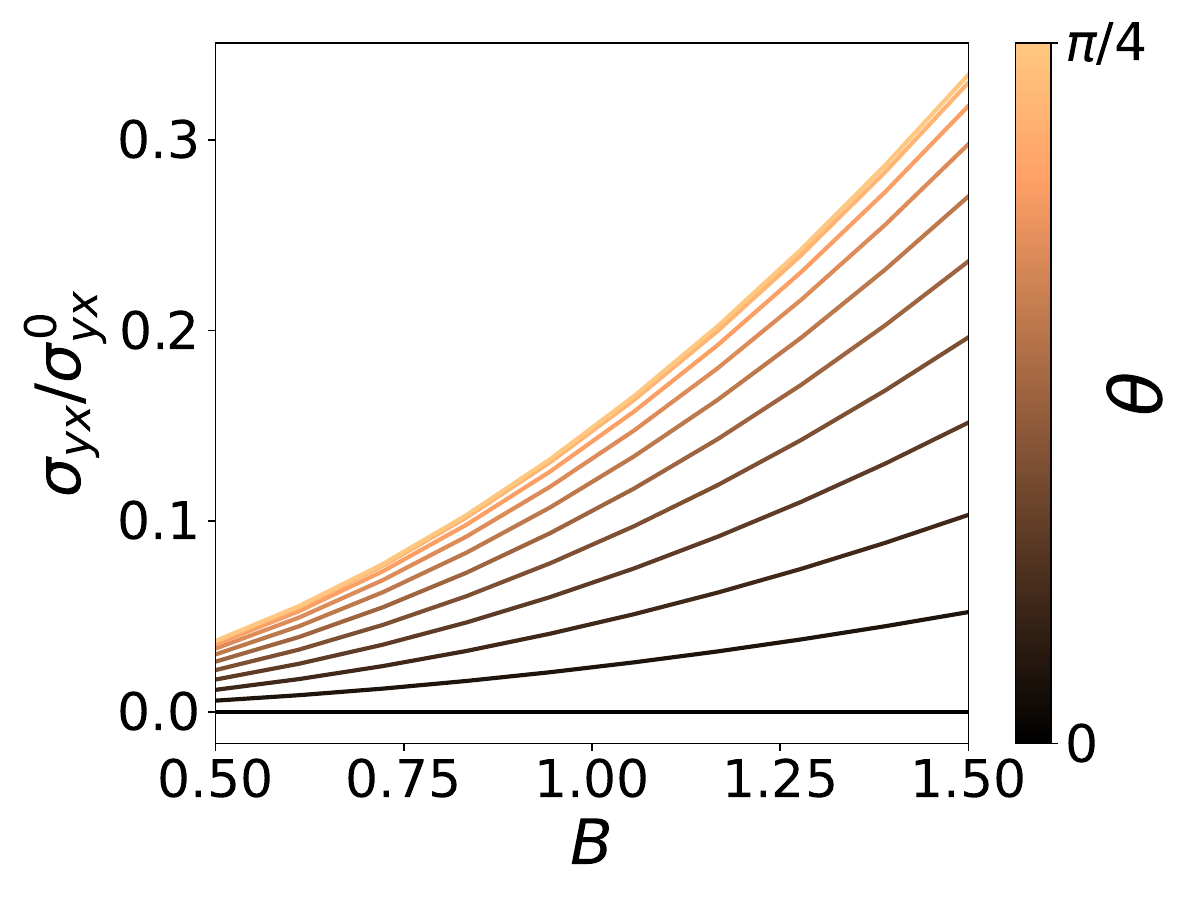}%
}\caption{Normalized planar Hall conductivity (see Eq.(\ref{2.1.2})) $\sigma_{yx}/\sigma_{yx}^0$ (with $\sigma^0_{yx} = \frac{\tau^* e^4}{8 \sqrt{2} \pi^2 \hbar m^{3/2} \epsilon_\alpha^{1/2}}$) as a function of (top) the angle $\theta$ between the magnetic field and current for field strengths $0.5\le B\le1.5$ (color scale), and (bottom) the field magnitude $B$ for angles $0\le\theta\le\pi/4$ (color scale). We can notice that the maximum value for the field strength occurs when $\theta = \pi/4$. Here, we have chosen $T = 30K$, $\mu = 50*\epsilon_\alpha$.}
\label{sig_linear}
\end{figure}

We now derive the charge and heat currents arising from each chiral center in the presence of an applied electric field and thermal gradient. 

Starting from the semiclassical equations of motion in the presence of Berry curvature, the charge and heat currents for carriers of chirality \(s\) is
\begin{eqnarray}
 \{\bm{j}^{s}_{e}, \bm{j}^s_Q\}=\int[dk]\,(D^s)^{-1}\{-e,(\epsilon_s-\mu)\}\dot{\bm r}\,\delta f^{s}_{k},
\end{eqnarray}
where \([dk]\equiv d^{3}k/(2\pi)^{3}\).  Employing Eq.\,\eqref{notation} we rewrite this as
\begin{eqnarray}
\{\bm{j}^{s}_{e}, \bm{j}^s_Q\}
  &=&\int[dk]\,(D^s)^{-1}\{-e,(\epsilon_s-\mu)\}\dot{\bm r}\bigl(f^{s}_{k}-f^{s}_{eq}\bigr)\nonumber\\
  &=&\int[dk]\sum_{n}(D^s)^{-1}\{-e,(\epsilon_s-\mu)\}\dot{\bm r}\,f^{(n)}_{k}.
\end{eqnarray}
Hence the currents decompose naturally into their linear and higher‑order parts.  The $n$‑th order current in the driving fields is given by
\begin{eqnarray}
\{\bm{j}^{s(n)}_{e}, \bm{j}^{s(n)}_Q\}
  =\int[dk]\,(D^s)^{-1}\{-e,(\epsilon_s-\mu)\}\dot{\bm r}\,f^{(n)}_{k}.
\label{2.1}
\end{eqnarray}
Eq.\,\eqref{2.1} will be the basis for evaluating the linear and nonlinear transport coefficients. The linear order analysis for these equations are presented in Appendix\,\ref{appendix: Appendix C} \cite{PhysRevB.108.045405}.



\begin{figure}[t]
    \hspace*{-0.3 cm}
    \subfloat{%
        \includegraphics[scale=0.4]{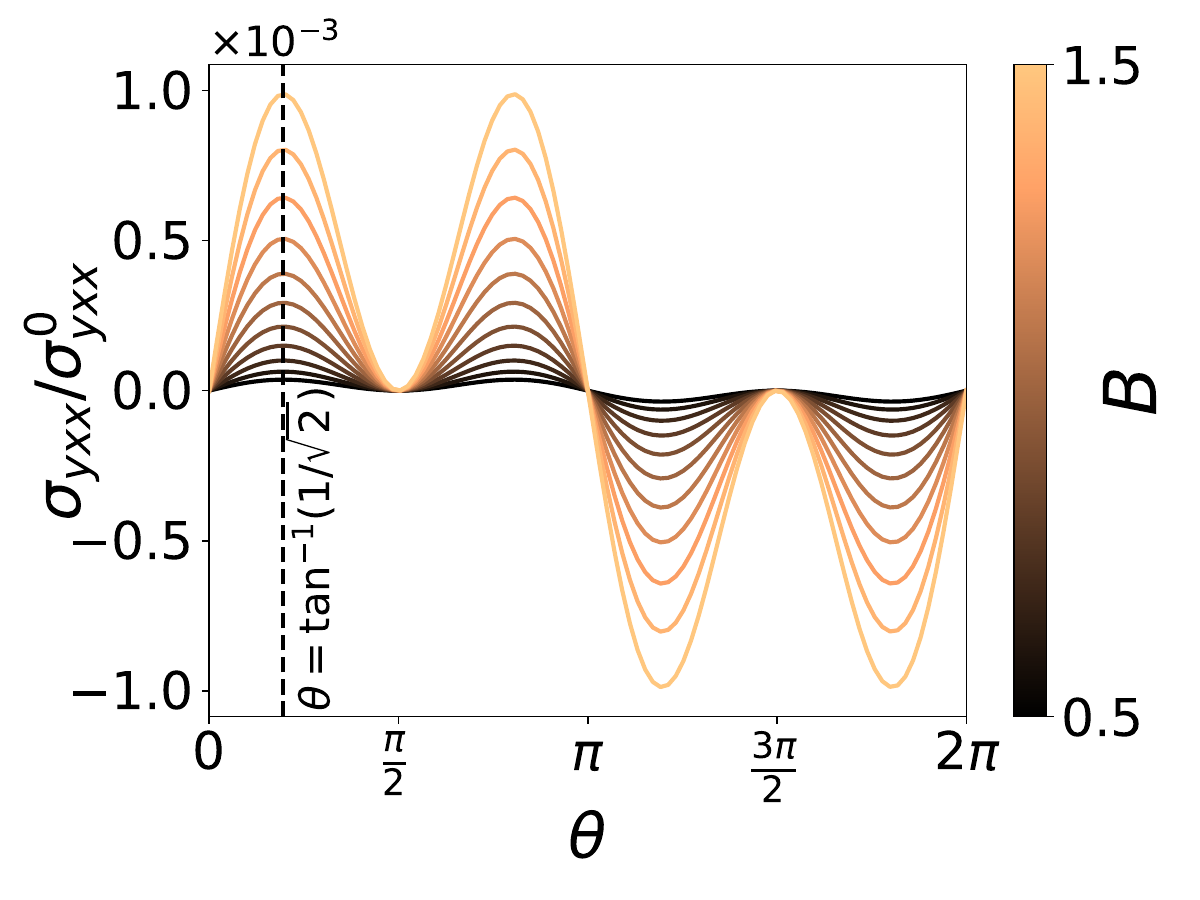}%
    }
    \hspace{0.5 cm}
    \subfloat{%
        \includegraphics[scale=0.4]{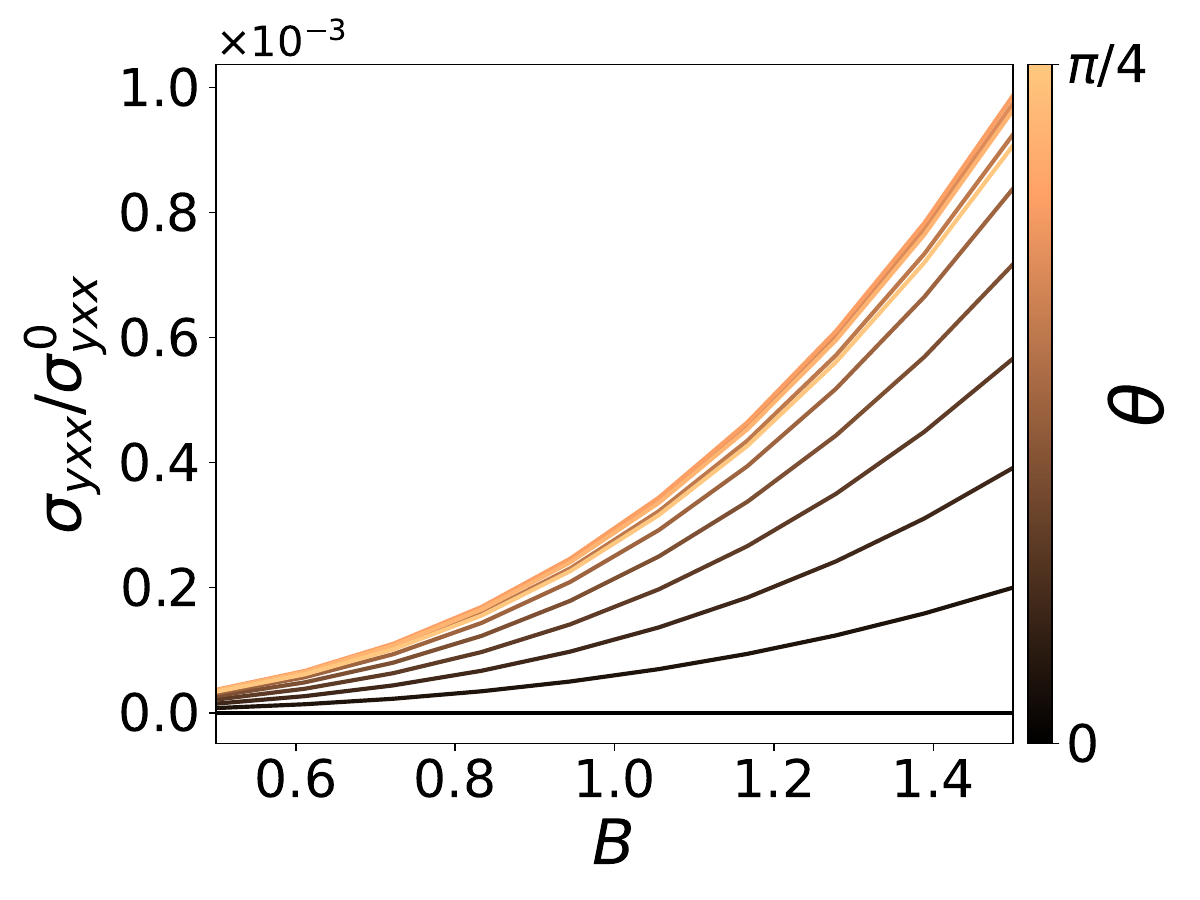}%
    }
    \caption{Non‑linear normalized planar Hall conductivity (see Eq.(\ref{2.2.6})) $\sigma_{yxx}/\sigma_{yxx}^{0}$ (with $\sigma^0_{yxx} = \frac{\tau_0 \tau^* e^6}{32 \pi^2 m^3 \epsilon_\alpha^2}$) as a function of (top) the in‑plane field angle $\theta$ for $0.5\le B\le1.5$ (color bar), with the linear‑response peak at $\theta=\pi/4$ indicated by a dashed line; and (bottom) the field magnitude $B$ for $0\le\theta\le\pi/4$ (color bar). Unlike the linear response (Fig.\ref{sig_linear}), the non‑linear conductivity reaches its maximum at $\theta = \tan^{-1}\!\bigl(1/\sqrt2\bigr)\,<\,\pi/4$. Here, we have chosen $T = 30K$, $\mu =50*\epsilon_\alpha$.}
    \label{sig_quad}
\end{figure}
The second order response to external perturbations is given by 
\begin{eqnarray}
\{\bm{j}^{s(2)}_{e}, \bm{j}^{s(2)}_Q\}
  =\int[dk]\,(D^s)^{-1}\{-e,(\epsilon_s-\mu)\}\dot{\bm r}\,f^{(2)}_{k},
\label{2.2.1}
\end{eqnarray}
where \(f^{(2)}_{k}\) is the quadratic correction obtained in Eq.\,\eqref{1.10}.
\noindent In analogy with Eqs.\,\eqref{1.7} and \eqref{1.8} we introduce
\begin{eqnarray}
\Sigma^{\nu}_{s}
  =\int[dk]\;\frac{\partial^{2}f_{eq}}{\partial\epsilon^{s2}}\,
     \frac{e}{\hbar}(\bm v^{s}\!\cdot\!\bm\Omega^{s})
     (\epsilon^{s}-\mu)^{\nu},
\label{2.2.3}
\end{eqnarray}
which allows Eq.\,\eqref{2.2.1} to be written compactly as
\begin{eqnarray}
\bm j^{(2)}_{e}
  &=&\frac{2eB\tau^{*}}{\tau_{\nu}}
    \Bigl[(\delta\mu^{s})^{2}\Sigma^{0}_{s}
         +2\delta\mu^{s}\frac{\delta T^{s}}{T}\bigl(C^{0}_{s}+\Sigma^{1}_{s}\bigr) \nonumber\\
  &&\hspace*{6.5em}
         +\Bigl(\frac{\delta T^{s}}{T}\Bigr)^{2}
          \bigl(2C^{1}_{s}+\Sigma^{2}_{s}\bigr)\Bigr]
\label{2.2.4}
\end{eqnarray}
\begin{eqnarray}
\bm j^{(2)}_{Q}
  &=&-\frac{2B\tau^{*}}{\tau_{\nu}}
    \Bigl[(\delta\mu^{s})^{2}\Sigma^{1}_{s}
         +2\delta\mu^{s}\frac{\delta T^{s}}{T}\bigl(C^{1}_{s}+\Sigma^{2}_{s}\bigr) \nonumber\\
  &&\hspace*{6.5em}
         +\Bigl(\frac{\delta T^{s}}{T}\Bigr)^{2}
          \bigl(2C^{2}_{s}+\Sigma^{3}_{s}\bigr)\Bigr]
\label{jQ}
\end{eqnarray}
The quadratic electric and thermal responses can be parameterized as  
\begin{eqnarray}
\bm j^{(2)}_{ea}
   =\sum_{b,c}\bigl[\sigma_{abc}E_{b}E_{c}
                    +\alpha_{abc}\nabla_{b}T\nabla_{c}T
                    -\beta_{abc}E_{b}\nabla_{c}T\bigr].
\label{quad_response}                    
\end{eqnarray}
\begin{eqnarray}
\bm j^{(2)}_{Qa}
   =\sum_{b,c}\bigl[\bar\sigma_{abc}E_{b}E_{c}
                    +l_{abc}\nabla_{b}T\nabla_{c}T
                    -\bar\beta_{abc}E_{b}\nabla_{c}T\bigr].
\label{quadQ_response}                    
\end{eqnarray}
These equations serve as the definitions for the various nonlinear charge and thermal response coefficients. In the case of coplanar Magnetic and applied driving fields, these coefficients include the nonlinear planar Hall coefficient ($\sigma_{yxx}$), the nonlinear planar Nernst coefficient ($\alpha_{yxx}$), the planar mixed thermoelectric coefficient ($\beta_{yxx}$), the planar nonlinear Ettingshausen coefficient ($\bar{\sigma}_{yxx}$), the planar nonlinear thermal Hall coefficient ($l_{yxx}$), and the planar thermal mixed thermoelectric coefficient ($\bar{\beta}_{yxx}$). We find that in addition to the purely electric and thermal gradient responses, in the presence of both an electric field and a thermal gradient, the chiral‑anomaly mechanism generates the additional response \(\beta_{abc}\), which encodes a mixed thermoelectric response to the charge current that appears only when an electric field and a thermal gradient act simultaneously. Similarly, in the presence of both driving fields (electric field and thermal gradient) we also find a mixed thermoelectric response tensor \(\bar\beta_{abc}\) that contributes to the heat current. Combining the nonlinear currents in Eqs.\,\eqref{2.2.4} and \,\eqref{jQ} with the corrections \(\delta\mu^{s}\) and \(\delta T^{s}\) from Eqs.\,\eqref{1.5}–\eqref{1.6} yields the second‑order planar transport tensors

\begin{eqnarray}
\sigma_{yxx}
 &=&\frac{\tau^{*}\tau_{\nu}}{2}\,e^{3}B^{3}\cos^{2}\theta\sin\theta
     \sum_{s}\!\Bigl[\Sigma^{0}_{s}(S^{12})^{2}(R^{01}_{12})^{2}
\nonumber\\
 &&\qquad\qquad
    +(2C^{1}_{s}+\Sigma^{2}_{s})(S^{01})^{2}(R^{10}_{10})^{2}
\nonumber\\
 &&\qquad\qquad
+2(C^{0}_{s}+\Sigma^{1}_{s})S^{12}S^{01}R^{01}_{12}R^{10}_{10}\Bigr],\label{2.2.6} \\[6pt]
\alpha_{yxx}
 &=&\frac{\tau^{*}\tau_{\nu}}{2}\frac{e}{T^{2}}B^{3}\cos^{2}\theta\sin\theta
     \sum_{s}\!\Bigl[\Sigma^{0}_{s}(S^{12})^{2}(R^{12}_{12})^{2}
\nonumber\\
 &&\qquad\qquad
    +(2C^{1}_{s}+\Sigma^{2}_{s})(S^{01})^{2}(R^{21}_{10})^{2}
\nonumber\\
 &&\qquad\qquad
    +2(C^{0}_{s}+\Sigma^{1}_{s})S^{12}S^{01}R^{12}_{12}R^{21}_{10}\Bigr],\label{2.2.7}
\end{eqnarray}

\begin{eqnarray}
\beta_{yxx}
 &=&-\frac{\tau^{*}\tau_{\nu}}{2}\frac{e^{2}}{T}B^{3}\cos^{2}\theta\sin\theta
     \sum_{s}\!\Bigl[\Sigma^{0}_{s}(S^{12})^{2}R^{01}_{12}R^{12}_{12}
\nonumber\\
 &&\qquad\qquad
    +(2C^{1}_{s}+\Sigma^{2}_{s})(S^{01})^{2}R^{10}_{10}R^{21}_{10}
\nonumber\\
    &&\hspace*{4.3em}
    +2(C^{0}_{s}+\Sigma^{1}_{s})S^{12}S^{01} \nonumber\\
 &&\hspace*{4.5em}\times\bigl(R^{01}_{12}R^{21}_{10}+R^{10}_{10}R^{12}_{12}\bigr)\Bigr]\label{2.2.8}
\end{eqnarray}

\begin{eqnarray}
\bar{\sigma}_{yxx}
 &=&\frac{\tau^{*}\tau_{\nu}}{2}e^{2}B^{3}\cos^{2}\theta\sin\theta
     \sum_{s}\!\Bigl[\Sigma^{1}_{s}(S^{12})^{2}(R^{01}_{12})^{2}
\nonumber\\
 &&\qquad\qquad
    +(2C^{2}_{s}+\Sigma^{3}_{s})(S^{01})^{2}(R^{10}_{10})^{2}
\nonumber\\
    &&\hspace*{4.3em}
    +2(C^{1}_{s}+\Sigma^{2}_{s})S^{12}S^{01}R^{01}_{12}R^{10}_{10}\Bigr]\label{sigmabaryxx}
\end{eqnarray}

\begin{eqnarray}
\bar{\beta}_{yxx}
 &=&\frac{\tau^{*}\tau_{\nu}}{2}\frac{e}{T}B^{3}\cos^{2}\theta\sin\theta
     \sum_{s}\!\Bigl[\Sigma^{1}_{s}(S^{12})^{2}R^{01}_{12}R^{12}_{12}
\nonumber\\
 &&\qquad\qquad
    +(2C^{2}_{s}+\Sigma^{3}_{s})(S^{01})^{2}R^{10}_{10}R^{21}_{10}
\nonumber\\
    &&\hspace*{4.3em}
    +2(C^{1}_{s}+\Sigma^{2}_{s})S^{12}S^{01} \nonumber\\
 &&\hspace*{6.5em}\times\bigl(R^{01}_{12}R^{21}_{10}+R^{10}_{10}R^{12}_{12}\bigr)\Bigr]\label{betabaryxx}
\end{eqnarray}

\begin{eqnarray}
    l_{yxx}
 &=&-\frac{\tau^{*}\tau_{\nu}}{2}\frac{1}{T^{2}}B^{3}\cos^{2}\theta\sin\theta
     \sum_{s}\!\Bigl[\Sigma^{1}_{s}(S^{12})^{2}(R^{12}_{12})^{2}
\nonumber\\
 &&\qquad\qquad
    +(2C^{2}_{s}+\Sigma^{3}_{s})(S^{01})^{2}(R^{21}_{10})^{2}
\nonumber\\
 &&\qquad\qquad
+2(C^{1}_{s}+\Sigma^{2}_{s})S^{12}S^{01}R^{12}_{12}R^{21}_{10}\Bigr],\label{lyxx}
\end{eqnarray}

where
\begin{eqnarray}
R^{\alpha\beta,s}_{\gamma\delta}
   &=&\frac{C^{\alpha}_{s}}{\mathcal{D}^{\gamma}_{s}}
      -\frac{C^{\beta}_{s}}{\mathcal{D}^{\delta}_{s}},\label{2.2.5}\\
S^{\alpha\beta,s}&=&\frac{\mathcal{D}^{\alpha}_{s}\mathcal{D}^{\beta}_{s}}
                         {\mathcal{D}^{0}_{s}\mathcal{D}^{2}_{s}+(\mathcal{D}^{1}_{s})^{2}},\label{R_ten}\\
C^{\nu}_{s}
  &=&\int[dk]\Bigl(-\frac{\partial f_{eq}}{\partial\epsilon^{s}}\Bigr)
     \frac{e}{\hbar}(\bm v^{s}\!\cdot\!\bm\Omega^{s})
     (\epsilon^{s}-\mu)^{\nu}.\label{non2.1.9}
\end{eqnarray}

\noindent The above nonlinear transverse planar transport coefficients scale as \(B^{3}\) and have a \(\sin\theta\cos^2\theta\) angular dependence. Consequently they exhibit an extrema at \(\theta = \arctan(1/\sqrt{2})\) [Fig.(\ref{sig_quad})]. This angle is smaller than extremum angle for the linear order case. This indicates that higher-order corrections, in $\bm{E}$ and $\bm{\nabla}T$, to the transverse planar transport coefficients reach their maximum at angles closer to the configuration where $\bm{E}$ and $\bm{\nabla}T$ are aligned with $\bm{B}$. We have not shown the plots for the other nonlinear planar transport coefficients because they have an identical angular and magnetic field strength scaling (see Eqs.(\ref{2.2.6}) - (\ref{lyxx})). We now proceed to compute the transverse planar transport coefficients driving the charge and heat currents arising from the chiral anomaly in three-dimensional spin orbit coupled metals. 

\begin{figure*}[t]
    \centering
    \subfloat{\includegraphics[width=0.3\linewidth,height=0.28\textheight,keepaspectratio]{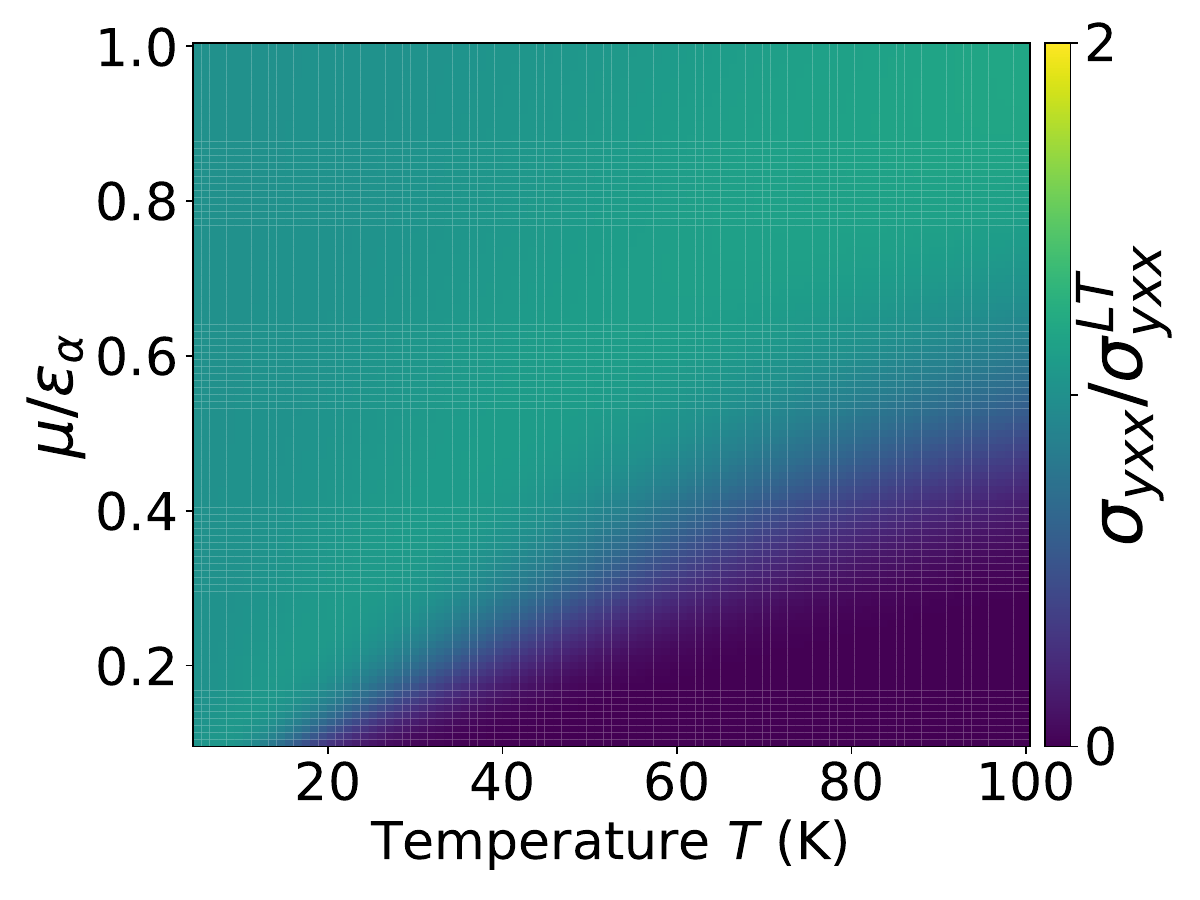}\label{fig:1}}
    \subfloat{\includegraphics[width=0.3\linewidth,height=0.28\textheight,keepaspectratio]{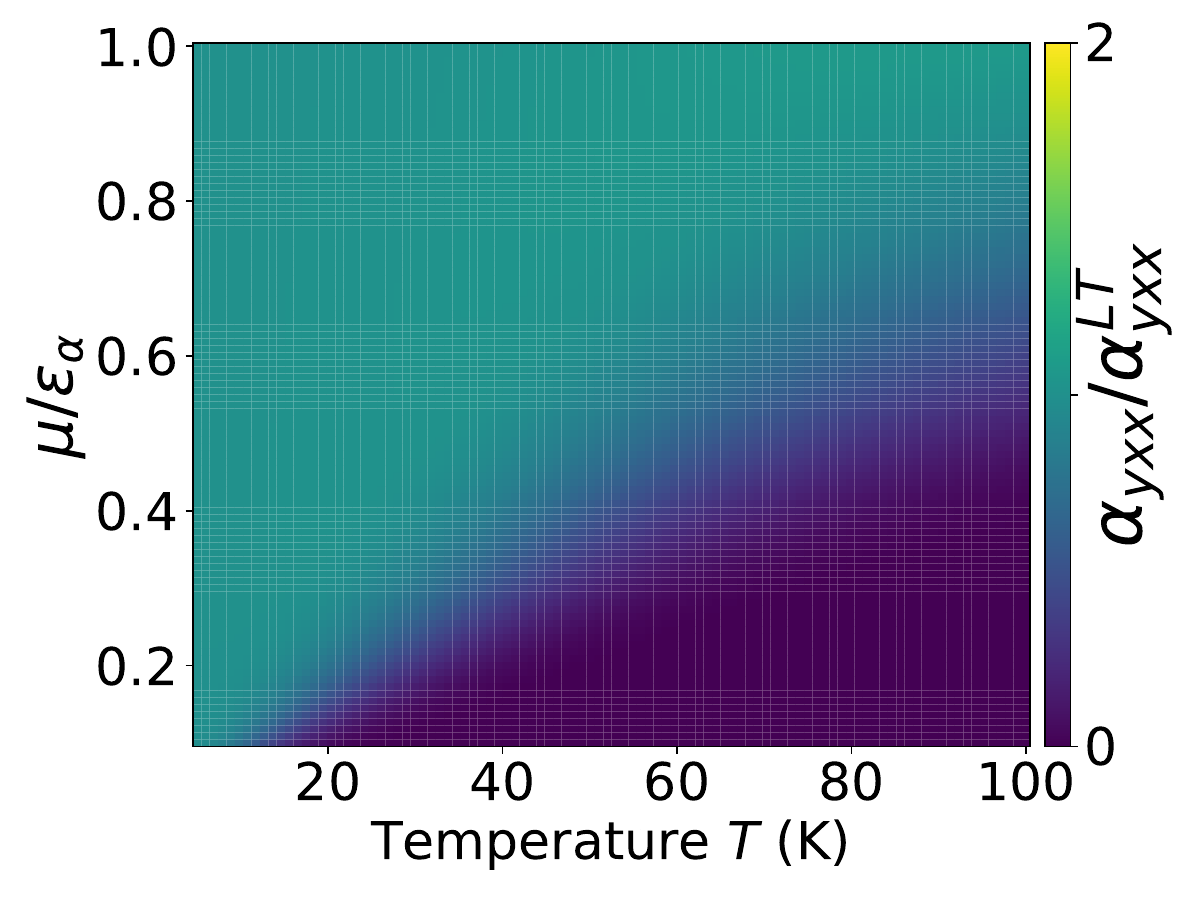}\label{fig:2}}
    \subfloat{\includegraphics[width=0.3\linewidth,height=0.28\textheight,keepaspectratio]{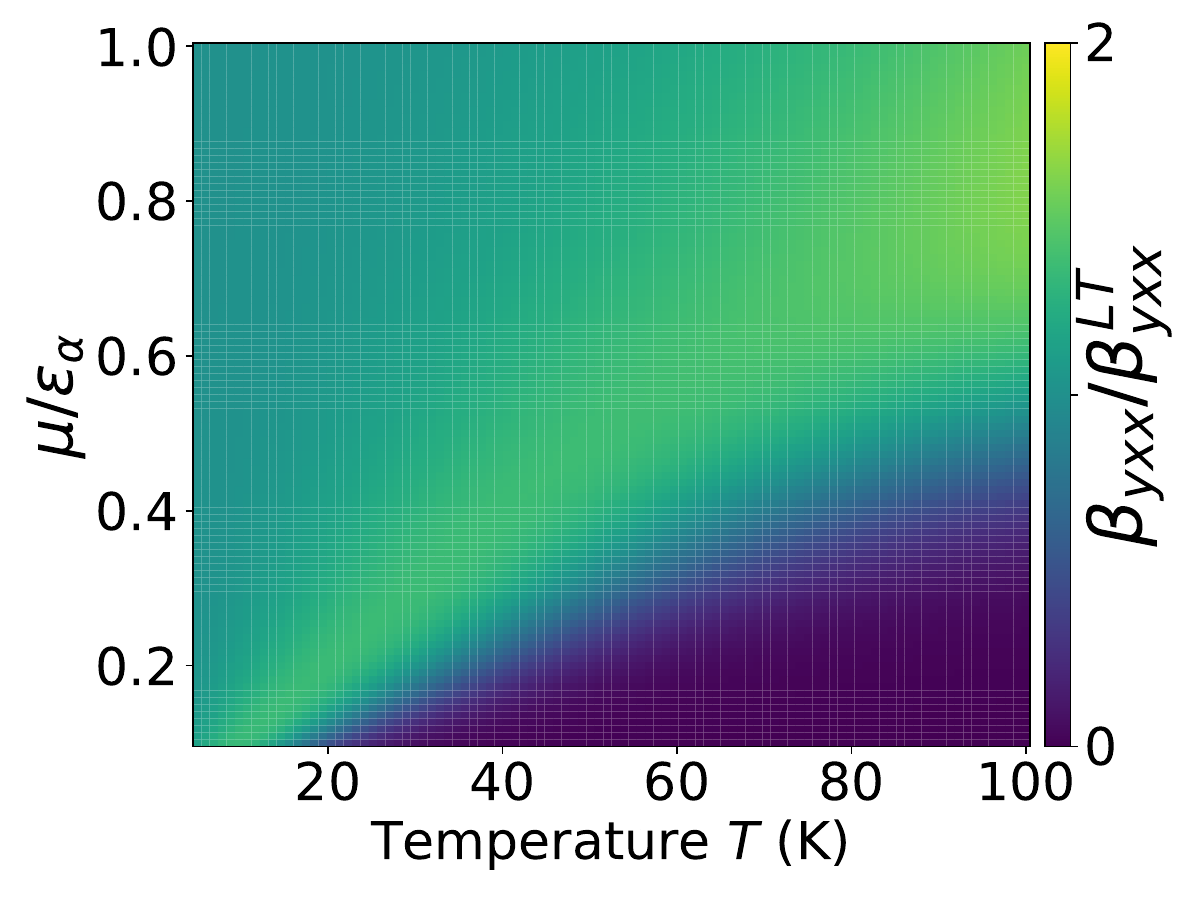}\label{fig:3}}
    \caption{Heatmaps showcasing the ratios of the nonlinear electric transverse response coefficients ($\sigma_{yxx}/\sigma_{yxx}^{LT}$, $\alpha_{yxx}/\alpha_{yxx}^{LT}$, and $\beta_{yxx}/\beta_{yxx}^{LT}$) to their low temperature forms as a function of temperature and chemical potential. A ratio of unity (sea-green color) denotes exact agreement with the low-T approximation. Left: Displays the breakdown of the low temperature limit of the nonlinear planar Hall conductivity (coefficient $\sigma_{yxx}$. See Eq.(\ref{2.2.6}) for $\sigma_{yxx}$, and Eq.(\ref{3.17}) for $\sigma_{yxx}^{LT}$). Center: Displays the breakdown of the low temperature limit of the planar Nernst coefficient (coefficient $\alpha_{yxx}$. See Eq.(\ref{2.2.7}) for $\alpha_{yxx}$, and Eq.(\ref{3.18}) for $\alpha_{yxx}^{LT}$). Right: Displays the breakdown of the low temperature limit of the transverse planar mixed thermoelectric transport coefficient (coefficient $\beta_{yxx}$. See Eq.(\ref{2.2.8}) for $\beta_{yxx}$, and Eq.(\ref{3.19}) for $\beta_{yxx}^{LT}$).} Unity (sea green color) indicates exact agreement with the analytic low‑$T$ form, while deviations (yellow to purple) from unity mark the breakdown of the expansion at low $\mu/\varepsilon_{\alpha}$ and elevated $T$. Here, we have chosen $B = 1T$, $\theta = \pi/4$.
    \label{Heatmaps_electric}
\end{figure*}

Below we present the analytic expressions for the fundamental moments entering the transport coefficients for our system (see Eq.(\ref{3.1})), as defined in Eqs.\,\eqref{2.2.3} and \eqref{non2.1.9}.

\begin{eqnarray}
[C^0_s, C^1_s, C^2_s] = -\frac{se}{4\pi^2\hbar^2}[F_0, \frac{1}{\beta}F_1, \frac{1}{\beta^2}F_2]
\label{3.6}
\end{eqnarray}
\begin{eqnarray}
[\Sigma^0_s, \Sigma^1_s, \Sigma^2_s, \Sigma^3_s] = -\frac{se}{4\pi^2\hbar^2}[G_0, \frac{1}{\beta}G_1, \frac{1}{\beta^2}G_2, \frac{1}{\beta^3}G_3]
\label{3.7}
\end{eqnarray}

where,
\begin{eqnarray}
F^0(x_s) = \frac{1}{1+e^{-x_s}}
\label{3.8}
\end{eqnarray}
\begin{eqnarray}
F^1(x_s) = \frac{x_s}{1+e^{x_s}} + \ln(1+e^{-x_s})
\label{3.9}
\end{eqnarray}
\begin{eqnarray}
F^2(x_s) =&& \frac{\pi^2}{3} - x(\frac{x_s}{1+e^{x_s}} + 2\ln(1+e^{-x_s})) \nonumber \\
&&+ 2Li_2(-e^{-x_s})
\label{3.10}
\end{eqnarray}
\begin{eqnarray}
G^0(x_s) = \frac{\beta e^{-x_s}}{(1+e^{-x_s})^2}
\label{3.11}
\end{eqnarray}
\begin{eqnarray}
G^1(x_s) =  -\frac{\beta x_se^{-x_s}}{(1+e^{-x_s})^2} + \frac{\beta}{1+e^{-x_s}}
\label{3.12}
\end{eqnarray}
\begin{eqnarray}
G^2(x_s) = && \frac{\beta x^2_se^{-x_s}}{(1+e^{-x_s})^2} + \frac{2\beta x_s}{1+e^{x_s}} \nonumber \\ && + 2\beta \ln(1+e^{-x_s})
\label{3.13}
\end{eqnarray}
\begin{eqnarray}
G^3(x_s) = &&\pi^2\beta + \frac{\beta x^3_se^{-x_s}}{(1+e^{-x_s})^2} - \frac{3\beta x_s^2}{1+e^{x_s}} \nonumber \\ &&+6\beta Li_2(-e^{-x_s}) - 6\beta x_s \ln(1+e^{-x_s})
\label{G^3}
\end{eqnarray}

\begin{figure*}[t]
    \centering
    \subfloat{\includegraphics[width=0.3\linewidth,height=0.28\textheight,keepaspectratio]{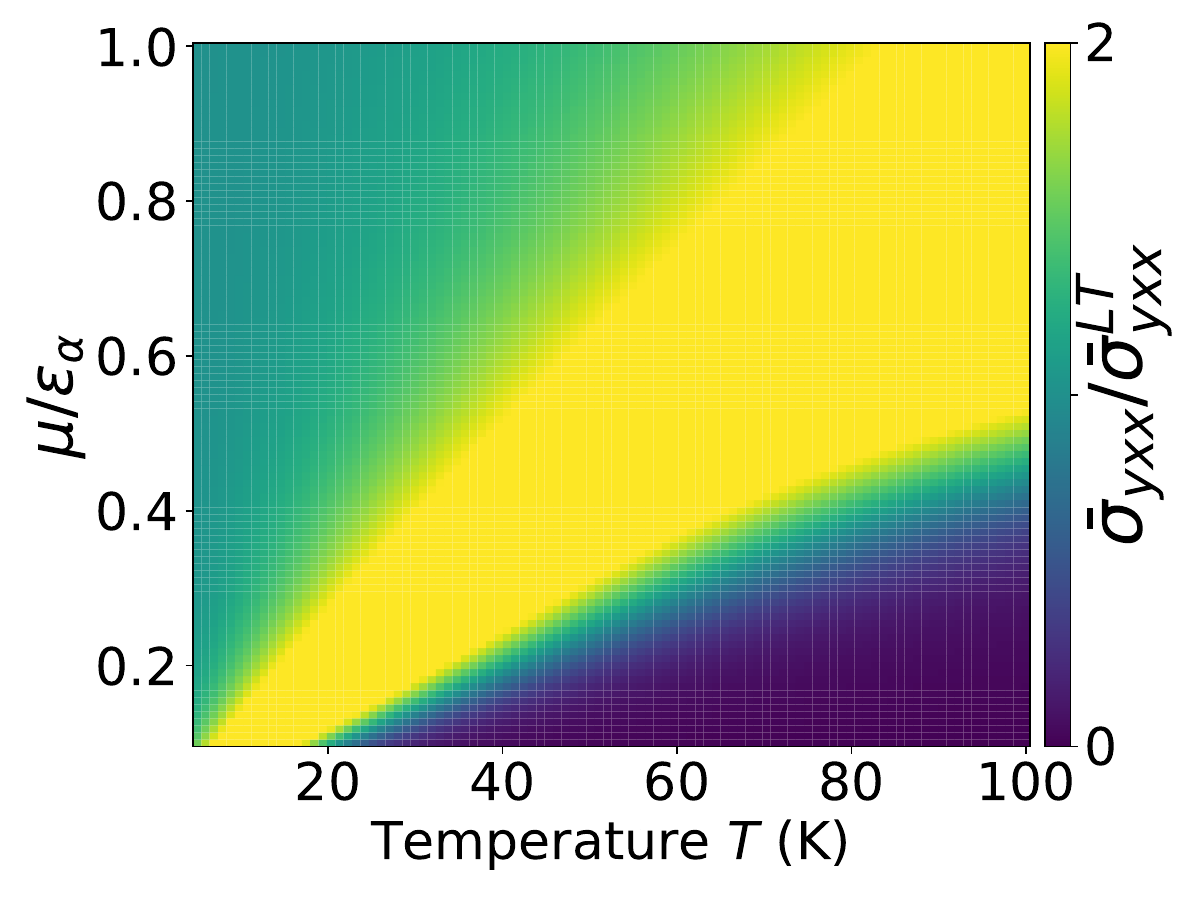}\label{fig:1}}
    \subfloat{\includegraphics[width=0.3\linewidth,height=0.28\textheight,keepaspectratio]{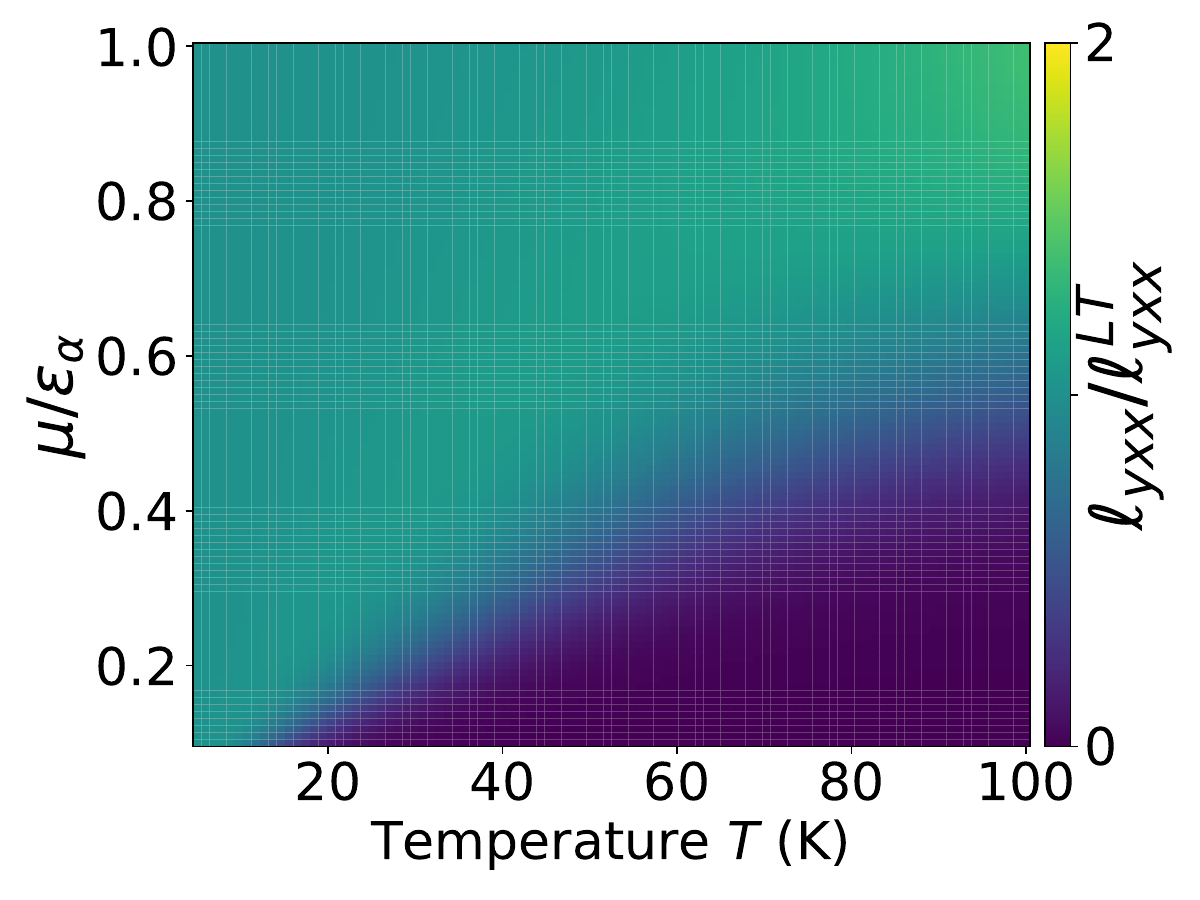}\label{fig:2}}
    \subfloat{\includegraphics[width=0.3\linewidth,height=0.28\textheight,keepaspectratio]{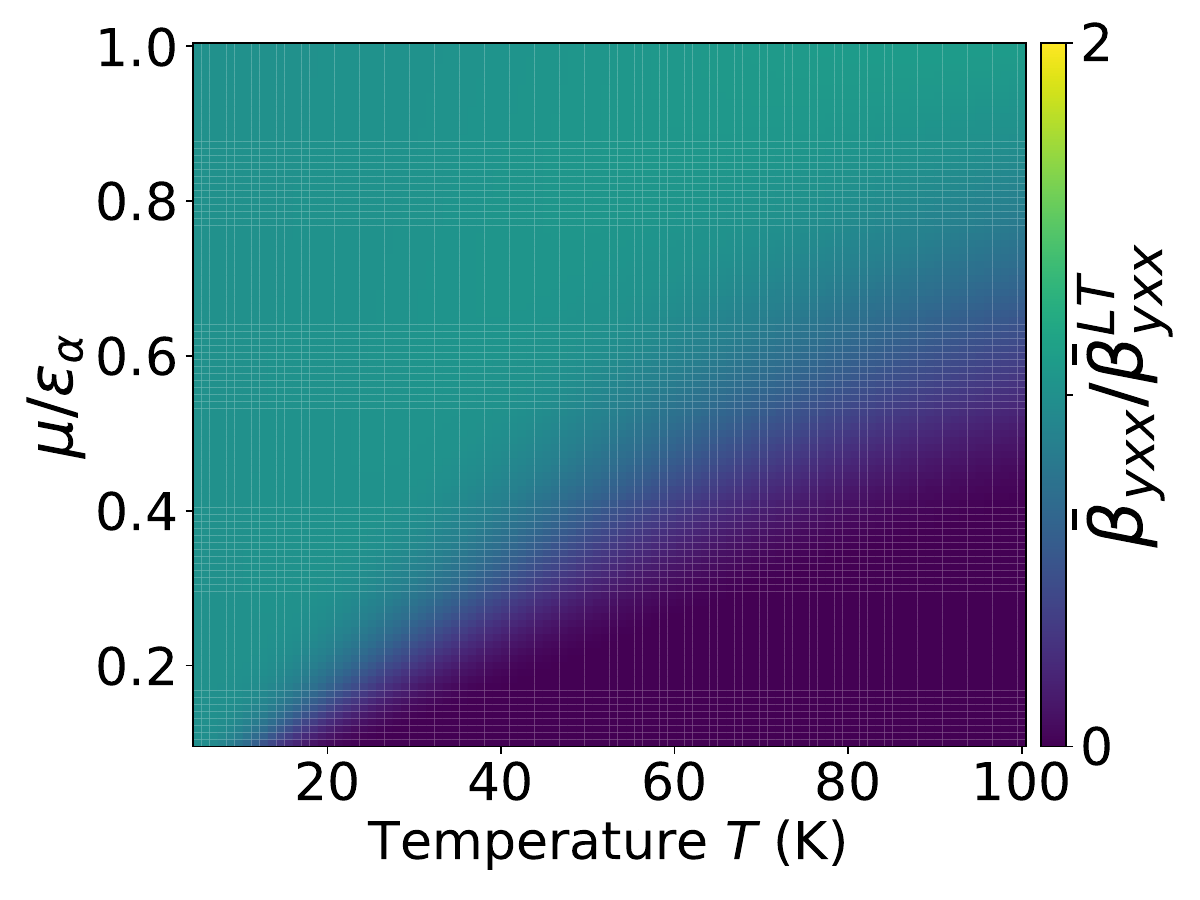}\label{fig:3}}
    \caption{Heatmaps showcasing the deviation of the ratio of the nonlinear thermal transverse response coefficients ($\bar{\sigma}_{yxx}/\bar{\sigma}_{yxx}^{LT}$, $l_{yxx}/l_{yxx}^{LT}$, and $\bar{\beta}_{yxx}/\bar{\beta}_{yxx}^{LT}$) to their low temperature forms from unity for large temperatures and low chemical potential (rescaled) values, thereby displaying the validity of the low temperature limits. Left: Displays the breakdown of the low temperature limit of the nonlinear planar Ettingshausen coefficient (coefficient $\bar{\sigma}_{yxx}$. See Eq.(\ref{sigmabaryxx}) for $\bar{\sigma}_{yxx}$, and Eq.(\ref{3.20}) for $\bar{\sigma}_{yxx}^{LT}$). Center: Displays the breakdown of the low temperature limit of the nonlinear planar thermal Hall conductivity (coefficient $l_{yxx}$. See Eq.(\ref{lyxx}) for $l_{yxx}$, and Eq.(\ref{3.21}) for $l_{yxx}^{LT}$). Right: Displays the breakdown of the low temperature limit of the nonlinear thermal transverse planar mixed thermoelectric coefficient (coefficient $\bar{\beta}_{yxx}$. See Eq.(\ref{betabaryxx}) for $\bar{\beta}_{yxx}$, and Eq.(\ref{betabarLT}) for $\bar{\beta}_{yxx}^{LT}$)}. Unity (sea green color) indicates exact agreement with the analytic low‑$T$ form, while deviations (yellow to purple) from unity mark the breakdown of the expansion at low $\mu/\varepsilon_{\alpha}$ and elevated $T$. Here, we have chosen $B = 1T$, $\theta = \pi/4$.
    \label{Heatmaps_thermal}
\end{figure*}

\noindent Here, \(x_{-1}=\beta\mu\) and \(x_{1}=\beta(\mu+\epsilon_{\alpha})\), and $Li_{2}(x)$ denotes the dilogarithm function.  In the low temperature limit \(\beta\mu\gg1\) (i.e.\ \(x_{s}\gg1\)), one finds
\(F^{0}\to1\), \(F^{2}\to\pi^{2}/3\), and \(G^{1}\to\beta\), while all other functions decay exponentially.  Accordingly, in the low‑temperature regime \(\beta\mu\gg1\), keeping only the leading terms in \(T\), the moments \(D^{\nu}_{s}\) reduce to:
\begin{eqnarray}
\begin{bmatrix}
         \mathcal{D}^0_s\\
         \mathcal{D}^1_s\\
         \mathcal{D}^2_s
     \end{bmatrix} =  -\frac{m^{\frac{3}{2}}\sqrt{\epsilon_{\alpha}}}{\sqrt{2}\pi^2\hbar^3}\begin{bmatrix}
         \left(\frac{(1-s\sqrt{1+\Tilde{\mu}})^2}{\sqrt{1+\Tilde{\mu}}}\right)F_0 \\ 
         \left(\frac{\Tilde{\mu}}{2\beta^2\epsilon_\alpha(1+\Tilde{\mu})^{\frac{3}{2}}}\right)F_2 \\
         \frac{1}{\beta^2}\left(\frac{(1-s\sqrt{1+\Tilde{\mu}})^2}{\sqrt{1+\Tilde{\mu}}}\right)F_2
     \end{bmatrix}
\label{3.14}
\end{eqnarray}

We now turn to the nonlinear planar transport coefficients arising from the chiral anomaly. We present some of the important features of the nonlinear response coefficients, more specifically, highlighting their their magnetic-field scaling and angular dependence. A new feature of our calculations in the nonlinear order (in driving fields) is the emergence of these mixed transport coefficients $\beta$ and $\bar\beta$ (see Eqs.(\ref{2.2.8}) and (\ref{betabaryxx})) which gives rise to transverse nonlinear charge and heat currents proportional to both $\bm{E}$ and $\bm{\nabla} T$ (see Eq.(\ref{quad_response})). These quantities can be experimentally demonstrated in the spin-orbit coupled metallic system using coplanar temperature gradient, electric, and magnetic fields simultaneously in the 3D system and measuring the charge and heat currents transverse to the driving fields proportional to both $\bm{E}$ and $\bm{\nabla} T$. All the nonlinear transport coefficients (see Eqs.(\ref{2.2.6}) - (\ref{betabaryxx})) obey the same angular and magnetic field strength dependence which we discuss below. \\

Unlike the linear order transport coefficients which vary quadratically with the magnetic field strength (see Fig.\ref{sig_linear} top) and exhibit a $\sin\theta\cos\theta$ dependence with extrema at $\theta = \pi/4$ (see Fig.\ref{sig_linear} bottom), 
the nonlinear counterparts scale as $B^{3}$ (see Fig.\ref{sig_quad} top) and follow a $\cos^{2}\theta\sin\theta$ dependence, reaching their maximum magnitude at $\theta = \tan^{-1}(1/\sqrt{2})$ (see Fig.\ref{sig_quad} bottom). 
This shift in angular position indicates that higher-order corrections in the driving electric and thermal fields enhance the planar response closer to configurations where the magnetic field is nearly aligned with the applied fields. 
The field dependence also distinguishes the nonlinear regime through a steeper cubic scaling, consistent with the fact that the nth-order chiral anomaly induced response in the applied driving fields varies as $B^{n+1}$. An \(n\)th‑order response coefficient also varies as \(\cos^{n}\theta\,\sin\theta\), placing its extremum at \(\theta=\arctan(1/\sqrt{n})\) and rapidly suppressing the amplitude once \(\bm B\) and \(\bm E\) deviate from parallel. We do however note that despite the closeness to parallel, the transverse planar response (in all orders in driving fields) does vanish when the magnetic field is perfectly aligned with the driving electric and thermal fields.\\

We now study the low temperature behavior of the transport coefficients in the nonlinear order.

\label{beta_bar_mu_theta_quad}

\begin{figure}[t]
\centering
\hspace*{-0.5cm}
\includegraphics[scale=0.35]{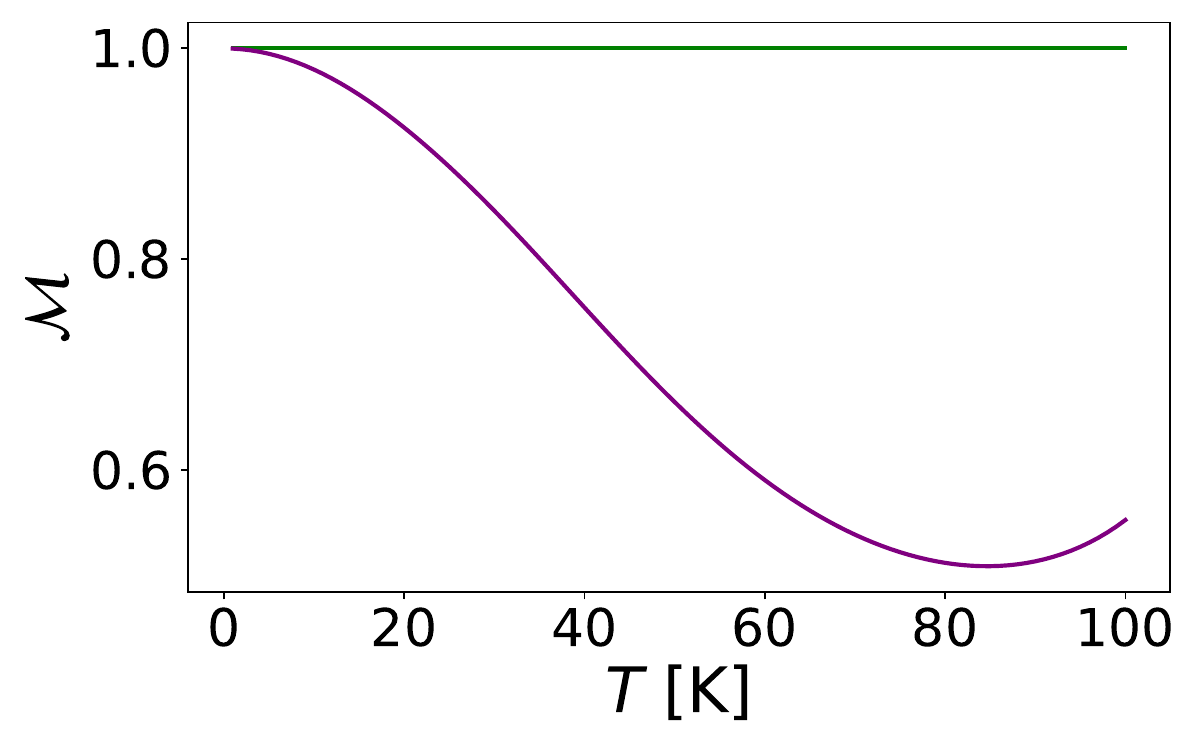}
\caption{Variation of the Mott number as a function of temperature. \textit{Solid purple line}: Numerical result for the Mott number $\mathcal{M} = \alpha_{yx}/\bigl(T\,\partial_\mu\sigma_{yx}\bigr)$ (see Eq.\eqref{2.1.2} for $\sigma_{yx}$ and Eq.\eqref{2.1.3} for $\alpha_{yx}$). \textit{Solid green line}: Expected value of the Mott number scaled to unity. The departure of $\mathcal{M}$ from unity at higher temperatures highlights the violation of the Mott relation. Here, we have chosen $\mu = 0.5*\epsilon_\alpha$, $B = 1T$, $\theta = \pi/4$.}
\label{Mott_linear}
\end{figure}

For understanding the low temperature behavior of the non-linear response coefficients in driving fields, we begin by substituting Eqs.\,\eqref{2.2.6}–\eqref{betabaryxx} into our Hamiltonian’s expressions. Retaining only the leading order terms as \(T\to0\), we find:
\begin{eqnarray}
\sigma^{LT}_{yxx} =&& (\tau^*\tau_\nu)\left(\frac{e^6}{32\pi^2m^3\epsilon_\alpha}\right)B^3\cos^2(\theta)\sin(\theta) \nonumber \\ 
&& \hspace{-0.3 cm} \times\left[\frac{12(1+\Tilde{\mu})^{\frac{1}{2}} + 40(1+\Tilde{\mu})^{\frac{3}{2}} + 12(1+\Tilde{\mu})^{\frac{5}{2}}}{\Tilde{\mu}^5}\right]
\label{3.17}
\end{eqnarray}
\begin{eqnarray}
\alpha^{LT}_{yxx} =&& (\tau^*\tau_\nu)\left(\frac{e^4k_B}{96m^3\epsilon_\alpha}\right)B^3\cos^2(\theta)\sin(\theta) \nonumber \\
&& \hspace{-0.3 cm} \times \left[\frac{12(1+\Tilde{\mu})^{\frac{1}{2}} + 40(1+\Tilde{\mu})^{\frac{3}{2}} + 12(1+\Tilde{\mu})^{\frac{5}{2}}}{\Tilde{\mu}^5}\right]
\label{3.18}
\end{eqnarray}
\begin{eqnarray}
\beta^{LT}_{yxx} =&& (\tau^*\tau_\nu)\left(\frac{e^5}{2\pi^2m^3T}\right)B^3\cos^2(\theta)\sin(\theta) \nonumber \\ 
&& \times \left[\frac{(1+\Tilde{\mu})^{\frac{1}{2}} + (1+\Tilde{\mu})^{\frac{3}{2}}}{\Tilde{\mu}^4}\right]
\label{3.19}
\end{eqnarray}
\begin{eqnarray}
\bar{\sigma}^{LT}_{yxx} =&& -\frac{\tau^*\tau_\nu}{8\pi^2}\left(\frac{e^5}{\epsilon_\alpha m^3}\right)B^3\cos^2(\theta)\sin(\theta) \nonumber \\ 
&& \hspace{-0.3 cm} \times\left[\frac{(2+\Tilde{\mu})(1+\Tilde{\mu})^{\frac{3}{2}}}{\Tilde{\mu}^4}\right]
\label{3.20}
\end{eqnarray}
\begin{eqnarray}
l^{LT}_{yxx} =&& -\frac{5\tau^*\tau_\nu}{24\pi^2}\left(\frac{e^3k^2_B}{\epsilon_\alpha m^3}\right)B^3\cos^2(\theta)\sin(\theta) \nonumber \\ 
&& \hspace{-0.3 cm} \times\left[\frac{(2+\Tilde{\mu})(1+\Tilde{\mu})^{\frac{3}{2}}}{\Tilde{\mu}^4}\right]
\label{3.21}
\end{eqnarray}
\begin{eqnarray}
\bar{\beta}^{LT}_{yxx} =&& -\frac{\tau^*\tau_\nu}{8\pi^2}\left(\frac{e^4k^2_BT}{\epsilon^2_\alpha m^3}\right)B^3\cos^2(\theta)\sin(\theta) \nonumber \\ 
&& \hspace{-0.3 cm} \times\left[\frac{3(1+\Tilde{\mu})^{\frac{1}{2}} + 10(1+\Tilde{\mu})^{\frac{3}{2}} + 3(1+\Tilde{\mu})^{\frac{5}{2}}}{\Tilde{\mu}^5}\right]
\label{betabarLT}
\end{eqnarray}

\begin{figure}[t]
\centering
\hspace*{-0.5cm}
\includegraphics[scale=0.35]{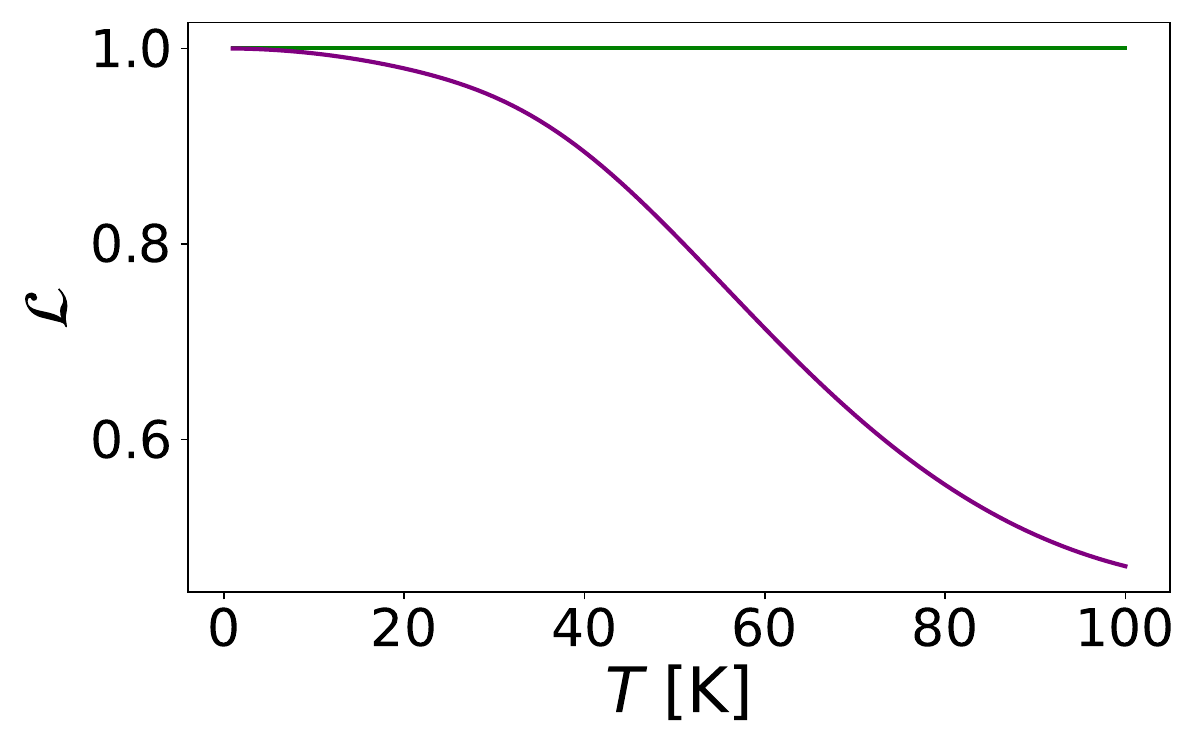}
\caption{Variation of the Lorenz number as a function of temperature. \textit{Solid purple line}: Numerical result for the Lorenz number $\mathcal{L} = \kappa_{yx}/\bigl(T\,\sigma_{yx}\bigr)$  (see Eq.\eqref{2.1.2} for $\sigma_{yx}$ and Eq.\eqref{kappalin} for $\kappa_{yx}$). \textit{Solid green line}: Expected value of the Lorenz number scaled to unity. The departure of $\mathcal{L}$ from unity at higher temperatures highlights the violation of the Wiedemann-Franz Law. Here, we have chosen $\mu = 0.5*\epsilon_\alpha$, $B = 1T$, $\theta = \pi/4$.}
\label{WF_linear}
\end{figure}

Turning to the charge and thermal mixed thermoelectric transport coefficients (see Eqs.(\ref{2.2.6}) - (\ref{lyxx})), Fig.\ref{Heatmaps_electric} and Fig.\ref{Heatmaps_thermal} show heatmaps of the ratios of the nonlinear transverse electric ($\sigma_{yxx}, \alpha_{yxx}, \beta_{yxx}$) and thermal transport coefficients ($\bar{\sigma}_{yxx}, l_{yxx}, \bar{\beta}_{yxx}$) respectively to their low temperature expansions (see Eqs.(\ref{3.17}) - (\ref{betabarLT})), as a function of the temperature $T$ and the scaled chemical potential $\mu/\epsilon_\alpha$. A ratio of unity (sea-green color) denotes exact agreement with the low-T expansion. 

In Fig.\ref{Heatmaps_electric} the left and center sub-figures describe the behavior of the of the nonlinear planar Hall and Nernst coefficients (see Eqs.(\ref{2.2.6}) and (\ref{2.2.7}) respectively) in comparison to their low temperature counterparts see Eqs.(\ref{3.17}) and (\ref{3.18}) respectively. A ratio of unity (sea-green color) denotes exact agreement with the low-T expansion. As we traverse the principal diagonal (low-T, high-$\mu$ region to high-T, low-$\mu$ region), we note that the ratio is very close to unity in the low-T, high-$\mu$ region, after which it first increases slightly above unity (grass green color) and then reduces below unity (blue color). The right sub-figure describes the behavior of the electric nonlinear planar mixed thermoelectric response coefficient (Eq. (\ref{3.19})) in comparison to its low-T counterpart. Traversing along the principal diagonal, we note that this ratio also starts off with a value close to unity, but the deviation above unity is slightly larger as compared to the other two nonlinear electric responses with the  ratio increasing in value to over 1.6 (yellow-green color) in the middle and taking values very close to 0 (blue color) at the high-T, low-$\mu$ region. In Fig.\ref{Heatmaps_thermal} we note similar features in the thermal heatmaps. The left sub-figure describes the nonlinear planar Ettingshausen coefficient (Eq.(\ref{sigmabaryxx})). Traversing along the principal diagonal, this ratio starts with a value close to unity in the low-T, high-$\mu$ region denoting agreement with the low-T expansion and along the middle takes values over 2 (yellow color). In the high-T, low-$\mu$ region, the takes values close to 0 (blue color). The center and right sub-figures denote the thermal nonlinear Hall and mixed thermoelectric coefficients respectively. Moving along the principal diagonal, in the low-T, high-$\mu$ region, the ratio takes values close to unity, while it starts decreasing to 0 as the temperature increases and chemical potential values decreases. These departures from the low-T limit also manifest as violations of the Wiedemann–Franz and Mott-type relations, providing a complementary perspective on this breakdown. We now present the low temperature linear order transport coefficients \cite{PhysRevB.108.045405} below to study the Wiedemann-Franz and Mott relations in these systems:\\

\begin{eqnarray}
\sigma^{LT}_{yx} =&& \left(\frac{\tau^*e^4}{4\sqrt{2}\pi^2\hbar m^{\frac{3}{2}}\sqrt{\epsilon_\alpha}}\right)B^2\cos(\theta)\sin(\theta) \nonumber \\
&&\times\left[\frac{(2+\Tilde{\mu})\sqrt{1+\Tilde{\mu}}}{\Tilde{\mu}^2}\right]
\label{3.15}
\end{eqnarray}
\begin{eqnarray}
\alpha^{LT}_{yx} =&& \left(\frac{\tau^*e^3k_B}{24\sqrt{2}\beta\hbar m^{\frac{3}{2}}\epsilon_\alpha^{\frac{3}{2}}}\right)B^2\cos(\theta)\sin(\theta) \nonumber \\ 
&&\times\left[\frac{\Tilde{\mu}^2 + 8\Tilde{\mu} + 8}{\Tilde{\mu}^3\sqrt{1+\Tilde{\mu}}}\right]
\label{3.16}
\end{eqnarray}
\begin{eqnarray}
\bar{\alpha}^{LT}_{yx} =&& \left(\frac{\tau^*e^3}{24\sqrt{2}\beta^2\hbar m^{\frac{3}{2}}\epsilon_\alpha^{\frac{3}{2}}}\right)B^2\cos(\theta)\sin(\theta) \nonumber \\ 
&&\times\left[\frac{\Tilde{\mu}^2 + 8\Tilde{\mu} + 8}{\Tilde{\mu}^3\sqrt{1+\Tilde{\mu}}}\right]
\label{alphabar}
\end{eqnarray}
\begin{eqnarray}
\kappa^{LT}_{yx} =&& \left(\frac{\tau^*e^2k_B}{24\sqrt{2}\beta\hbar m^{\frac{3}{2}}\sqrt{\epsilon_\alpha}}\right)B^2\cos(\theta)\sin(\theta) \nonumber \\
&&\times\left[\frac{(2+\Tilde{\mu})\sqrt{1+\Tilde{\mu}}}{\Tilde{\mu}^2}\right]
\label{kappa}
\end{eqnarray}

To further understand the validity of the low temperature expansion, Fig.\ref{Mott_linear} shows a plot comparing the Mott number ($\mathcal{M} = \frac{\alpha}{T\partial_\mu\sigma}$) rescaled to unity in the low temperature limit, computed numerically (purple line) with its expected value (green line). A Mott number value close to unity indicates agreement with the Mott relation, while a deviation signals a breakdown of this approximation. We can see that the Mott relation is not valid for higher temperatures ($> 10K$) and the non-linear deviation is apparent at fairly low temperatures (in comparison to $\mu/k_B\approx300K$ for the parameters used) (approximately $10\%$ deviation at $20K$ from the ideal value). Similarly, Fig.\ref{WF_linear}
shows the violation of the Wiedemann-Franz law ($\kappa = LT\sigma$, with $L$ being the Lorenz number) at low temperatures (approximately $10\%$ deviation at $40K$ from the ideal value). The nonlinear case shows further interesting results. 


\begin{figure}[t]
\centering
\hspace{-0.5cm}
\includegraphics[scale=0.35]{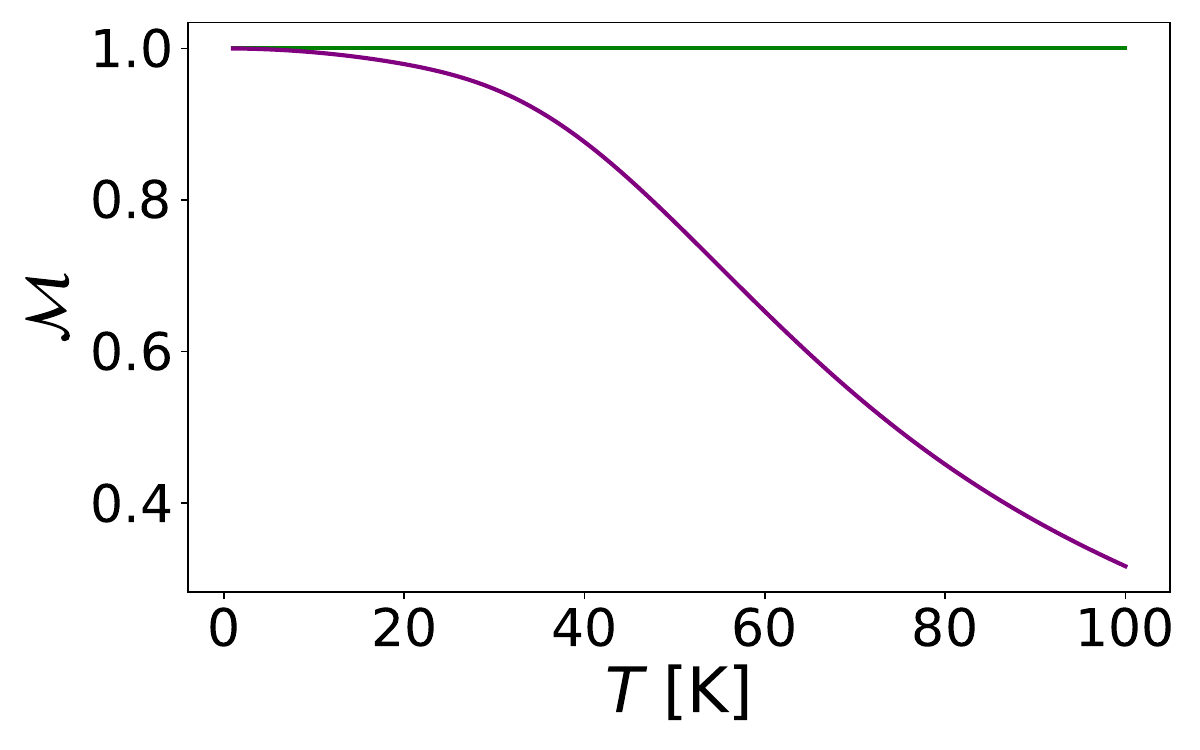}
\caption{Variation of the nonlinear Mott number as a function of temperature. \textit{Solid purple line}: Numerical result for the nonlinear Mott number $\mathcal{M} = \alpha_{yxx}/\,\sigma_{yxx}$  (see Eq.\eqref{2.2.6} for $\sigma_{yxx}$ and Eq.\eqref{2.2.7} for $\alpha_{yxx}$). \textit{Solid green line}: Expected value of the nonlinear Mott number scaled to unity. The departure of $\mathcal{M}$ from unity at higher temperatures highlights the violation of the nonlinear Mott relation. Here, we have chosen $\mu = 0.5*\epsilon_\alpha$, $B = 1T$, $\theta = \pi/4$.}
\label{Mott_quad}
\end{figure}

Substituting the exact low‑\(T\) expressions from Eqs.\,\eqref{3.17} and \eqref{3.18} shows
\begin{eqnarray}
\alpha^{LT}_{yxx}
  &=& \biggl(\frac{\pi^{2}k_{B}^{2}}{3e^{2}}\biggr)\sigma^{LT}_{yxx}
      ,
\end{eqnarray}
confirming the validity of the nonlinear analog of the Mott relation \cite{PhysRevB.105.125131, PhysRevResearch.2.032066}. The same conclusion is revealed from Fig.\ref{Mott_quad}, which shows the normalized Mott number 
\(\mathcal{M}=\alpha_{yxx}/\sigma_{yxx}\) as a function of temperature.  As predicted, 
\(\mathcal{M}\) remains constant at low \(T\) and then deviates non-linearly at higher \(T\) ($> 30K$). In the nonlinear case however, the nonlinear Mott relation is valid for temperatures higher than its linear counterpart (approximately $10\%$ deviation at $40K$ from the ideal value). \\

However, even in the low temperature limit, the non-linear analog of the Weidemann-Franz law shown in \cite{PhysRevB.105.125131, PhysRevResearch.2.032066} is violated. We can see this result by considering the ratio $\mathcal{W} = l^{LT}_{yxx}/(T^2\frac{\partial \sigma_{yxx}^{LT}}{\partial\mu})$ (see Eq.\eqref{3.17} for $\sigma_{yxx}^{LT}$ and Eq.\eqref{3.21} for $l_{yxx}^{LT}$). This quantity does not obey the non-linear analog of the Wiedemann-Franz law:
\begin{eqnarray}
l^{LT}_{yxx} = -\frac{7\pi^4k_B^4T^2}{15e^2} \frac{\partial \sigma_{yxx}^{LT}}{\partial \mu} 
\label{nonlinWiede}
\end{eqnarray}

Clearly, this ratio in Eq.\eqref{wtfwf} is neither system independent (due to the dependence on $\Tilde{\mu}$ and $\epsilon_\alpha$) nor temperature independent. All other combinations of these quantities will yield results that have system dependence. This implies the absence of a non-linear analog of the Wiedemann-Franz law in this system. We want to emphasize that the violation of the Wiedemann-Franz law and Mott relations, both in the linear and non-linear orders in driving fields goes beyond the failure for the conditions required for Sommerfeld expansion. \\

\balance
\begin{widetext}

Computing this ratio using Eqs. (\ref{3.17}) and (\ref{3.21}), we find:
\begin{eqnarray}
    \mathcal{W} = -\frac{20k_B^2\epsilon_\alpha}{3e^3T^2} \frac{\Tilde{\mu}^2(2+\Tilde{\mu})(1+\Tilde{\mu})}{6\Tilde{\mu}(1+\Tilde{\mu})^{-1} + 60\Tilde{\mu} + 30\Tilde{\mu}(1+\Tilde{\mu}) - 60 - 200(1+\Tilde{\mu}) - 60(1+\Tilde{\mu})^2}
\label{wtfwf}    
\end{eqnarray}

\end{widetext}

Mathematically, these violations for the low temperature approximations occur due to the exponential relation of the fundamental moments (Eqs. (\ref{3.8}) - (\ref{3.13})) with temperature and chemical potential, in contrast to the polynomial dependence assumed in the conventional Wiedemann-Franz law and Mott relations derived via the Sommerfeld expansion \cite{Girvin_Yang_2019}. The precise physical origin of this behavior remains an open question, which we plan to investigate further in future work. We also stress that no simple Wiedemann–Franz or Mott‑type proportionality can link the mixed thermoelectric coefficients \(\beta\) and \(\bar{\beta}\) to the other nonlinear planar transport coefficients.
\section{Experimental proposal}\label{Exptprop}

In this section, we outline a feasible experimental setup and a strategy to probe the predicted second-order responses in three-dimensional spin-orbit coupled metals. As discussed earlier, all three fields, i.e., $\mathbf{E}$, $\nabla T$, and $\mathbf{B}$, are applied in-plane (see Fig.\ref{Schematic}). The transverse voltage is generated between midpoints on opposite $y$-edges of the sample, labeled $V^+$ and $V^-$, which yields the signal $V_y = V^+ - V^-$. A heater and thermal sink at opposite ends establish a gradient in temperature ($\nabla T$) along the $x$-direction, while the electric field is applied along the $x$-direction. Furthermore, an in-plane rotation of $\mathbf{B}$, with angle $\theta$ defined with respect to the $x$-axis, enables angular scans of the nonlinear response (see Fig.\ref{sig_quad}, top). 
Each of the predicted nonlinear transport coefficients can be individually isolated using harmonic detection \cite{kang2019nonlinear, yasuda2017current, tiwari2021giant} and the following modulation schemes:
(i) \textit{nonlinear planar Hall coefficient ($\sigma_{yxx}$):} We apply an a.c. electric field $\mathbf{E}(t) \sim \sin(\omega t)$, and measure transverse voltage $V_y$ at frequency $2\omega$ using a lock-in amplifier. The Hall voltage $V_y^{(2\omega)} \sim \sigma_{yxx}^{(2)} E_x^2$, showing cubic scaling with $B$ and angular dependence $\sim \sin\theta \cos^2\theta$ (see Eq.\,\eqref{2.2.6} and Fig.\ref{sig_quad}), isolates the $E^2$ contribution. (ii) \textit{nonlinear planar Nernst coefficient $(\alpha_{yxx})$:}  We apply an a.c. thermal gradient via modulated heating power at frequency $\omega$: $\nabla T(t) \sim \sin(\omega t)$, and measure $V_y$ at $2\omega$ to isolate the $(\nabla T)^2$ term. A transverse thermoelectric voltage, scaling quadratically with heating amplitude and cubic in $B$, with the same angular dependence as above, isolates this response (see Eq.\eqref{2.2.7}). 
(c) \textit{mixed nonlinear coefficient ($\beta_{yxx}$):} (a) We simultaneously modulate the electric field at frequency $\omega_1$ and heater at $\omega_2$, both along the x-direction (as shown in Fig.\ref{Schematic}). (b) The $E \cdot \nabla T$ term generates a voltage component at sum or difference frequency: $\omega_1 \pm \omega_2$. One can then use a lock-in amplifier at $\omega_1 + \omega_2$ or $\omega_1 - \omega_2$ to extract the mixed response. This signal is expected to be odd under either $\mathbf{E}$ or $\mathbf{\nabla T}$ reversal but even under both. The signal scales cubic in the magnetic field strength ($\propto B^3$). In all three cases, one can further scan the in-plane angle $\theta$ of the magnetic field. To measure the heat currents, one may detect the transverse temperature difference generated under a longitudinal drive using nanoscale thermometers placed on opposite side edges of the sample \cite{childs2015nanoscale, sadat2012high}. Alternatively, microcalorimetry \cite{calvet2016recent} or scanning thermal microscopy \cite{https://doi.org/10.1002/adfm.201900892, 10.1063/5.0091494} can be used to spatially resolve the temperature profile and infer the associated transverse heat flow. Such measurements would provide complementary signatures of the nonlinear thermoelectric and thermal response functions and help verify the broader symmetry constraints and Onsager relations. Our predicted angular dependence $\sim \sin \theta \cos^2 \theta$ provides a stringent diagnostic for the underlying mechanisms (see Eqs.\,\eqref{2.2.6}-\eqref{2.2.8}). Reversing the directions of $\mathbf{E}$, or $\mathbf{\nabla T}$ independently allows separation of various contributions and verification of symmetry constraints. Note that to ensure a measurable response within the semiclassical regime, the modulation frequency must satisfy $ \omega \tau \ll 1$, where $\tau$ is the effective relaxation time. For typical clean metallic systems, the relaxation times range from $\tau\sim 0.1\text{--}1\,\text{ps}$, corresponding to scattering rates of $\sim 1\text{--}10\,\text{THz} $ \cite{ashcroft1976solid}. Using a modulation frequency in the range $\sim 10\text{--}500\,\text{Hz}$, which is common for lock-in measurements \cite{cultrera2019calibration, nam2018probe}, we find that $\omega \tau \sim 10^{-6}\text{--}10^{-3}$, which comfortably satisfies the semiclassical criteria. Thus, under realistic experimental conditions, the system operates well within the semiclassical regime, ensuring that the second-order response can be reliably attributed to relaxation processes via intraband and interband scattering, without involving quantum coherence or high-frequency dynamics.



\section{Conclusion}
In this paper, we have investigated nonlinear planar transport phenomena driven by the chiral anomaly in three-dimensional spin-orbit coupled metallic systems, employing a semiclassical Boltzmann transport formalism up to second order in the applied fields. Unlike conventional Weyl semimetals, these systems host distinct Fermi surfaces characterized by opposite Berry curvature fluxes without explicit band degeneracies at the Fermi energy. Our theoretical framework highlights novel anomaly driven transport responses, which can be experimentally accessed through nonlinear planar Hall (coefficient $\sigma_{abc}$, see Eq.(\ref{quad_response}) and Eq.(\ref{2.2.6}), nonlinear planar thermal Hall (coefficient $l_{abc}$, see Eq.(\ref{quadQ_response}) and Eq.(\ref{lyxx})), nonlinear planar Nernst (coefficient $\alpha_{abc}$, see Eq.(\ref{quad_response}) and Eq.(\ref{2.2.7})), nonlinear planar Ettingshausen (coefficient $\bar{\sigma}_{abc}$, see Eq.(\ref{quadQ_response}) and Eq.(\ref{sigmabaryxx})), and planar mixed thermoelectric (coefficient $\beta_{abc}$, see Eq.(\ref{quad_response}) and Eq.(\ref{2.2.8})) and planar thermal mixed thermoelectric (coefficient $\bar{\beta}_{abc}$, see Eq.(\ref{quadQ_response}) and Eq.(\ref{betabaryxx})) response coefficients.

Within this framework, for the paper to be self-contained, we have rederived the thermoelectric coefficients in the linear response regime, which agree with the results in Ref.~[\cite{PhysRevB.108.045405}]. For instance, the behavior of the linear transverse planar coefficients $(\sigma_{yx}, \alpha_{yx}, \bar\alpha_{yx},\kappa_{yx})$ display the expected $\sin\theta \cos\theta$ angular dependence (see Fig.\ref{sig_linear}, top), scale quadratically with magnetic field strength (see Fig.\ref{sig_linear}, bottom), and already show signs of both Mott relation (see Fig.\ref{Mott_linear}) and Wiedemann-Franz law violation (see Fig.\ref{WF_linear}) at very low temperatures where the Sommerfeld approximation should remain valid ($T \ll \mu/k_B \approx300K$).

The main focus of our work is in the nonlinear regime. A new feature of our calculations in the nonlinear order is the emergence of mixed thermoelectric and thermal mixed thermoelectric transport coefficients $\beta$ and $\bar{\beta}$ (see Eqs.(\ref{quad_response}) and (\ref{quadQ_response}) respectively) which gives rise to transverse charge and heat currents proportional to both $\bm{E}$ and $\bm{\nabla} T$ (see Eqs.(\ref{quad_response}) and (\ref{quadQ_response}) respectively). These quantities can be experimentally demonstrated in the spin-orbit coupled metallic system in the presence of chiral anomaly by having a coplanar temperature gradient, electric field, and magnetic field simultaneously in the 3D system and measuring the charge and heat currents transverse to the applied driving fields proportional to both $\bm{E}$ and $\bm{\nabla} T$. These nonlinear planar responses exhibit a distinct angular dependence, with peaks shifted from $\pi/4$ in the linear response regime (see Fig.\ref{sig_linear}, top) to a smaller angle  of $tan^{-1}(1/\sqrt{2})$ (see Fig.\ref{sig_quad}, top). Moreover, the nonlinear planar transverse responses display a cubic scaling with magnetic field strength (see Fig.\ref{sig_quad}, bottom and Eqs.(\ref{2.2.6})-(\ref{lyxx})) as opposed to a quadratic scaling with the magnetic field strength as shown by the linear planar transverse responses (see Fig.\ref{sig_linear}, bottom). Finally, we observe significant deviations from the nonlinear Mott relation (see Fig.\ref{Mott_quad}) at temperatures well below the regime where the polynomial expansion of the nonlinear transverse planar transport tensor should remain valid ($T \ll \mu/k_B \approx300K$). The same low-temperature violation behavior of both the Wiedemann-Franz law (see Fig.\ref{WF_linear}) and Mott relation (see Fig.\ref{Mott_linear}) are observed in the linear response regime as well at temperatures well below the regime where the polynomial expansion for the linear order transport tensors should remain valid ($T \ll \mu/k_B \approx300K$). Mathematically, these violations occur due to the exponential temperature and chemical potential dependence of the moments of the generalized density and energy-velocity integrals (as defined in Eqs.(\ref{1.8}), (\ref{1.7}), and (\ref{2.2.3})) that enter the linear (see Eqs.(\ref{2.1.2})-(\ref{kappalin})) and nonlinear transverse planar transport tensors (see Eqs.(\ref{2.2.6})-(\ref{lyxx})). The nonlinear analog of the Wiedemann-Franz law (see Eq.(\ref{nonlinWiede})) is violated even by the low temperature expansions of the corresponding transport coefficients (see Eq.(\ref{wtfwf})).

We note that the effective Hamiltonian considered in this paper given in Eq.~(15) is closely related to that of a Weyl semimetal, with the conventional tilt term replaced by a rotationally symmetric contribution proportional to $k^2$ \cite{PhysRevB.103.245119, PhysRevB.97.041101}. Using Eq.\eqref{3.4} we note that the off-diagonal components of the Berry Curvature Dipole (BCD) density $d_{ab}$ $(a\neq b)$ (defined by, $d_{ab} = \partial_a\Omega_{b}$) exhibits $ab$-symmetry, which forces the off-diagonal components of the BCD tensor, defined as the Brillouin zone integral of the BCD density ($d_{ab}$), to vanish. We therefore obtain that only the diagonal components of the BCD tensor are nonzero, and consequently the BCD contribution to the nonlinear transverse current vanishes (unless the Hamiltonian possesses a rotationally asymmetric tilt term similar to that in the case of tilted  WSMs in \cite{PhysRevB.103.245119, PhysRevB.97.041101}). Moreover, since the system preserves time reversal symmetry, we expect the Quantum Metric Dipole (QMD) contributions to vanish as well. As shown in the Supplementary section of Ref.[\onlinecite{PhysRevLett.133.106701}], the quantum metric dipole given by the integral of the first order derivative of the quantum metric vanishes if the system preserves inversion of time reveral symmetry. Since the Hamiltonian given in Eq.~(15) preserves time reversal, we expect that the QMD contributions to the non-linear transverse current vanish in the quadratic order. 



Our theoretical predictions can be directly tested through planar transport measurements under the simultaneous application of electric fields and thermal gradients (as discussed in Section \ref{Exptprop}). Such experiments would not only validate our theoretical results but also establish signatures of chiral anomaly driven transport in three-dimensional, spin-orbit coupled metallic systems.

\section{Acknowledgment}
S.T., R.G.G and B.B.R. acknowledge support from SC Quantum, ARO W911NF2210247 and ONR-N000142312061. G.S. was supported by ANRF-SERB
Core Research Grant CRG/2023/005628. G.S. thanks Azaz Ahmad and Gautham Varma K. for useful discussions. BBR acknowledges support from the National Center for Transportation Cybersecurity and Resiliency (TraCR) (a US Department of Transportation National University Transportation Center) headquartered at Clemson University, Clemson, South Carolina, USA. Any opinions, findings, conclusions, and recommendations expressed in this material are those of the author(s) and do not necessarily reflect the views of TraCR. The US Government assumes no liability for the contents or use thereof.

\appendix
\section{CALCULATION OF THE FIRST AND SECOND ORDER PERTURBATION TERMS}
\label{appendix: Appendix A}

To calculate the correction terms, let's start with the collision term in the transport equation, 

\begin{align}
I_{coll}=& -\left(\frac{\delta f_k^s}{\tau^*}\right) \nonumber \\
&-\left(\frac{f_{eq}(\epsilon^{s}, \mu^{s}, T^{s}) - f_{eq}(\epsilon^{s}, \mu^{\bar{s}}, T^{\bar{s}})}{\tau_\nu}\right)
\label{A1}
\end{align}

Using the definition of $\delta T^s$ and $\delta \mu^s$ on the second term in the collision equation we obtain,

\begin{flalign}
\hspace{-1cm}
f_{eq}(\epsilon^{s}, \mu^{\bar{s}}, T^{\bar{s}}) = f_{eq}(\epsilon^{s}, \mu^{s} - 2\delta\mu^{s}, T^{s} - 2\delta T^s) \nonumber \\
\implies f_{eq}(\epsilon^{s}, \mu^{\bar{s}}, T^{\bar{s}}) = f_{eq}(\epsilon^{s}, \mu^{s}, T^{s}) - 2\left(\frac{\partial f_{eq}}{\partial \mu}\right)\delta\mu^s  \nonumber \\
 - 2\left(\frac{\partial f_{eq}}{\partial T}\right)\delta T^s + \begin{bmatrix}
         \delta\mu^s & \delta T^s\\
     \end{bmatrix}\begin{bmatrix}
         \frac{\partial^2f_{eq}}{\partial\mu^2} & \frac{\partial^2f_{eq}}{\partial\mu\partial T} \\
         \frac{\partial^2f_{eq}}{\partial\mu\partial T} & \frac{\partial^2f_{eq}}{\partial T^2}
     \end{bmatrix}\begin{bmatrix}
         \delta\mu^s\\
         \delta T^s
     \end{bmatrix}
\label{A2}
\end{flalign}

The second and third terms on the right-hand side correspond to the first-order deviation ($f_D^{(1)}$) of $f_{eq}(\epsilon^{s}, \mu^{\bar{s}}, T^{\bar{s}})$ from $f_{eq}(\epsilon^{s}, \mu^{s}, T^{s})$, and the third term, also called the Hessian, corresponds to the second-order deviation ($f_D^{(2)}$). Now, using the fact that $\frac{\partial f_{eq}}{\partial \mu} = -\frac{\partial f_{eq}}{\partial \epsilon^s}$ and $\frac{\partial f_{eq}}{\partial T} = -\frac{\partial f_{eq}}{\partial \epsilon^s}\left(\frac{\epsilon - \mu}{T}\right)$, we obtain,

\begin{align}
f_D^{(1)} = 2\left(\frac{\partial f_{eq}}{\partial \epsilon^s}\right)\delta\mu^s + 2\left(\frac{\partial f_{eq}}{\partial \epsilon^s}\right)\left(\frac{\epsilon-\mu}{T}\right)\delta T^s
\label{A3}
\end{align}
\begin{align}
f_D^{(2)} =& 2\left(\frac{\partial^2f_{eq}}{\partial \epsilon^{s2}}\right)(\delta\mu^s)^2 + 4\left[\frac{1}{T}\left(\frac{\partial f_{eq}}{\partial \epsilon^{s}}\right) + \left(\frac{\epsilon^s - \mu}{T}\right) \right.\nonumber \\
&\left.\times\left(\frac{\partial^2f_{eq}}{\partial \epsilon^{s2}}\right)\right](\delta\mu^s)(\delta T^s)+ 2\left[2\left(\frac{\epsilon^s - \mu}{T^2}\right) \right. \nonumber \\
& \left. \times\left(\frac{\partial f_{eq}}{\partial \epsilon^{s}}\right) + \left(\frac{\epsilon^s - \mu}{T}\right)^2\left(\frac{\partial^2f_{eq}}{\partial \epsilon^{s2}}\right)\right](\delta T^{s})^2
\label{A4}
\end{align}

Thus, the correction equations can be collectively written as,

\begin{align}
&D\left(\bm{v}^s + \frac{e}{\hbar}(\bm{v}^s\cdot\bm{\Omega}^s)\bm{B}\right)\cdot \left(e\bm{E} + \left(\frac{\epsilon^s - \mu}{T}\right)\bm{\nabla} T\right) \nonumber \\ 
&\times \left(-\frac{\partial (f_{eq}^s + f^{(1)}_k + f^{(2)}_k + ...)}{\partial \epsilon^s}\right) = -\left(\frac{\delta (f^{(1)}_k + f^{(2)}_k + ...)}{\tau^*}\right) \nonumber \\
& - \left(\frac{f_{eq}(\epsilon^{s}, \mu^{s}, T^{s}) - f_{eq}(\epsilon^{s}, \mu^{\bar{s}}, T^{\bar{s}})}{\tau_\nu}\right)
\label{A5}
\end{align}

We obtain the correction terms by comparing terms of even order on both sides of the above equation. Now, to obtain the values of $\delta\mu^s$ and $\delta T^s$, we only consider the first-order contribution in equation (\ref{A5}). 
\begin{align}
&D\left(\bm{v}^s + \frac{e}{\hbar}(\bm{v}^s\cdot\bm{\Omega}^s)\bm{B}\right)\cdot \left(e\bm{E} + \left(\frac{\epsilon^s - \mu}{T}\right)\bm{\nabla} T\right)\left(-\frac{\partial f_{eq}^s}{\partial \epsilon^s}\right) \nonumber \\
&= -\left(\frac{\delta f^{(1)}_k}{\tau^*}\right) - \frac{2}{\tau_\nu}\left(\frac{\partial f_{eq}}{\partial \epsilon^s}\right)\left[\delta \mu^s + \left(\frac{\epsilon^s - \mu}{T}\right)\delta T^s\right]
\label{A6}
\end{align}

Integrating this equation over the Brillouin zone and using the definitions (\ref{1.7}) and (\ref{1.8}), we have,
\begin{align}
-\frac{\tau_\nu}{2}\left(e\bm{E}.\bm{\Lambda}^0_s + \frac{\bm{\nabla} T}{T}.\bm{\Lambda}^1_s\right) = (\delta\mu^s)D^0_s + \left(\frac{\delta T^s}{T}\right)D^1_s
\label{A7}
\end{align}

\noindent The integral of the first term on the right-hand side of equation (\ref{A6}) vanishes since it is the integration over all the perturbations in the Brillouin zone. We can obtain an equation similar to (\ref{A7}) by multiplying equation (\ref{A6}) by $(\epsilon^s - \mu)$.
\begin{eqnarray}
\hspace{-1cm}
-\frac{\tau_\nu}{2}\left(e\bm{E}.\bm{\Lambda}^1_s + \frac{\bm{\nabla} T}{T}.\bm{\Lambda}^2_s\right) = (\delta\mu^s)D^1_s + \left(\frac{\delta T^s}{T}\right)D^2_s
\label{A8}
\end{eqnarray}
Solving for $\delta\mu^s$ and $\delta T^s$, we finally obtain,
\begin{align}
    \delta\mu^s = & -\frac{\tau_\nu}{2}\left(\frac{D^1_sD^2_s}{D^0_sD^2_s - (D^1_s)^2}\right)\left[\left(\frac{\bm{\Lambda}^0_s}{D^1_s} - \frac{\bm{\Lambda}^1_s}{D^2_s}\right)\cdot e\bm{E} \right. \nonumber\\
    &\left. + \left(\frac{\bm{\Lambda}^1_s}{D^1_s} - \frac{\bm{\Lambda}^2_s}{D^2_s}\right)\cdot \frac{\bm{\nabla} T}{T}\right] 
\label{A9}
\end{align}
\begin{align}
    \frac{\delta T^s}{T} =& -\frac{\tau_\nu}{2}\left(\frac{D^0_sD^1_s}{D^0_sD^2_s - (D^1_s)^2}\right)\left[\left(\frac{\bm{\Lambda}^1_s}{D^1_s} - \frac{\bm{\Lambda}^0_s}{D^0_s}\right)\cdot e\bm{E} \right. \nonumber \\
    &\left.+ \left(\frac{\bm{\Lambda}^2_s}{D^1_s} - \frac{\bm{\Lambda}^1_s}{D^0_s}\right)\cdot \frac{\bm{\nabla} T}{T}\right]
\label{A10}
\end{align}

Finally, from equation \ref{A6}, we obtain the first-order correction term as follows:
\begin{align}
f^{(1)}_k =& -\tau^*D\left[\bm{v}^s + \frac{e}{\hbar}(\bm{v}^s\cdot\bm{\Omega}^s)\bm{B}\right]\cdot\left[e\bm{E} + \left(\frac{\epsilon^s - \mu}{T}\right)\bm{\nabla} T\right]\nonumber\\
 \times \left(\frac{\partial f_{eq}}{\partial\epsilon^s}\right) &-2\left(\frac{\tau^{*}}{\tau_{\nu}}\right)\left(\frac{\partial f_{eq}}{\partial\epsilon^s}\right)\times\left[\delta\mu^s + \left(\frac{\epsilon^s- \mu}{T}\right)\delta T^s\right]
\label{A11}
\end{align}

\balance

\begin{widetext}
\setcounter{equation}{12}
\markboth{}{}
\noindent Now, to obtain the second order correction, the correction equation is simply modified into,
\begin{align}
&D\left(\bm{v}^s + \frac{e}{\hbar}(\bm{v}^s\cdot\bm{\Omega}^s)\bm{B}\right)\cdot \left(e\bm{E} + \left(\frac{\epsilon^s - \mu}{T}\right)\bm{\nabla} T\right) \
\left(-\frac{\partial f^{(1)}_k}{\partial \epsilon^s}\right)
= -\left(\frac{\delta f^{(2)}_k}{\tau^*}\right) - \frac{2}{\tau_\nu}\left[\left(\frac{\partial^2f_{eq}}{\partial \epsilon^{s2}}\right)(\delta\mu^s)^2 \right. \nonumber \\
& \left. + 2\left(\frac{1}{T}\left(\frac{\partial f_{eq}}{\partial \epsilon^{s}}\right) + \left(\frac{\epsilon^s - \mu}{T}\right)\left(\frac{\partial^2f_{eq}}{\partial \epsilon^{s2}}\right)\right)(\delta\mu^s)(\delta T^s)\right]
- \frac{2}{\tau_\nu}\left[2\left(\frac{\epsilon^s - \mu}{T^2}\right)\left(\frac{\partial f_{eq}}{\partial \epsilon^{s}}\right) \right. 
 \left. + \left(\frac{\epsilon^s - \mu}{T}\right)^2\left(\frac{\partial^2f_{eq}}{\partial \epsilon^{s2}}\right)\right](\delta T^{s})^2
\label{A12}
\end{align}
\noindent Using equations (\ref{A11}) and (\ref{A12}), we obtain the second-order corrections,

\begin{align}
f^{(2)}_k =& +(\tau^*)^2D^2\left[\tilde{\bm{v}}^s.\left(e\bm{E} + \left(\frac{\epsilon^s - \mu}{T}\right)\bm{\nabla} T\right)\right] \times \left(\tilde{\bm{v}}^s.\left(\frac{\bm{\nabla} T}{T}\right)\right)\left(\frac{\partial f_{eq}}{\partial\epsilon^s}\right) \nonumber \\
&+ (\tau^*)^2D^2\left[\tilde{\bm{v}}^s.\left(e\bm{E} + \left(\frac{\epsilon^s - \mu}{T}\right)\bm{\nabla} T\right)\right]^2\left(\frac{\partial^2 f_{eq}}{\partial\epsilon^{s2}}\right) \nonumber \\
&+ 2\left(\frac{(\tau^*)^2}{\tau_\nu}\right)D\left[\left(\frac{\delta T^s}{T}\right)\left(\frac{\partial f_{eq}}{\partial\epsilon^s}\right) \right.\left. + \left(\delta\mu^s + \left(\frac{\epsilon^s - \mu}{T}\right)\delta T^s\right)\left(\frac{\partial^2 f_{eq}}{\partial\epsilon^{s2}}\right)\right] \nonumber \\
&-2\left(\frac{\tau^*}{\tau_\nu}\right)\left[\left(\frac{\partial^2 f_{eq}}{\partial\epsilon^{s2}}\right)(\delta\mu^s)^2 + \frac{2}{T}\left(\left(\frac{\partial f_{eq}}{\partial\epsilon^{s}}\right) \right.\right. \left.\left.+ (\epsilon^s - \mu)\left(\frac{\partial^2 f_{eq}}{\partial\epsilon^{s2}}\right)\right)(\delta\mu^s)(\delta T^s)\right] \nonumber \\
&-2\left(\frac{\tau^*}{\tau_\nu}\right)\left[\frac{2}{T^2}(\epsilon^s - \mu)\left(\frac{\partial f_{eq}}{\partial\epsilon^{s}}\right) \right. \left. + \frac{1}{T^2}(\epsilon^s - \mu)^2\left(\frac{\partial^2 f_{eq}}{\partial\epsilon^{s2}}\right)\right](\delta T^s)^2
\label{A13}
\end{align}

\noindent Here, $\tilde{\bm{v^s}} = \bm{v^s} + \frac{e}{\hbar}(\bm{v}^s.\bm{\Omega}^s)\bm{B}$. In the chiral limit, where $\tau_\nu \gg \tau^*$, we have,

\begin{align}
f^{(2)}_k = & -2\left(\frac{\tau^*}{\tau_\nu}\right)\left[\left(\frac{\partial^2 f_{eq}}{\partial\epsilon^{s2}}\right)(\delta\mu^s)^2 + \right. \left.\frac{2}{T}\left(\left(\frac{\partial f_{eq}}{\partial\epsilon^{s}}\right) + (\epsilon^s - \mu)\left(\frac{\partial^2 f_{eq}}{\partial\epsilon^{s2}}\right)\right)(\delta\mu^s)(\delta T^s)\right] \nonumber\\
&-2\left(\frac{\tau^*}{\tau_\nu}\right)\left[\frac{2}{T^2}(\epsilon^s - \mu)\left(\frac{\partial f_{eq}}{\partial\epsilon^{s}}\right) + \right.\left.\frac{1}{T^2}(\epsilon^s - \mu)^2\left(\frac{\partial^2 f_{eq}}{\partial\epsilon^{s2}}\right)\right](\delta T^s)^2
\label{A15}
\end{align}

\noindent This equation will be used to derive the current and the various transport coefficients associated with it. 

\end{widetext}

\begin{table*}[t]
\centering
\caption{Comparison of linear order (in driving fields) response coefficients between 3D spin–orbit coupled (SOC) metals and Weyl semimetals (WSMs) (Ref.~\cite{PhysRevResearch.2.013088}).}
\begin{tabular}{|c|c|c|}
\hline
\textbf{Quantity} & \textbf{SOC metals} & \textbf{Weyl semimetals (Ref.~\cite{PhysRevResearch.2.013088})} \\
\hline
$\sigma_{yx}$ vs $\mu$  & scaling from Eq. (\ref{B1}) & $\sigma_{yx}^{WSM} \propto 1/\mu^{2}$ \\
\hline
$\alpha_{yx}$ vs $\mu$ & scaling from Eq. (\ref{B2}) & $ \alpha_{yx}^{WSM} \propto 1/\mu^{3}$ \\
\hline
$\sigma_{yx}/\alpha_{yx}$ vs B & $\propto B^{2}$ & $\propto B^{2}$ \\
\hline
$\sigma_{yx}/\alpha_{yx}$ vs $\theta$ & $\propto \cos{\theta}\sin{\theta}$  & $\propto \cos{\theta}\sin{\theta}$  \\
\hline
\end{tabular}
\label{tab:comparison}
\end{table*}

\section{LINEAR ORDER CHARGE TRANSPORT}

\label{appendix: Appendix C}

We provide a derivation of the linear order transport coefficients in this section. Substituting the expressions for velocity (Eq. (\ref{1.1})) and Eq.(\ref{A11}) into the current equations (Eq. (\ref{2.1})), we obtain,

\begin{eqnarray}
\bm j^{s(1)}_{e}
  &=&-\frac{2e\tau^{*}}{\tau_{\nu}}\int\!\frac{d^{3}k}{(2\pi)^{3}}
     \bigl[\delta\mu^{s}+(\epsilon^{s}-\mu)\tfrac{\delta T^{s}}{T}\bigr]\nonumber\\
  &&\times\Bigl[\frac{e}{\hbar}(\bm v^{s}\!\cdot\!\bm\Omega^{s})\bm B\Bigr]
     \Bigl(-\frac{\partial f^{s}_{eq}}{\partial\epsilon^{s}}\Bigr).
\label{2.1.8}
\end{eqnarray}
\begin{eqnarray}
\bm j^{s(1)}_{Q}
  &=&\frac{2\tau^{*}}{\tau_{\nu}}\int\!\frac{d^{3}k}{(2\pi)^{3}}
     \bigl[\delta\mu^{s}+(\epsilon^{s}-\mu)\tfrac{\delta T^{s}}{T}\bigr]\nonumber\\
  &&\times (\epsilon^{s}-\mu)\Bigl[\frac{e}{\hbar}(\bm v^{s}\!\cdot\!\bm\Omega^{s})\bm B\Bigr]
     \Bigl(-\frac{\partial f^{s}_{eq}}{\partial\epsilon^{s}}\Bigr).
\label{appcurrQ}
\end{eqnarray}

 \noindent For concreteness we work in the planar geometry  
\(\bm E =E_{x}\hat{\bm x}\),  
\(\nabla T=\nabla_{x}T\,\hat{\bm x}\),  
and a magnetic field confined to the \(x\!-\!y\) plane,  
\(\bm B=B\cos\theta\,\hat{\bm x}+B\sin\theta\,\hat{\bm y}\),  
with \(\theta\) measured from the \(x\)-axis [Fig.(\ref{Schematic})].  
We focus on \(j^{(1)}_{ey}\) and \(j^{(1)}_{Qy}\) as the primary interest of this work lies in the chiral anomaly induced planar Hall (coefficient $\sigma_{yx}$, see \eqref{linresp}), planar Nernst (coefficient $\alpha_{yx}$, see \eqref{linresp}) effects, both of which manifest through the transverse charge current, and the chiral anomaly induced planar Ettingshausen (coefficient $\bar{\alpha}_{yx}$, see \eqref{linrespQ}), planar thermal Hall (coefficient $\kappa_{yx}$, see \eqref{linrespQ}) effects, both of which manifest through the transverse thermal current. Writing the linear response relations as
\begin{eqnarray}
\bm j^{(1)}_{a}=\sum_{b}\bigl[\sigma_{ab}E_{b}-\alpha_{ab}\nabla_{b}T\bigr]
\label{linresp}
\end{eqnarray}
\begin{eqnarray}
\bm j^{(1)}_{Qa}=\sum_{b}\bigl[\bar\alpha_{ab}E_{b}-\kappa_{ab}\nabla_{b}T\bigr]
\label{linrespQ}
\end{eqnarray}
and inserting Eqs.\,\eqref{A9} and \eqref{A10} into Eqs.\,\eqref{2.1.8} and \,\eqref{appcurrQ}, and using the definitions \eqref{linresp} and \eqref{linrespQ} yields \cite{PhysRevB.108.045405}
\begin{eqnarray}
\sigma_{yx} &=& -\tau^{*}e^{2}B^{2}\sin\theta\cos\theta\,
   \sum_{s}\Bigl[S^{12,s}R^{01,s}_{12}C^{0}_{s} \nonumber\\
&&\hspace*{4.3em}+\,S^{01,s}R^{12,s}_{12}C^{1}_{s}\Bigr]
\label{2.1.2}
\end{eqnarray}
\begin{eqnarray}
\alpha_{yx} &=& -\tau^{*}\frac{e}{T}B^{2}\sin\theta\cos\theta\,
   \sum_{s}\Bigl[S^{12,s}R^{12,s}_{12}C^{0}_{s} \nonumber\\
&&\hspace*{4.3em}+\,S^{01,s}R^{21,s}_{10}C^{1}_{s}\Bigr]
\label{2.1.3}
\end{eqnarray}
\begin{eqnarray}
\bar\alpha_{yx} &=& \tau^{*}eB^{2}\sin\theta\cos\theta\,
   \sum_{s}\Bigl[S^{12,s}R^{01,s}_{12}C^{1}_{s} \nonumber\\
&&\hspace*{4.3em}+\,S^{01,s}R^{12,s}_{12}C^{2}_{s}\Bigr]
\label{alphabarlin}
\end{eqnarray}
\begin{eqnarray}
\kappa_{yx} &=& \tau^{*}\frac{1}{T}B^{2}\sin\theta\cos\theta\,
   \sum_{s}\Bigl[S^{12,s}R^{12,s}_{12}C^{1}_{s} \nonumber\\
&&\hspace*{4.3em}+\,S^{01,s}R^{21,s}_{10}C^{2}_{s}\Bigr]
\label{kappalin}
\end{eqnarray}

where
\begin{eqnarray}
R^{\alpha\beta,s}_{\gamma\delta}
   &=&\frac{C^{\alpha}_{s}}{\mathcal{D}^{\gamma}_{s}}
      -\frac{C^{\beta}_{s}}{\mathcal{D}^{\delta}_{s}},\label{2.2.5}\\
S^{\alpha\beta,s}&=&\frac{\mathcal{D}^{\alpha}_{s}\mathcal{D}^{\beta}_{s}}
                         {\mathcal{D}^{0}_{s}\mathcal{D}^{2}_{s}+(\mathcal{D}^{1}_{s})^{2}},\label{R_ten}\\
C^{\nu}_{s}
  &=&\int[dk]\Bigl(-\frac{\partial f_{eq}}{\partial\epsilon^{s}}\Bigr)
     \frac{e}{\hbar}(\bm v^{s}\!\cdot\!\bm\Omega^{s})
     (\epsilon^{s}-\mu)^{\nu}.\label{2.1.9}
\end{eqnarray}

\section{COMPARISON WITH WEYL SEMIMETALS}

\label{appendix: Appendix B}

We now provide a comparison between the results for the three dimensional spin orbit coupled (SOC) system and those for a Weyl semimetal (WSM). Contrasting our results with those for WSMs in Ref.~\cite{PhysRevResearch.2.013088}, the key phenomenological distinction lies in the dependence on chemical potential. The analytic expressions for the linear-order electric and thermal conductivities in the 3D SOC system are given by Eqs.~(\ref{3.15}) and (\ref{3.16}), whereas the corresponding results for the WSM are given in Eqs.~(\textcolor{red}{17}) and (\textcolor{red}{18}) in Ref.~\cite{PhysRevResearch.2.013088}.

For completeness, the low temperature expansions of the linear-order electric and thermal conductivities for 3D SOC metals are provided below:
\begin{eqnarray}
\sigma^{3D}_{yx} =&& \left(\frac{\tau^*e^4}{4\sqrt{2}\pi^2\hbar m^{\frac{3}{2}}\sqrt{\epsilon_\alpha}}\right)B^2\cos(\theta)\sin(\theta) \nonumber \\
&&\times\left[\frac{(2+\Tilde{\mu})\sqrt{1+\Tilde{\mu}}}{\Tilde{\mu}^2}\right]
\label{B1}
\end{eqnarray}
\begin{eqnarray}
\alpha^{3D}_{yx} =&& \left(\frac{\tau^*e^3k_B}{24\sqrt{2}\beta\hbar m^{\frac{3}{2}}\epsilon_\alpha^{\frac{3}{2}}}\right)B^2\cos(\theta)\sin(\theta) \nonumber \\ 
&&\times\left[\frac{\Tilde{\mu}^2 + 8\Tilde{\mu} + 8}{\Tilde{\mu}^3\sqrt{1+\Tilde{\mu}}}\right]
\label{B2}
\end{eqnarray}
The corresponding linear-order conductivities for a WSM (taken directly from Ref.~\cite{PhysRevResearch.2.013088}, with the corresponding Hamiltonian in Eq.~(\textcolor{red}{13})) are: 
\begin{eqnarray}
\sigma^{WSM}_{yx} =&& \left(\frac{\tau e^4v_F^3}{16\pi^2\hbar} \right)B^2\cos(\theta)\sin(\theta) \nonumber \\
&&\times\left[\frac{1}{\mu^2}\right]
\label{B3}
\end{eqnarray}
\begin{eqnarray}
\alpha^{WSM}_{yx} =&& \left(\frac{\tau^*e^3v_F^3k_B}{24\beta\hbar}\right)B^2\cos(\theta)\sin(\theta) \nonumber \\ 
&&\times\left[\frac{1}{\mu^3}\right]
\label{B4}
\end{eqnarray}
Table~\ref{tab:comparison} summarizes the similarities and differences for the linear order (in driving fields) response coefficients between the two systems. While the angular and magnetic field dependencies for the electric and thermal conductivities are similar due to the common semiclassical formalism, the chemical potential scaling provides a clear point of distinction. To the best of our knowledge, second-order response calculations in the driving electric field and thermal gradient for WSMs within the double scattering formalism have not been reported. 

\bibliography{references}

@article{Hasan2010,
  author = {Hasan, M. Zahid and Kane, Charles L.},
  title = {Colloquium: Topological insulators},
  journal = {Rev. Mod. Phys.},
  volume = {82},
  number = {4},
  pages = {3045--3067},
  year = {2010},
  doi = {10.1103/RevModPhys.82.3045},
  url = {https://doi.org/10.1103/RevModPhys.82.3045}
}

@article{Qi2011,
  author = {Qi, Xiao-Liang and Zhang, Shou-Cheng},
  title = {Topological insulators and superconductors},
  journal = {Rev. Mod. Phys.},
  volume = {83},
  number = {4},
  pages = {1057--1110},
  year = {2011},
  doi = {10.1103/RevModPhys.83.1057},
  url = {https://doi.org/10.1103/RevModPhys.83.1057}
}

@article{Armitage2018,
  author = {Armitage, N. P. and Mele, E. J. and Vishwanath, A.},
  title = {Weyl and Dirac semimetals in three-dimensional solids},
  journal = {Rev. Mod. Phys.},
  volume = {90},
  number = {1},
  pages = {015001},
  year = {2018},
  doi = {10.1103/RevModPhys.90.015001},
  url = {https://doi.org/10.1103/RevModPhys.90.015001}
}

@article{Burkov2014,
  author = {Burkov, A. A.},
  title = {Chiral anomaly and diffusive magnetotransport in Weyl metals},
  journal = {Phys. Rev. Lett.},
  volume = {113},
  pages = {247203},
  year = {2014},
  doi = {10.1103/PhysRevLett.113.247203},
  url = {https://doi.org/10.1103/PhysRevLett.113.247203}
}

@article{Young2012,
  author = {Young, Steve M. and Zaheer, Safwan and Teo, Jeffrey C. Y. and Kane, Charles L. and Mele, Eugene J. and Rappe, Andrew M.},
  title = {Dirac Semimetal in Three Dimensions},
  journal = {Phys. Rev. Lett.},
  volume = {108},
  pages = {140405},
  year = {2012},
  doi = {10.1103/PhysRevLett.108.140405},
  url = {https://doi.org/10.1103/PhysRevLett.108.140405}
}

@article{Wang2012,
  author = {Wang, Zhijun and Sun, Yiyuan and Chen, Xing-Qiu and Franchini, Cesare and Xu, Gang and Weng, Hongming and Dai, Xi and Fang, Zhong},
  title = {Dirac semimetal and topological phase transitions in A$_3$Bi (A=Na, K, Rb)},
  journal = {Phys. Rev. B},
  volume = {85},
  pages = {195320},
  year = {2012},
  doi = {10.1103/PhysRevB.85.195320},
  url = {https://doi.org/10.1103/PhysRevB.85.195320}
}

@article{Fu2008,
  author = {Fu, Liang and Kane, C. L.},
  title = {Superconducting Proximity Effect and Majorana Fermions at the Surface of a Topological Insulator},
  journal = {Phys. Rev. Lett.},
  volume = {100},
  pages = {096407},
  year = {2008},
  doi = {10.1103/PhysRevLett.100.096407},
  url = {https://doi.org/10.1103/PhysRevLett.100.096407}
}

@article{Sharma2019,
  author = {Sharma, Girish and Tewari, Sumanta},
  title = {Transverse thermopower in Dirac and Weyl semimetals},
  journal = {Phys. Rev. B},
  volume = {100},
  number = {19},
  pages = {195113},
  year = {2019},
  doi = {10.1103/PhysRevB.100.195113},
  url = {https://doi.org/10.1103/PhysRevB.100.195113}
}

@article{Jia:2016wal,
    author = "Jia, Shuang and Xu, Su-Yang and Hasan, M. Zahid",
    title = "{Weyl semimetals, Fermi arcs and chiral anomalies}",
    doi = "10.1038/nmat4787",
    journal = "Nat. Mater.",
    volume = "15",
    number = "11",
    pages = "1140--1144",
    year = "2016"
}

@book{Peskin:1995ev,
    author = "Peskin, Michael E. and Schroeder, Daniel V.",
    title = "{An Introduction to quantum field theory}",
    doi = "10.1201/9780429503559",
    isbn = "978-0-201-50397-5, 978-0-429-50355-9, 978-0-429-49417-8",
    publisher = "Addison-Wesley",
    address = "Reading, USA",
    year = "1995"
}

@article{PhysRevB.110.115436,
  title = {Stabilizing topological superconductivity in disordered spin-orbit coupled semiconductor-superconductor heterostructures},
  author = {Roy, Binayyak B. and Jaiswal, R. and Stanescu, Tudor D. and Tewari, Sumanta},
  journal = {Phys. Rev. B},
  volume = {110},
  issue = {11},
  pages = {115436},
  numpages = {20},
  year = {2024},
  month = {Sep},
  publisher = {American Physical Society},
  doi = {10.1103/PhysRevB.110.115436},
  url = {https://link.aps.org/doi/10.1103/PhysRevB.110.115436}
}

@article{Sato_2017,
   title={Topological superconductors: a review},
   volume={80},
   ISSN={1361-6633},
   url={http://dx.doi.org/10.1088/1361-6633/aa6ac7},
   DOI={10.1088/1361-6633/aa6ac7},
   number={7},
   journal={Reports on Progress in Physics},
   publisher={IOP Publishing},
   author={Sato, Masatoshi and Ando, Yoichi},
   year={2017},
   month=may, pages={076501} 
}

@article{Yangnature,
   title={Classification of stable three-dimensional Dirac semimetals with nontrivial topology},
   volume={5},
   url={ https://doi.org/10.1038/ncomms5898},
   DOI={ https://doi.org/10.1038/ncomms5898},
   number={4898},
   journal={Nat. Commun..},
   author={Yang, BJ and Nagaosa, N.},
   year={2014}
}

@article{doi:10.1080/01422419808240874,
author = {A. D. Dolgov},
title = {Baryogenesis, 30 years after},
journal = {Surveys in High Energy Physics},
volume = {13},
number = {1-3},
pages = {83--117},
year = {1998},
publisher = {Taylor \& Francis},
doi = {10.1080/01422419808240874},
URL = {https://doi.org/10.1080/01422419808240874},
eprint = {https://doi.org/10.1080/01422419808240874
}}

@article{PhysRevX.5.031023,
  title = {Observation of the Chiral-Anomaly-Induced Negative Magnetoresistance in 3D Weyl Semimetal TaAs},
  author = {Huang, Xiaochun and Zhao, Lingxiao and Long, Yujia and Wang, Peipei and Chen, Dong and Yang, Zhanhai and Liang, Hui and Xue, Mianqi and Weng, Hongming and Fang, Zhong and Dai, Xi and Chen, Genfu},
  journal = {Phys. Rev. X},
  volume = {5},
  issue = {3},
  pages = {031023},
  numpages = {9},
  year = {2015},
  month = {Aug},
  publisher = {American Physical Society},
  doi = {10.1103/PhysRevX.5.031023},
  url = {https://link.aps.org/doi/10.1103/PhysRevX.5.031023}
}

@article{PhysRevResearch.2.033511,
  title = {Chiral anomalies induced transport in Weyl metals in quantizing magnetic field},
  author = {Das, Kamal and Singh, Sahil Kumar and Agarwal, Amit},
  journal = {Phys. Rev. Res.},
  volume = {2},
  issue = {3},
  pages = {033511},
  numpages = {11},
  year = {2020},
  month = {Sep},
  publisher = {American Physical Society},
  doi = {10.1103/PhysRevResearch.2.033511},
  url = {https://link.aps.org/doi/10.1103/PhysRevResearch.2.033511}
}

@article{NIELSEN1981219,
title = {A no-go theorem for regularizing chiral fermions},
journal = {Phys. Lett. B},
volume = {105},
number = {2},
pages = {219-223},
year = {1981},
issn = {0370-2693},
doi = {https://doi.org/10.1016/0370-2693(81)91026-1},
url = {https://www.sciencedirect.com/science/article/pii/0370269381910261},
author = {H.B. Nielsen and M. Ninomiya}
}

@article{NIELSEN1983389,
title = {The Adler-Bell-Jackiw anomaly and Weyl fermions in a crystal},
journal = {Phys. Lett. B},
volume = {130},
number = {6},
pages = {389-396},
year = {1983},
issn = {0370-2693},
doi = {https://doi.org/10.1016/0370-2693(83)91529-0},
url = {https://www.sciencedirect.com/science/article/pii/0370269383915290},
author = {H.B. Nielsen and Masao Ninomiya}
}

@article{PhysRevLett.107.186806,
  title = {Chern Semimetal and the Quantized Anomalous Hall Effect in ${\mathrm{HgCr}}_{2}{\mathrm{Se}}_{4}$},
  author = {Xu, Gang and Weng, Hongming and Wang, Zhijun and Dai, Xi and Fang, Zhong},
  journal = {Phys. Rev. Lett.},
  volume = {107},
  issue = {18},
  pages = {186806},
  numpages = {5},
  year = {2011},
  month = {Oct},
  publisher = {American Physical Society},
  doi = {10.1103/PhysRevLett.107.186806},
  url = {https://link.aps.org/doi/10.1103/PhysRevLett.107.186806}
}

@article{PhysRevB.84.235126,
  title = {Topological nodal semimetals},
  author = {Burkov, A. A. and Hook, M. D. and Balents, Leon},
  journal = {Phys. Rev. B},
  volume = {84},
  issue = {23},
  pages = {235126},
  numpages = {14},
  year = {2011},
  month = {Dec},
  publisher = {American Physical Society},
  doi = {10.1103/PhysRevB.84.235126},
  url = {https://link.aps.org/doi/10.1103/PhysRevB.84.235126}
}

@article{PhysRevLett.107.127205,
  title = {Weyl Semimetal in a Topological Insulator Multilayer},
  author = {Burkov, A. A. and Balents, Leon},
  journal = {Phys. Rev. Lett.},
  volume = {107},
  issue = {12},
  pages = {127205},
  numpages = {4},
  year = {2011},
  month = {Sep},
  publisher = {American Physical Society},
  doi = {10.1103/PhysRevLett.107.127205},
  url = {https://link.aps.org/doi/10.1103/PhysRevLett.107.127205}
}

@article{PhysRevB.83.205101,
  title = {Topological semimetal and Fermi-arc surface states in the electronic structure of pyrochlore iridates},
  author = {Wan, Xiangang and Turner, Ari M. and Vishwanath, Ashvin and Savrasov, Sergey Y.},
  journal = {Phys. Rev. B},
  volume = {83},
  issue = {20},
  pages = {205101},
  numpages = {9},
  year = {2011},
  month = {May},
  publisher = {American Physical Society},
  doi = {10.1103/PhysRevB.83.205101},
  url = {https://link.aps.org/doi/10.1103/PhysRevB.83.205101}
}

@article{PhysRevB.105.L180303,
  title = {Chiral anomaly in noncentrosymmetric systems induced by spin-orbit coupling},
  author = {Cheon, Suik and Cho, Gil Young and Kim, Ki-Seok and Lee, Hyun-Woo},
  journal = {Phys. Rev. B},
  volume = {105},
  issue = {18},
  pages = {L180303},
  numpages = {6},
  year = {2022},
  month = {May},
  publisher = {American Physical Society},
  doi = {10.1103/PhysRevB.105.L180303},
  url = {https://link.aps.org/doi/10.1103/PhysRevB.105.L180303}
}

@article{Gao_2022,
doi = {10.1088/0256-307X/39/2/021101},
url = {https://dx.doi.org/10.1088/0256-307X/39/2/021101},
year = {2022},
month = {feb},
publisher = {Chinese Physical Society and IOP Publishing Ltd},
volume = {39},
number = {2},
pages = {021101},
author = {Lan-Lan Gao and Xu-Guang Huang},
title = {Chiral Anomaly in Non-Relativistic Systems: Berry Curvature and Chiral Kinetic Theory},
journal = {Chin. Phys. Lett.}
}

@article{PhysRevB.108.045405,
  title = {Chiral anomalies in three-dimensional spin-orbit coupled metals: Electrical, thermal, and gravitational anomalies},
  author = {Das, Sunit and Das, Kamal and Agarwal, Amit},
  journal = {Phys. Rev. B},
  volume = {108},
  issue = {4},
  pages = {045405},
  numpages = {14},
  year = {2023},
  month = {Jul},
  publisher = {American Physical Society},
  doi = {10.1103/PhysRevB.108.045405},
  url = {https://link.aps.org/doi/10.1103/PhysRevB.108.045405}
}

@article{He_2021,
   title={Kramers Weyl semimetals as quantum solenoids and their applications in spin-orbit torque devices},
   volume={4},
   ISSN={2399-3650},
   url={http://dx.doi.org/10.1038/s42005-021-00564-w},
   DOI={10.1038/s42005-021-00564-w},
   number={1},
   journal={Commun.. Phys.},
   publisher={Springer Science and Business Media LLC},
   author={He, Wen-Yu and Xu, Xiao Yan and Law, K. T.},
   year={2021},
   month=mar }

@article{PhysRevB.91.134401,
  title = {Transport theory of metallic $B20$ helimagnets},
  author = {Kang, Jian and Zang, Jiadong},
  journal = {Phys. Rev. B},
  volume = {91},
  issue = {13},
  pages = {134401},
  numpages = {9},
  year = {2015},
  month = {Apr},
  publisher = {American Physical Society},
  doi = {10.1103/PhysRevB.91.134401},
  url = {https://link.aps.org/doi/10.1103/PhysRevB.91.134401}
}

@article{PhysRevB.78.144511,
  title = {Effects of impurities on the upper critical field ${H}_{c2}$ in superconductors without inversion symmetry},
  author = {Samokhin, K. V.},
  journal = {Phys. Rev. B},
  volume = {78},
  issue = {14},
  pages = {144511},
  numpages = {8},
  year = {2008},
  month = {Oct},
  publisher = {American Physical Society},
  doi = {10.1103/PhysRevB.78.144511},
  url = {https://link.aps.org/doi/10.1103/PhysRevB.78.144511}
}

@article{492de033-74af-3791-807c-34957550404d,
 ISSN = {0003486X, 19398980},
 URL = {http://www.jstor.org/stable/1971423},
 author = {R. J. DiPerna and P. L. Lions},
 journal = {Annals of Mathematics},
 number = {2},
 pages = {321--366},
 publisher = {[Annals of Mathematics, Trustees of Princeton University on Behalf of the Annals of Mathematics, Mathematics Department, Princeton University]},
 title = {On the Cauchy Problem for Boltzmann Equations: Global Existence and Weak Stability},
 urldate = {2024-10-01},
 volume = {130},
 year = {1989}
}

@book{Girvin_Yang_2019, place={Cambridge}, title={Modern Condensed Matter Physics}, publisher={Cambridge University Press}, author={Girvin, Steven M. and Yang, Kun}, year={2019}}

@book{de1984non,
  title={Non-equilibrium Thermodynamics},
  author={de Groot, S.R. and Mazur, P.},
  isbn={9780486647418},
  lccn={84007956},
  series={Dover Books on Physics},
  url={https://books.google.com/books?id=HFAIv43rlGkC},
  year={1984},
  publisher={Dover Publications}
}

@article{PhysRevResearch.2.013088,
  title = {Thermal and gravitational chiral anomaly induced magneto-transport in Weyl semimetals},
  author = {Das, Kamal and Agarwal, Amit},
  journal = {Phys. Rev. Res.},
  volume = {2},
  issue = {1},
  pages = {013088},
  numpages = {9},
  year = {2020},
  month = {Jan},
  publisher = {American Physical Society},
  doi = {10.1103/PhysRevResearch.2.013088},
  url = {https://link.aps.org/doi/10.1103/PhysRevResearch.2.013088}
}

@article{PhysRevB.105.125131,
  title = {Chiral anomaly induced nonlinear Nernst and thermal Hall effects in Weyl semimetals},
  author = {Zeng, Chuanchang and Nandy, Snehasish and Tewari, Sumanta},
  journal = {Phys. Rev. B},
  volume = {105},
  issue = {12},
  pages = {125131},
  numpages = {11},
  year = {2022},
  month = {Mar},
  publisher = {American Physical Society},
  doi = {10.1103/PhysRevB.105.125131},
  url = {https://link.aps.org/doi/10.1103/PhysRevB.105.125131}
}

@article{RevModPhys.82.1959,
  title = {Berry phase effects on electronic properties},
  author = {Xiao, Di and Chang, Ming-Che and Niu, Qian},
  journal = {Rev. Mod. Phys.},
  volume = {82},
  issue = {3},
  pages = {1959--2007},
  numpages = {0},
  year = {2010},
  month = {Jul},
  publisher = {American Physical Society},
  doi = {10.1103/RevModPhys.82.1959},
  url = {https://link.aps.org/doi/10.1103/RevModPhys.82.1959}
}

@article{PhysRevB.103.045105,
  title = {Nonlinear Hall effect in Weyl semimetals induced by chiral anomaly},
  author = {Li, Rui-Hao and Heinonen, Olle G. and Burkov, Anton A. and Zhang, Steven S.-L.},
  journal = {Phys. Rev. B},
  volume = {103},
  issue = {4},
  pages = {045105},
  numpages = {11},
  year = {2021},
  month = {Jan},
  publisher = {American Physical Society},
  doi = {10.1103/PhysRevB.103.045105},
  url = {https://link.aps.org/doi/10.1103/PhysRevB.103.045105}
}

@article{PhysRevB.104.205124,
  title = {Chiral anomaly induced nonlinear Hall effect in semimetals with multiple Weyl points},
  author = {Nandy, Snehasish and Zeng, Chuanchang and Tewari, Sumanta},
  journal = {Phys. Rev. B},
  volume = {104},
  issue = {20},
  pages = {205124},
  numpages = {7},
  year = {2021},
  month = {Nov},
  publisher = {American Physical Society},
  doi = {10.1103/PhysRevB.104.205124},
  url = {https://link.aps.org/doi/10.1103/PhysRevB.104.205124}
}

@article{Bauer_2024,
   title={Quantum description of Fermi arcs in Weyl semimetals in a magnetic field},
   volume={6},
   ISSN={2643-1564},
   url={http://dx.doi.org/10.1103/PhysRevResearch.6.043201},
   DOI={10.1103/physrevresearch.6.043201},
   number={4},
   journal={Phys. Rev. Research},
   publisher={American Physical Society (APS)},
   author={Bauer, Tim and Buccheri, Francesco and De Martino, Alessandro and Egger, Reinhold},
   year={2024},
   month=nov }

@article{PhysRevB.107.L241101,
  title = {Evidence for three-dimensional Dirac conical bands in TlBiSSe by optical and magneto-optical spectroscopy},
  author = {Le Mardel\'e, F. and Wyzula, J. and Mohelsky, I. and Nasrallah, S. and Loh, M. and Ben David, S. and Toledano, O. and Tolj, D. and Novak, M. and Eguchi, G. and Paschen, S. and Bari\ifmmode \check{s}\else \v{s}\fi{}i\ifmmode \acute{c}\else \'{c}\fi{}, N. and Chen, J. and Kimura, A. and Orlita, M. and Rukelj, Z. and Akrap, Ana and Santos-Cottin, D.},
  journal = {Phys. Rev. B},
  volume = {107},
  issue = {24},
  pages = {L241101},
  numpages = {6},
  year = {2023},
  month = {Jun},
  publisher = {American Physical Society},
  doi = {10.1103/PhysRevB.107.L241101},
  url = {https://link.aps.org/doi/10.1103/PhysRevB.107.L241101}
}

@article{Wieder2020,
  author    = {Benjamin J. Wieder and Zhijun Wang and Jennifer Cano and Xi Dai and Leslie M. Schoop and Barry Bradlyn and B. Andrei Bernevig},
  title     = {Strong and fragile topological Dirac semimetals with higher-order Fermi arcs},
  journal   = {Nat. Commun..},
  year      = {2020},
  volume    = {11},
  number    = {1},
  pages     = {627},
  doi       = {10.1038/s41467-020-14443-5},
  url       = {https://doi.org/10.1038/s41467-020-14443-5},
  issn      = {2041-1723}
}

@article{RevModPhys.81.109,
  title = {The electronic properties of graphene},
  author = {Castro Neto, A. H. and Guinea, F. and Peres, N. M. R. and Novoselov, K. S. and Geim, A. K.},
  journal = {Rev. Mod. Phys.},
  volume = {81},
  issue = {1},
  pages = {109--162},
  numpages = {0},
  year = {2009},
  month = {Jan},
  publisher = {American Physical Society},
  doi = {10.1103/RevModPhys.81.109},
  url = {https://link.aps.org/doi/10.1103/RevModPhys.81.109}
}

@article{Sato2011,
  author    = {T. Sato and Kouji Segawa and K. Kosaka and S. Souma and K. Nakayama and K. Eto and T. Minami and Yoichi Ando and T. Takahashi},
  title     = {Unexpected mass acquisition of Dirac fermions at the quantum phase transition of a topological insulator},
  journal   = {Nat. Phys.},
  year      = {2011},
  volume    = {7},
  number    = {11},
  pages     = {840--844},
  doi       = {10.1038/nphys2058},
  url       = {https://doi.org/10.1038/nphys2058},
  issn      = {1745-2481}
}

@article{Liu_2014,
   title={Discovery of a Three-Dimensional Topological Dirac Semimetal, Na
            3
            Bi},
   volume={343},
   ISSN={1095-9203},
   url={http://dx.doi.org/10.1126/science.1245085},
   DOI={10.1126/science.1245085},
   number={6173},
   journal={Science},
   publisher={American Association for the Advancement of Science (AAAS)},
   author={Liu, Z. K. and Zhou, B. and Zhang, Y. and Wang, Z. J. and Weng, H. M. and Prabhakaran, D. and Mo, S.-K. and Shen, Z. X. and Fang, Z. and Dai, X. and Hussain, Z. and Chen, Y. L.},
   year={2014},
   month=feb, pages={864–867} }

@article{Jeon_2014,
   title={Landau quantization and quasiparticle interference in the three-dimensional Dirac semimetal Cd3As2},
   volume={13},
   ISSN={1476-4660},
   url={http://dx.doi.org/10.1038/nmat4023},
   DOI={10.1038/nmat4023},
   number={9},
   journal={Nature Materials},
   publisher={Springer Science and Business Media LLC},
   author={Jeon, Sangjun and Zhou, Brian B. and Gyenis, Andras and Feldman, Benjamin E. and Kimchi, Itamar and Potter, Andrew C. and Gibson, Quinn D. and Cava, Robert J. and Vishwanath, Ashvin and Yazdani, Ali},
   year={2014},
   month=jun, pages={851–856} }

@article{PhysRevB.89.235127,
  title = {Weyl and Dirac semimetals with ${\mathbb{Z}}_{2}$ topological charge},
  author = {Morimoto, Takahiro and Furusaki, Akira},
  journal = {Phys. Rev. B},
  volume = {89},
  issue = {23},
  pages = {235127},
  numpages = {13},
  year = {2014},
  month = {Jun},
  publisher = {American Physical Society},
  doi = {10.1103/PhysRevB.89.235127},
  url = {https://link.aps.org/doi/10.1103/PhysRevB.89.235127}
}

@article{PhysRevLett.108.266802,
  title = {Multi-Weyl Topological Semimetals Stabilized by Point Group Symmetry},
  author = {Fang, Chen and Gilbert, Matthew J. and Dai, Xi and Bernevig, B. Andrei},
  journal = {Phys. Rev. Lett.},
  volume = {108},
  issue = {26},
  pages = {266802},
  numpages = {5},
  year = {2012},
  month = {Jun},
  publisher = {American Physical Society},
  doi = {10.1103/PhysRevLett.108.266802},
  url = {https://link.aps.org/doi/10.1103/PhysRevLett.108.266802}
}

@article{PhysRevLett.108.046602,
  title = {Charge Transport in Weyl Semimetals},
  author = {Hosur, Pavan and Parameswaran, S. A. and Vishwanath, Ashvin},
  journal = {Phys. Rev. Lett.},
  volume = {108},
  issue = {4},
  pages = {046602},
  numpages = {5},
  year = {2012},
  month = {Jan},
  publisher = {American Physical Society},
  doi = {10.1103/PhysRevLett.108.046602},
  url = {https://link.aps.org/doi/10.1103/PhysRevLett.108.046602}
}

@article{Kitaev2003,
  author = {Kitaev, A. Y.},
  journal = {Annals. Phys.},
  volume = {303},
  pages = {2},
  year = {2003},
}

@article{Nayak2008,
  title = {Non-Abelian anyons and topological quantum computation},
  author = {Nayak, Chetan and Simon, Steven H. and Stern, Ady and Freedman, Michael and Das Sarma, Sankar},
  journal = {Rev. Mod. Phys.},
  volume = {80},
  issue = {3},
  pages = {1083--1159},
  numpages = {0},
  year = {2008},
  month = {Sep},
  publisher = {American Physical Society},
  doi = {10.1103/RevModPhys.80.1083},
  url = {https://link.aps.org/doi/10.1103/RevModPhys.80.1083}
}

@article{Schober_2024,
   title={Chiral Anomaly and Dynamos from Inhomogeneous Chemical Potential Fluctuations},
   volume={132},
   ISSN={1079-7114},
   url={http://dx.doi.org/10.1103/PhysRevLett.132.065101},
   DOI={10.1103/physrevlett.132.065101},
   number={6},
   journal={Phys. Rev. Lett.},
   publisher={American Physical Society (APS)},
   author={Schober, Jennifer and Rogachevskii, Igor and Brandenburg, Axel},
   year={2024},
   month=feb }

@article{PhysRevD.22.3080,
  title = {Equilibrium parity-violating current in a magnetic field},
  author = {Vilenkin, Alexander},
  journal = {Phys. Rev. D},
  volume = {22},
  issue = {12},
  pages = {3080--3084},
  numpages = {0},
  year = {1980},
  month = {Dec},
  publisher = {American Physical Society},
  doi = {10.1103/PhysRevD.22.3080},
  url = {https://link.aps.org/doi/10.1103/PhysRevD.22.3080}
}

@article{Wang2022,
  author    = {Mudi Wang and Qiyun Ma and Shan Liu and Ruo-Yang Zhang and Lei Zhang and Manzhu Ke and Zhengyou Liu and C. T. Chan},
  title     = {Observation of boundary induced chiral anomaly bulk states and their transport properties},
  journal   = {Nat. Commun..},
  year      = {2022},
  volume    = {13},
  number    = {1},
  pages     = {5916},
  doi       = {10.1038/s41467-022-33447-x},
  url       = {https://doi.org/10.1038/s41467-022-33447-x},
  issn      = {2041-1723}
}

@article{Arouca_2022,
   title={Quantum Field Theory Anomalies in Condensed Matter Physics},
   ISSN={2590-1990},
   url={http://dx.doi.org/10.21468/SciPostPhysLectNotes.62},
   DOI={10.21468/scipostphyslectnotes.62},
   journal={SciPost Physics Lecture Notes},
   publisher={Stichting SciPost},
   author={Arouca, Rodrigo and Cappelli, Andrea and Hansson, Hans},
   year={2022},
   month=sep }

@article{PhysRevB.108.L161106,
  title = {Gravitational anomaly in the ferrimagnetic topological Weyl semimetal NdAlSi},
  author = {Tanwar, Pardeep Kumar and Ahmad, Mujeeb and Alam, Md Shahin and Yao, Xiaohan and Tafti, Fazel and Matusiak, Marcin},
  journal = {Phys. Rev. B},
  volume = {108},
  issue = {16},
  pages = {L161106},
  numpages = {7},
  year = {2023},
  month = {Oct},
  publisher = {American Physical Society},
  doi = {10.1103/PhysRevB.108.L161106},
  url = {https://link.aps.org/doi/10.1103/PhysRevB.108.L161106}
}

@article{Ong_2021,
   title={Experimental signatures of the chiral anomaly in Dirac–Weyl semimetals},
   volume={3},
   ISSN={2522-5820},
   url={http://dx.doi.org/10.1038/s42254-021-00310-9},
   DOI={10.1038/s42254-021-00310-9},
   number={6},
   journal={Nat. Rev. Phys.},
   publisher={Springer Science and Business Media LLC},
   author={Ong, N. P. and Liang, Sihang},
   year={2021},
   month=may, pages={394–404} }

@article{LI2016107,
title = {Chiral magnetic effect in condensed matter systems},
journal = {Nucl. Phys. A},
volume = {956},
pages = {107-111},
year = {2016},
note = {The XXV International Conference on Ultrarelativistic Nucleus-Nucleus Collisions: Quark Matter 2015},
issn = {0375-9474},
doi = {https://doi.org/10.1016/j.nuclphysa.2016.03.055},
url = {https://www.sciencedirect.com/science/article/pii/S0375947416300410},
author = {Qiang Li and Dmitri E. Kharzeev}
}

@article{PhysRevResearch.2.032066,
  title = {Fundamental relations for anomalous thermoelectric transport coefficients in the nonlinear regime},
  author = {Zeng, Chuanchang and Nandy, Snehasish and Tewari, Sumanta},
  journal = {Phys. Rev. Res.},
  volume = {2},
  issue = {3},
  pages = {032066},
  numpages = {6},
  year = {2020},
  month = {Sep},
  publisher = {American Physical Society},
  doi = {10.1103/PhysRevResearch.2.032066},
  url = {https://link.aps.org/doi/10.1103/PhysRevResearch.2.032066}
}

@article{varma2024magnetotransport,
  title={Magnetotransport in spin-orbit coupled noncentrosymmetric and Weyl metals},
  author={Varma, Gautham and Ahmad, Azaz and Tewari, Sumanta and Sharma, Gargee},
  journal={Phys. Rev. B},
  volume={109},
  number={16},
  pages={165114},
  year={2024},
  publisher={APS}
}

@article{ahmad2023longitudinal,
  title={Longitudinal magnetoconductance and the planar Hall conductance in inhomogeneous Weyl semimetals},
  author={Ahmad, Azaz and Raman, Karthik V and Tewari, Sumanta and Sharma, Gargee},
  journal={Phys. Rev. B},
  volume={107},
  number={14},
  pages={144206},
  year={2023},
  publisher={APS}
}

@article{ahmad2021longitudinal,
  title={Longitudinal magnetoconductance and the planar Hall effect in a lattice model of tilted Weyl fermions},
  author={Ahmad, Azaz and Sharma, Gargee},
  journal={Phys. Rev. B},
  volume={103},
  number={11},
  pages={115146},
  year={2021},
  publisher={APS}
}

@article{sharma2017nernst,
  title={Nernst effect in Dirac and inversion-asymmetric Weyl semimetals},
  author={Sharma, Gargee and Moore, Christopher and Saha, Subhodip and Tewari, Sumanta},
  journal={Phys. Rev. B},
  volume={96},
  number={19},
  pages={195119},
  year={2017},
  publisher={APS}
}

@article{sharma2023decoupling,
  title={Decoupling intranode and internode scattering in Weyl fermions},
  author={Sharma, Gargee and Nandy, Snehasish and Raman, Karthik V and Tewari, Sumanta},
  journal={Phys. Rev. B},
  volume={107},
  number={11},
  pages={115161},
  year={2023},
  publisher={APS}
}

@article{goswami2015optical,
  title={Optical activity as a test for dynamic chiral magnetic effect of weyl semimetals},
  author={Goswami, Pallab and Sharma, Gargee and Tewari, Sumanta},
  journal={Phys. Rev. B},
  volume={92},
  number={16},
  pages={161110},
  year={2015},
  publisher={APS}
}

@article{sharma2016nernst,
  title={Nernst and magnetothermal conductivity in a lattice model of Weyl fermions},
  author={Sharma, Gargee and Goswami, Pallab and Tewari, Sumanta},
  journal={Phys. Rev. B},
  volume={93},
  number={3},
  pages={035116},
  year={2016},
  publisher={APS}
}

@article{nandy2017chiral,
  title={Chiral anomaly as the origin of the planar Hall effect in Weyl semimetals},
  author={Nandy, S and Sharma, Gargee and Taraphder, A and Tewari, Sumanta},
  journal={Phys. Rev. Letters},
  volume={119},
  number={17},
  pages={176804},
  year={2017},
  publisher={APS}
}

@article{sharma2017chiral,
  title={Chiral anomaly and longitudinal magnetotransport in type-II Weyl semimetals},
  author={Sharma, Gargee and Goswami, Pallab and Tewari, Sumanta},
  journal={Phys. Rev. B},
  volume={96},
  number={4},
  pages={045112},
  year={2017},
  publisher={APS}
}

@article{sharma2020sign,
  title={Sign of longitudinal magnetoconductivity and the planar Hall effect in Weyl semimetals},
  author={Sharma, Gargee and Nandy, S and Tewari, Sumanta},
  journal={Phys. Rev. B},
volume={102},
number={20},
pages={205107},
year={2020},
publisher={APS}
}

@article{ahmad2024geometry,
  title={Geometry, anomaly, topology, and transport in Weyl fermions},
  author={Ahmad, Azaz and Varma, Gautham and Sharma, Gargee},
  journal={Journal of Physics: Condensed Matter},
  volume={37},
  number={4},
  pages={043001},
  year={2024},
  publisher={IOP Publishing}
}

@article{ahmad2025chiral,
  title={Chiral anomaly induced nonlinear Hall effect in three-dimensional chiral fermions},
  author={Ahmad, Azaz and K, Gautham Varma and Sharma, Gargee},
  journal={Phys. Rev. B},
  volume={111},
  number={3},
  pages={035138},
  year={2025},
  publisher={APS}
}

@book{volovik2003universe,
  title={The universe in a helium droplet},
  author={Volovik, Grigory E},
  volume={117},
  year={2003},
  publisher={Oxford University Press on Demand}
}

@article{xu2011chern,
  title={Chern semimetal and the quantized anomalous Hall effect in HgCr 2 Se 4},
  author={Xu, Gang and Weng, Hongming and Wang, Zhijun and Dai, Xi and Fang, Zhong},
  journal={Phys. Rev. Lett.},
  volume={107},
  number={18},
  pages={186806},
  year={2011},
  publisher={APS}
}

@article{zyuzin2012weyl,
  title={Weyl semimetal with broken time reversal and inversion symmetries},
  author={Zyuzin, AA and Wu, Si and Burkov, AA},
  journal={Phys. Rev. B},
  volume={85},
  number={16},
  pages={165110},
  year={2012},
  publisher={APS}
}

@article{son2013chiral,
  title={Chiral anomaly and classical negative magnetoresistance of Weyl metals},
  author={Son, DT and Spivak, BZ},
  journal={Phys. Rev. B},
  volume={88},
  number={10},
  pages={104412},
  year={2013},
  publisher={APS}
}

@article{zyuzin2017magnetotransport,
  title={Magnetotransport of Weyl semimetals due to the chiral anomaly},
  author={Zyuzin, Vladimir A},
  journal={Phys. Rev. B},
  volume={95},
  number={24},
  pages={245128},
  year={2017},
  publisher={APS}
}

@article{kim2014boltzmann,
  title={Boltzmann equation approach to anomalous transport in a Weyl metal},
  author={Kim, Ki-Seok and Kim, Heon-Jung and Sasaki, M},
  journal={Phys. Rev. B},
  volume={89},
  number={19},
  pages={195137},
  year={2014},
  publisher={APS}
}

@article{he2014quantum,
  title={Quantum transport evidence for the three-dimensional Dirac semimetal phase in Cd 3 As 2},
  author={He, LP and Hong, XC and Dong, JK and Pan, J and Zhang, Z and Zhang, J and Li, SY},
  journal={Phys. Rev. Lett.},
  volume={113},
  number={24},
  pages={246402},
  year={2014},
  publisher={APS}
}

@article{cortijo2016linear,
  title={Linear magnetochiral effect in Weyl semimetals},
  author={Cortijo, Alberto},
  journal={Phys. Rev. B},
  volume={94},
  number={24},
  pages={241105},
  year={2016},
  publisher={APS}
}

@article{lundgren2014thermoelectric,
  title={Thermoelectric properties of Weyl and Dirac semimetals},
  author={Lundgren, Rex and Laurell, Pontus and Fiete, Gregory A},
  journal={Phys. Rev. B},
  volume={90},
  number={16},
  pages={165115},
  year={2014},
  publisher={APS}
}

@article{goswami2015axial,
  title={Axial anomaly and longitudinal magnetoresistance of a generic three-dimensional metal},
  author={Goswami, Pallab and Pixley, JH and Sarma, S Das},
  journal={Phys. Rev. B},
  volume={92},
  number={7},
  pages={075205},
  year={2015},
  publisher={APS}
}

@article{xiong2015evidence,
  title={Evidence for the chiral anomaly in the Dirac semimetal Na3Bi},
  author={Xiong, Jun and Kushwaha, Satya K and Liang, Tian and Krizan, Jason W and Hirschberger, Max and Wang, Wudi and Cava, Robert Joseph and Ong, Nai Phuan},
  journal={Science},
  volume={350},
  number={6259},
  pages={413--416},
  year={2015},
  publisher={American Association for the Advancement of Science}
}

@article{hirschberger2016chiral,
  title={The chiral anomaly and thermopower of Weyl fermions in the half-Heusler GdPtBi},
  author={Hirschberger, Max and Kushwaha, Satya and Wang, Zhijun and Gibson, Quinn and Liang, Sihang and Belvin, Carina A and Bernevig, Bogdan Andrei and Cava, Robert Joseph and Ong, Nai Phuan},
  journal={Nat. Mater.},
  volume={15},
  number={11},
  pages={1161--1165},
  year={2016},
  publisher={Nature Publishing Group}
}

@article{bednik2020magnetotransport,
  title={Magnetotransport and internodal tunnelling in Weyl semimetals},
  author={Bednik, G and Tikhonov, KS and Syzranov, SV},
  journal={Phys. Rev. Research},
  volume={2},
  number={2},
  pages={023124},
  year={2020},
  publisher={APS}
}

@article{knoll2020negative,
  title={Negative longitudinal magnetoconductance at weak fields in Weyl semimetals},
  author={Knoll, Andy and Timm, Carsten and Meng, Tobias},
  journal={Phys. Rev. B},
  volume={101},
  number={20},
  pages={201402},
  year={2020},
  publisher={APS}
}

@article{zhong2015optical,
  title={Optical gyrotropy from axion electrodynamics in momentum space},
  author={Zhong, Shudan and Orenstein, Joseph and Moore, Joel E},
  journal={Phys. Rev. Lett.},
  volume={115},
  number={11},
  pages={117403},
  year={2015},
  publisher={APS}
}

@article{goswami2013axionic,
  title={Axionic field theory of (3+ 1)-dimensional Weyl semimetals},
  author={Goswami, Pallab and Tewari, Sumanta},
  journal={Phys. Rev. B},
  volume={88},
  number={24},
  pages={245107},
  year={2013},
  publisher={APS}
}

@article{kundu2020magnetotransport,
  title={Magnetotransport of Weyl semimetals with tilted Dirac cones},
  author={Kundu, Anirban and Siu, Zhuo Bin and Yang, Hyunsoo and Jalil, Mansoor BA},
  journal={New J. Phys.},
  volume={22},
  number={8},
  pages={083081},
  year={2020},
  publisher={IOP Publishing}
}

@article{li2016chiral,
  title={Chiral magnetic effect in ZrTe 5},
  author={Li, Qiang and Kharzeev, Dmitri E and Zhang, Cheng and Huang, Yuan and Pletikosi{\'c}, I and Fedorov, AV and Zhong, RD and Schneeloch, JA and Gu, GD and Valla, T},
  journal={Nat. Phys.},
  volume={12},
  number={6},
  pages={550--554},
  year={2016},
  publisher={Nature Publishing Group}
}

@article{liang2015ultrahigh,
  title={Ultrahigh mobility and giant magnetoresistance in the Dirac semimetal Cd 3 As 2},
  author={Liang, Tian and Gibson, Quinn and Ali, Mazhar N and Liu, Minhao and Cava, Robert Joseph and Ong, Nai Phuan},
  journal={Nat. Mater.},
  volume={14},
  number={3},
  pages={280--284},
  year={2015},
  publisher={Nature Publishing Group}
}

@article{zhang2016signatures,
  title={Signatures of the Adler--Bell--Jackiw chiral anomaly in a Weyl fermion semimetal},
  author={Zhang, Cheng-Long and Xu, Su-Yang and Belopolski, Ilya and Yuan, Zhujun and Lin, Ziquan and Tong, Bingbing and Bian, Guang and Alidoust, Nasser and Lee, Chi-Cheng and Huang, Shin-Ming and others},
  journal={Nat. Commun..},
  volume={7},
  number={1},
  pages={1--9},
  year={2016},
  publisher={Nature Publishing Group}
}

@article{Saha2018,
  author    = {Subhodip Saha and Sumanta Tewari},
  title     = {Anomalous Nernst effect in type-II Weyl semimetals},
  journal   = {The Eur. Phys. J. B},
  year      = {2018},
  volume    = {91},
  number    = {1},
  pages     = {4},
  doi       = {10.1140/epjb/e2017-80437-4},
  url       = {https://doi.org/10.1140/epjb/e2017-80437-4},
  issn      = {1434-6036}
}

@article{PhysRevB.96.195119,
  title = {Nernst effect in Dirac and inversion-asymmetric Weyl semimetals},
  author = {Sharma, Gargee and Moore, Christopher and Saha, Subhodip and Tewari, Sumanta},
  journal = {Phys. Rev. B},
  volume = {96},
  issue = {19},
  pages = {195119},
  numpages = {9},
  year = {2017},
  month = {Nov},
  publisher = {American Physical Society},
  doi = {10.1103/PhysRevB.96.195119},
  url = {https://link.aps.org/doi/10.1103/PhysRevB.96.195119}
}

@article{sau2010generic,
  title={Generic New Platform for Topological Quantum Computation Using Semiconductor Heterostructures},
  author={Sau, Jay D and Lutchyn, Roman M and Tewari, Sumanta and Das Sarma, Sankar},
  journal={Phys. Rev. Lett.},
  volume={104},
  number={4},
  pages={040502},
  year={2010},
  publisher={APS}
}

@incollection{sau2021topological,
  title={Topological superconductivity in spin-orbit-coupled semiconducting nanowires},
  author={Sau, Jay and Tewari, Sumanta},
  booktitle={Semiconductors and Semimetals},
  volume={108},
  pages={125--194},
  year={2021},
  publisher={Elsevier}
}

@article{jackiw2008axial,
  title={Axial anomaly},
  author={Jackiw, Roman W},
  journal={Scholarpedia},
  volume={3},
  number={10},
  pages={7302},
  year={2008}
}

@misc{claude1980quantum,
  title={Quantum field theory},
  author={Claude, Itzykson and Bernard, Zuber Jean},
  year={1980},
  publisher={McGraw Hill Book Company}
}

@article{huang2015observation,
  title={Observation of the chiral-anomaly-induced negative magnetoresistance in 3D Weyl semimetal TaAs},
  author={Huang, Xiaochun and Zhao, Lingxiao and Long, Yujia and Wang, Peipei and Chen, Dong and Yang, Zhanhai and Liang, Hui and Xue, Mianqi and Weng, Hongming and Fang, Zhong and others},
  journal={Phys. Rev. X},
  volume={5},
  number={3},
  pages={031023},
  year={2015},
  publisher={APS}
}

@article{wang2016gate,
  title={Gate-tunable negative longitudinal magnetoresistance in the predicted type-II Weyl semimetal WTe2},
  author={Wang, Yaojia and Liu, Erfu and Liu, Huimei and Pan, Yiming and Zhang, Longqiang and Zeng, Junwen and Fu, Yajun and Wang, Miao and Xu, Kang and Huang, Zhong and others},
  journal={Nat. Commun..},
  volume={7},
  number={1},
  pages={13142},
  year={2016},
  publisher={Nature Publishing Group UK London}
}

@article{dos2016search,
  title={On the search for the chiral anomaly in Weyl semimetals: the negative longitudinal magnetoresistance},
  author={Dos Reis, RD and Ajeesh, MO and Kumar, N and Arnold, F and Shekhar, C and Naumann, M and Schmidt, M and Nicklas, M and Hassinger, E},
  journal={New J. Phys.},
  volume={18},
  number={8},
  pages={085006},
  year={2016},
  publisher={IOP Publishing}
}

@article{arnold2016negative,
  title={Negative magnetoresistance without well-defined chirality in the Weyl semimetal TaP},
  author={Arnold, Frank and Shekhar, Chandra and Wu, Shu-Chun and Sun, Yan and Dos Reis, Ricardo Donizeth and Kumar, Nitesh and Naumann, Marcel and Ajeesh, Mukkattu O and Schmidt, Marcus and Grushin, Adolfo G and others},
  journal={Nat. Commun..},
  volume={7},
  number={1},
  pages={11615},
  year={2016},
  publisher={Nature Publishing Group UK London}
}

@article{wu2018probing,
  title={Probing the chiral anomaly by planar Hall effect in Dirac semimetal Cd 3 As 2 nanoplates},
  author={Wu, Min and Zheng, Guolin and Chu, Weiwei and Liu, Yequn and Gao, Wenshuai and Zhang, Hongwei and Lu, Jianwei and Han, Yuyan and Zhou, Jianhui and Ning, Wei and others},
  journal={Phys. Rev. B},
  volume={98},
  number={16},
  pages={161110},
  year={2018},
  publisher={APS}
}

@article{deng2019quantum,
  title={Quantum oscillations of the positive longitudinal magnetoconductivity: A fingerprint for identifying Weyl semimetals},
  author={Deng, Ming-Xun and Qi, GY and Ma, R and Shen, R and Wang, Rui-Qiang and Sheng, L and Xing, DY},
  journal={Phys. Rev. Lett.},
  volume={122},
  number={3},
  pages={036601},
  year={2019},
  publisher={APS}
}

@article{nielsen1983adler,
  title={The Adler-Bell-Jackiw anomaly and Weyl fermions in a crystal},
  author={Nielsen, Holger Bech and Ninomiya, Masao},
  journal={Phys. Lett. B},
  volume={130},
  number={6},
  pages={389--396},
  year={1983},
  publisher={Elsevier}
}

@article{aji2012adler,
  title={Adler-Bell-Jackiw anomaly in Weyl semimetals: Application to pyrochlore iridates},
  author={Aji, Vivek},
  journal={Phys. Rev. B—Condensed Matter and Materials Physics},
  volume={85},
  number={24},
  pages={241101},
  year={2012},
  publisher={APS}
}

@article{parameswaran2014probing,
  title={Probing the chiral anomaly with nonlocal transport in three-dimensional topological semimetals},
  author={Parameswaran, SA and Grover, T and Abanin, DA and Pesin, DA and Vishwanath, A},
  journal={Phys. Rev. X},
  volume={4},
  number={3},
  pages={031035},
  year={2014},
  publisher={APS}
}

@article{burkov2015negative,
  title={Negative longitudinal magnetoresistance in Dirac and Weyl metals},
  author={Burkov, AA},
  journal={Phys. Rev. B},
  volume={91},
  number={24},
  pages={245157},
  year={2015},
  publisher={APS}
}

@article{ong2021experimental,
  title={Experimental signatures of the chiral anomaly in Dirac--Weyl semimetals},
  author={Ong, NP and Liang, Sihang},
  journal={Nat. Rev. Phys.},
  volume={3},
  number={6},
  pages={394--404},
  year={2021},
  publisher={Nature Publishing Group UK London}
}

@misc{ahmad2024nonlinearanomaloushalleffect,
      title={Nonlinear anomalous Hall effect in three-dimensional chiral fermions}, 
      author={Azaz Ahmad and Gautham Varma K. and Gargee Sharma},
      year={2024},
      eprint={2409.02985},
      archivePrefix={arXiv},
      primaryClass={cond-mat.mes-hall},
      url={https://arxiv.org/abs/2409.02985}, 
}

@article{ashcroft1976solid,
  title={Solid State Physics},
  author={Ashcroft, NW},
  journal={Thomson Learning},
  volume={39},
  year={1976}
}

@article{PhysRevB.103.245119,
  title = {Nonlinear transport in Weyl semimetals induced by Berry curvature dipole},
  author = {Zeng, Chuanchang and Nandy, Snehasish and Tewari, Sumanta},
  journal = {Phys. Rev. B},
  volume = {103},
  issue = {24},
  pages = {245119},
  numpages = {12},
  year = {2021},
  month = {Jun},
  publisher = {American Physical Society},
  doi = {10.1103/PhysRevB.103.245119},
  url = {https://link.aps.org/doi/10.1103/PhysRevB.103.245119}
}

@article{PhysRevB.97.041101,
  title = {Berry curvature dipole in Weyl semimetal materials: An ab initio study},
  author = {Zhang, Yang and Sun, Yan and Yan, Binghai},
  journal = {Phys. Rev. B},
  volume = {97},
  issue = {4},
  pages = {041101},
  numpages = {6},
  year = {2018},
  month = {Jan},
  publisher = {American Physical Society},
  doi = {10.1103/PhysRevB.97.041101},
  url = {https://link.aps.org/doi/10.1103/PhysRevB.97.041101}
}

@article{PhysRevLett.133.106701,
  title = {Quantum Geometry Induced Nonlinear Transport in Altermagnets},
  author = {Fang, Yuan and Cano, Jennifer and Ghorashi, Sayed Ali Akbar},
  journal = {Phys. Rev. Lett.},
  volume = {133},
  issue = {10},
  pages = {106701},
  numpages = {7},
  year = {2024},
  month = {Sep},
  publisher = {American Physical Society},
  doi = {10.1103/PhysRevLett.133.106701},
  url = {https://link.aps.org/doi/10.1103/PhysRevLett.133.106701}
}

@article{https://doi.org/10.1002/adfm.201900892,
author = {Zhang, Yun and Zhu, Wenkai and Hui, Fei and Lanza, Mario and Borca-Tasciuc, Theodorian and Muñoz Rojo, Miguel},
title = {A Review on Principles and Applications of Scanning Thermal Microscopy (SThM)},
journal = {Advanced Functional Materials},
volume = {30},
number = {18},
pages = {1900892},
doi = {https://doi.org/10.1002/adfm.201900892},
url = {https://advanced.onlinelibrary.wiley.com/doi/abs/10.1002/adfm.201900892},
eprint = {https://advanced.onlinelibrary.wiley.com/doi/pdf/10.1002/adfm.201900892},
year = {2020}
}

@article{10.1063/5.0091494,
    author = {Bodzenta, Jerzy and Kaźmierczak-Bałata, Anna},
    title = {Scanning thermal microscopy and its applications for quantitative thermal measurements},
    journal = {Journal of Applied Physics},
    volume = {132},
    number = {14},
    pages = {140902},
    year = {2022},
    month = {10},
    issn = {0021-8979},
    doi = {10.1063/5.0091494},
    url = {https://doi.org/10.1063/5.0091494},
    eprint = {https://pubs.aip.org/aip/jap/article-pdf/doi/10.1063/5.0091494/20035311/140902\_1\_5.0091494.pdf},
}

@book{calvet2016recent,
  title={Recent progress in microcalorimetry},
  author={Calvet, Edouard and Prat, Henri},
  year={2016},
  publisher={Elsevier}
}

@book{childs2015nanoscale,
  title={Nanoscale thermometry and temperature measurement},
  author={Childs, Peter},
  year={2015}
}

@article{sadat2012high,
  title={High resolution resistive thermometry for micro/nanoscale measurements},
  author={Meyhofer, E},
  journal={Review of Scientific Instruments},
  volume={83},
  number={8},
  year={2012},
  publisher={AIP Publishing}
}

@article{kang2019nonlinear,
  title={Nonlinear anomalous Hall effect in few-layer WTe2},
  author={Kang, Kaifei and Li, Tingxin and Sohn, Egon and Shan, Jie and Mak, Kin Fai},
  journal={Nature materials},
  volume={18},
  number={4},
  pages={324--328},
  year={2019},
  publisher={Nature Publishing Group UK London}
}

@article{tiwari2021giant,
  title={Giant c-axis nonlinear anomalous Hall effect in Td-MoTe2 and WTe2},
  author={Tiwari, Archana and Chen, Fangchu and Zhong, Shazhou and Drueke, Elizabeth and Koo, Jahyun and Kaczmarek, Austin and Xiao, Cong and Gao, Jingjing and Luo, Xuan and Niu, Qian and others},
  journal={Nature communications},
  volume={12},
  number={1},
  pages={2049},
  year={2021},
  publisher={Nature Publishing Group UK London}
}

@article{yasuda2017current,
  title={Current-nonlinear Hall effect and spin-orbit torque magnetization switching in a magnetic topological insulator},
  author={Yasuda, Kenji and Tsukazaki, Atsushi and Yoshimi, Ryutaro and Kondou, Kouta and Takahashi, KS and Otani, Y and Kawasaki, M and Tokura, Y},
  journal={Physical review letters},
  volume={119},
  number={13},
  pages={137204},
  year={2017},
  publisher={APS}
}

@article{nam2018probe,
  title={How to probe the spin contribution to momentum relaxation in topological insulators},
  author={Nam, Moon-Sun and Williams, Benjamin H and Chen, Yulin and Contera, Sonia and Yao, Shuhua and Lu, Minghui and Chen, Yan-Feng and Timco, Grigore A and Muryn, Christopher A and Winpenny, Richard EP and others},
  journal={Nature communications},
  volume={9},
  number={1},
  pages={56},
  year={2018},
  publisher={Nature Publishing Group UK London}
}

@inproceedings{cultrera2019calibration,
  title={Calibration of lock-in amplifiers in the low-frequency range},
  author={Cultrera, Alessandro and Tran, Ngoc Thanh Mai and D’Elia, Vincenzo and Ortolano, Massimo and Callegaro, Luca and others},
  booktitle={Proceedings of the 23rd IMEKO TC4 International Symposium Electrical \& Electronic Measurements Promote Industry},
  volume={4},
  pages={122--125},
  year={2019}
}
\end{document}